\documentclass[a4paper,11pt]{article}
\usepackage{graphicx,amssymb,amsmath,amsfonts,epsfig}
\usepackage[usenames,dvipsnames]{color}
\usepackage{subfigure}
\title{ Black Hole versus Naked Singularity via Axial Perturbation }
\author{Parthapratim Pradhan\footnote{pppradhan77@gmail.com}\\ 
\textit{Department of Physics}\\
\textit{Hiralal Mazumdar Memorial College For Women}\\
{Dakshineswar, Kolkata-700035, India}}
\date{today}

\begin{document}

\maketitle

\begin{abstract}
We differentiate non-extremal black hole, \emph{extremal} black hole  and \emph{naked singularity} via metric 
perturbations for Reissner-Nordstr\"{o}m spacetime. First we study the axial perturbations for \emph{extremal} 
Reissner-Nordstr\"{o}m black hole and compute the effective potential due to these perturbations. 
Then we study the axial perturbations for the naked singularity case and compute the effective potential. 
We show that for the non-extremal black hole, \emph{the effective potential outside the event horizon~($r_{+}$) is real and positive. While in between Cauchy horizon~($r_{-}$) and event horizon~($r_{-}<r<r_{+}$) 
the effective potential is negative.}  For the \emph{extremal black hole, the effective potential is always positive}.  Also for \emph{naked singularity, the effective potential  is positive.}
From the effective potential diagram, we show that the structure of effective potentials for extremal BH looks like a potential barrier outside the horizon. While for non-extremal BH, 
the structure of the effective potential looks like a \emph{potential well} rather than a 
potential barrier. For NS, the structure of the effective potentials is 
\emph{neither a potential barrier nor a potential well. Preferably it looks like an exponential decay function}. 
We observe that an effective potential barrier's geometric construction due to axial perturbations could distinguish between the non-extremal black hole,  extremal black hole, and naked singularity. The stability of the extremal BH has been discussed.
\end{abstract}

\vspace{2.5cm}
PACS numbers: 04.70.Bw \hspace{0.3cm} 04.25.Nx \hspace{0.3cm} 04.40.Nr
\newpage

\textheight 25 cm

\section{Introduction}
The fundamental difference between a black hole~(BH) and a naked singularity~(NS) is that the former object has a horizon structure while the latter object doesn't have any horizon structure. Again a BH can be categorized into two types: non-extremal BH and extremal BH. A non-extremal BH is characterized by a Hawking temperature, while extremal BH is characterized by \emph{zero} Hawking temperature. In the case of  NS, the Hawking temperature is completely undefined since it is horizonless. Now the question is how to differentiate a BH from an NS? In this work, we will try to understand the difference between BH~( non-extremal BH \& extremal BH) and NS via perturbative formalism. 
We have considered here only axial~(sometimes called odd-parity) perturbations.

The Reissner-Nordstr\"{o}m~(RN) solution is a spherically symmetric solution of the coupled 
Einstein and Maxwell system. It represents a BH having mass $M$ and charge $Q_{\ast}$. The fact that the RN
BH consist of two horizons. One is the event horizon or outer horizon, and the other one is the Cauchy horizon or inner
horizon. When two horizons are coincident, we will get extremal BH. In general relativity, a spherically symmetric charged BH solution can be obtained when $M^2>Q_{\ast}^2$. When $M^2=Q_{\ast}^2$, one gets an extremal BH solution. 
Finally, when $M^2<Q_{\ast}^2$, we get NS, which is visible to the external observer.

The perturbation formalism of non-extremal~($M^2>Q_{\ast}^2$) RN BH was first introduced
by Moncrief~\cite{mon74,mon74a,mon75} and then  Zerilli~\cite{zer74}.  The author in Ref.~\cite{mon74} 
used the Hamiltonian variational principle for computations of Perturbation equations, and the author in Ref.~\cite{zer74} decomposed the gravitational and electromagnetic field perturbation equations using vector and tensor harmonics. After that Chandrasekhar~\cite{sc79,sc79a} studied the same problem 
via metric perturbations. He derived one-dimensional wave equations that govern the RN BH's axial perturbations via the procedure of metric perturbations.
There are several related work we should mention here that have studied perturbations of RN BH in 
different formalism~\cite{sc,fro,nam78,nam78a,mat79,
sand79,sand79a,hart82,price72,bicak72,bur95,bur99,ori95,ori97,and98,his81,fer05,hod12}.

However, in the present work we shall \emph{focus} on  particularly the gravitational axial 
perturbations of $M^2=Q_{\ast}^2$ BH and  $Q_{\ast}^2>M^2$ spacetime. 
The case $M^2>Q_{\ast}^2$ is already been studied in Ref.~\cite{sc}. First we will study the 
gravitational axial perturbations separately for $M^2=Q_{\ast}^2$ BH. Then we will 
differentiate  $M^2>Q_{\ast}^2$ BH, $M^2=Q_{\ast}^2$ BH and $Q_{\ast}^2>M^2$ spacetime 
visually by plotting the effective potential diagram. 
The effective potential that we have used should be derived via metric perturbations. 
The formalism and procedure that we have used here could be found in Ref.~\cite{sc}.  In what follows, in the next section, we have described the basic formalism of metric perturbations for extremal RN BH. In  Sec.~(\ref{axx}), 
we have studied the \emph{axial} perturbations for extremal case. In  Sec.~(\ref{axn}), 
 we have examined the same for NS case. 
And in each case we have plotted the effective potential barrier diagram to differentiate 
$M^2>Q_{\ast}^2$ BH, $M^2=Q_{\ast}^2$ BH and $Q_{\ast}^2>M^2$. Finally, we have given the 
conclusions in Sec~\ref{con}.

Why is it important to study the perturbations of $M^2=Q_{\ast}^2$ BH   separately?
There are several good reasons to study them. \emph{First}, $M^2=Q_{\ast}^2$ BHs  are special class of BHs.
They are playing a key role in classical gravity as well as in quantum gravity. They can be treated as a theoretical toy in quantum gravity to explore the characteristics of quantum gravity. 

\emph{Second}, the $M^2=Q_{\ast}^2$ BHs are often called as supersymmetric. They are invariant under several supercharges. They saturate the BPS~(Bogomolny-Prasad-Sommerfield) bound. They are stable. 
Their BH entropy has been calculated in string theory. 
They have no Hawking temperature; therefore, they do not Hawking radiate. 

\emph{Third}, the fact that $M^2=Q_{\ast}^2$ BHs don't have any trapped surfaces. They don't have any bifurcation two
sphere~(${\cal S}^2$). The proper distance between the outer and inner horizons in the extremal RN
BH is infinite even though the vanishing coordinate distance. Typically, a BH is defined to be extreme if it has zero surface gravity, which ``measures the equilibrium temperature for the thermal distribution
of the radiation''~\cite{pla11}. Another exciting feature of $M^2=Q_{\ast}^2$ BH is that it has 
Couch-Torrence symmetry~\cite{couch84} that means it is conformally invariant under a spatial inversion while its non-extremal counterparts don't have such a feature.

Distinguishing the BH and NS have been started by a pioneer work of de Felice~\cite{defelice78}
(See also\cite{calvini78}) where the author showed the classical instability of an NS. More precisely, he observed for a Kerr BH  that there is ``a sharp discontinuity from positive to negative values when $a\rightarrow M$''. In fact, in the regime $m<a<1.008m$, there exists stable circular orbits with negative energy with respect to $\infty$. 
That indicates if accretion happens then, a Kerr naked singularity slows down to the BH state. 
This is one of the motivations behind the present work. Before going to axisymmetric spacetime, it is crucial to see what happens in the spherically spacetime? That's why we have considered extremal RN BH. Do we see any interesting feature perturbatively in the effective
potential when one has to be taken this condition $M^2\rightarrow Q_{\ast}^2$? This forms the basis of our work. 
There are several important work~\cite{pug11a,pug11b,cha17,cha17a}; we should mention here that has to be considered the difference between  NS and BH for various types of BH. 
Recently, in~\cite{raj19}, the authors differentiated BHs from NSs by using the images of accretion disks. In~\cite{pul07,dem99}, it was pointed out that NS is unstable under linearized perturbations. While in~\cite{dem99}, it was pointed out that NS is unstable in the context of the gravitational collapse of a scalar field.

We find that for non-extremal BH, the effective potential outside the event horizon is real and positive.  While in between Cauchy horizon and event horizon ($r_{-}<r<r_{+}$) 
the effective potential is negative.  For extremal BH, the effective potential is always positive.  
For NS, the effective potential is also positive. 
Alternatively, from the effective potential diagram, we can argue that the structure of effective potentials for extremal BH looks like a potential barrier outside the horizon and the potentials
 $V_{1}^{(-)}$ and $V_{2}^{(-)}$  are \emph{real and positive} everywhere outside the event horizon.      
For non-extremal BH, the structure of the effective potentials look like a 
\emph{potential well} rather than a potential barrier and potentials are \emph{negative} 
in the region $r_{-}<r<r_{+}$.  While for NS, the structure of the effective potentials is 
\emph{neither a potential barrier nor a potential well. Rather it looks like an exponential decay function}. These are the prime differences between non-extremal 
BH, extremal BH, and NS. Using these features, one could differentiate between non-extremal BH, extremal BH, and NS in RN geometry.

%The structure of the paper is as follows. In Sec.~(\ref{mt}), we describe the metric perturbations 
%of extremal RN spacetime. In Sec.~(\ref{axx}), we analyze the axial  perturbations 
%of extremal RN spacetime. In Sec.~(\ref{axn}), we discuss the result of axial perturbations 
%in NS case of the said BH. Finally, in Sec.~(\ref{con}), we have given the conclusions and 
%future outlook.

%\cite{pug11a,pug11b,pug13,pug19,pug20,cha17,cha17a}

\section{\label{mt} The metric perturbations in Extremal Reissner-Nordstr\"{o}m spacetime}
The metric perturbations of non-extremal Reissner-Nordstr\"{o}m BH have been studied
in detail by S. Chandrasekhar~\cite{sc79}. Here, we are interested to study the  perturbation of 
\emph{ extremal BH and NS} following the 
approach of S. Chandrasekhar~\cite{sc}. We have used the notation and convention throughout the work
follwing the Ref. ~\cite{sc}. We begin by writing the unperturbed extremal Reissner-Nordstr\"{o}m 
Spacetime as 
\begin{eqnarray}
\, ds^{2} &=& e^{2\nu}\left(~ dx^{0}\right)^{2}-e^{2\psi}\left(~ dx^{1} \right)
^{2}-e^{2\mu _{2}}\left(~ dx^{2}\right)^{2}-e^{2\mu _{3}}\left(dx^{3}\right)^{2} \nonumber\\
          &=& e^{2\nu}\left(~ dx^{0}\right)^{2}-e^{2\mu _{2}}\left(~ dx^{2}\right)^{2}-r^{2}\, d\Omega^{2},
          ~ \label{metric}
\end{eqnarray} 
where the co-ordinates are denoted as 
\begin{eqnarray}
x^{0} &\Longleftrightarrow & t \nonumber\\
x^{1} &\Longleftrightarrow & \varphi \nonumber\\
x^{2} &\Longleftrightarrow & r \nonumber\\
x^{3} &\Longleftrightarrow & \theta,~ \label{cod}
\end{eqnarray}
$d\Omega^2=e^{2\psi}\left(dx^{1} \right)^{2}+e^{2\mu _{3}}\left(dx^{3}\right)^{2}$ 
is the line element on the unit two sphere of the metric and the metric coefficients are 
\begin{eqnarray}
 e^{2\nu} &=& e^{-2\mu_{2}}=\left(1-\frac{M}{r} \right)^2 \equiv \frac{\Upsilon}{r^2}  \nonumber\\
 e^{\mu_{3}} &=& r \nonumber\\
 e^{\psi} &=& r \sin\theta \nonumber\\
 \omega &=& q_{2}=q_{3}=0 \nonumber\\
 \Upsilon &=& r^2 e^{2\nu}.~~ \label{def}
\end{eqnarray}
Here $r$ is Schwarzschild like radial coordinates and having circumference $2\pi r$. To study the perturbations 
of any spherically symmetric system one should restrict on axisymmetric modes of perturbations. The fact that 
the equations which govern the perturbations of a spherically symmetric spacetime could be separable in all four 
coordinates $t$, $r$, $\theta$ and $\varphi$. Since the extremal RN metric~(\ref{metric}) is static and 
spherically symmetric. Once it is perturbed, the spacetime becomes time dependent and axially symmetric. 
The most general time dependent and axially symmetric spacetime can be written as
\begin{eqnarray}
ds^{2} = e^{2\nu}\left(~dx^{0}\right)^{2}-e^{2\psi}\left(~dx^{1}-\omega~dx^{0}-q_{2}\, dx^{2}-q_{3}\, dx^{3}\right)^{2}
 -e^{2\mu_{2}}\left(~dx^{2}\right)^{2}-e^{2\mu_{3}} \left(~dx^{3}\right)^{2}~\label{metric1}
\end{eqnarray}
This is a perturbed extremal RN spacetime. 
The above spacetime  metric~(\ref{metric1}) contains seven functions: 
$\nu, \psi, \mu_{2}, \mu_{3}, \omega, q_{2}, q_{3}$. Again they are 
functions of $t, x^{2}$ and  $x^{3}$ only. Whereas the Einstein's equation contains only six independent equations for 
the metric components therefore it could not be possible to determine seven functions arbitrarily hence there is a 
constraint which is a possible combinations of six independent function. This constraint  of the metric 
coefficients becomes an indentity: 
$$
(\omega_{,2}-q_{2,0})_{,3}-(\omega_{,3}-q_{3,0})_{,2}+(q_{2,3}-q_{3,2})_{,0}.
$$
It should be noted that in unperturbed extremal RN metric~(\ref{metric}), the axial metric components 
$\omega= q_{2}=q_{3}=0$ while in perturbed extremal RN metric, these components are not equal to zero. 
This is the basic difference between perturbed spacetime and unperturbed spacetime. 

Note that a general perturbation of the extremal RN spacetime should result in $\omega, q_{2}$ and 
$q_{3}$ are small quantities of the first order. While the functions $\nu, \psi, \mu_{2}, \mu_{3}$ 
are having the small increments like $\delta\nu, \delta\psi, \delta\mu_{2}, \delta\mu_{3}$. There are two possible kinds of metric perturbations in extremal RN spacetime. One is called polar~(even parity) 
perturbation, and another one is called axial~(or odd parity) perturbation. This strictly depends upon the transformation of the parameter $\varphi\rightarrow -\varphi$ on the metric. Simply, the perturbation leading to non-vanishing values of  $\omega, q_{2}$ and $q_{3}$ are called axial perturbations. 
The perturbation leading to small increments of $\delta\nu, \delta\psi, \delta\mu_{2}, \delta\mu_{3}$ 
are called polar perturbations.

Now we shall linearize the Maxwell equations. This can be grouped into two sets of four equations. In first 
set, we write the four odd equations.  
\begin{eqnarray}
\left(\frac{r^2 \sin\theta}{\sqrt{\Upsilon}}~ F_{12} \right)_{,3} + \left( r^2 \sin\theta~ F_{31} \right)_{, 2} &=& 0\\
\left(\sin\theta \sqrt{\Upsilon}~ F_{01} \right)_{,2} + \left( \frac{r^2 \sin\theta}{\sqrt{\Upsilon}} ~F_{12} \right)_{, 0} &=& 0\\
\left(\sin\theta \sqrt{\Upsilon} ~F_{01} \right)_{,3} + \left( r^2 \sin\theta~ F_{13} \right)_{, 0} &=& 0
\end{eqnarray}
$$
\left(\frac{r^2}{\sqrt{\Upsilon}} ~ F_{01} \right)_{,0} + \left(\sqrt{\Upsilon}~ F_{12} \right)_{,2} 
+ {F_{13}}_{,3}
$$ 
\begin{eqnarray}
&=&  r^2 \sin\theta~ F_{02}\, Q_{02} + \frac{r^2 \sin\theta}{\sqrt{\Upsilon}} ~ F_{03}\, Q_{03} 
-\sin\theta \sqrt{\Upsilon}~F_{23}\, Q_{23} ~\label{maxodd}
\end{eqnarray}

In the second set, we write the four even equations. They are represented by 
\begin{eqnarray}
\left( r^2 \sin\theta\, F_{02} \right)_{,2} + \left(\frac{r^2 \sin\theta}{\sqrt{\Upsilon}}\, F_{03} \right)_{, 3} &=& 0 \\
-\left(\sin\theta\, \sqrt{\Upsilon}\, F_{23} \right)_{,2} + \left(\frac{r^2 \sin\theta}{\sqrt{\Upsilon}}\, F_{03}\right)_{, 0} 
&=& 0\\
\left(\sin\theta\, \sqrt{\Upsilon}\, F_{23} \right)_{,3} + \left( r^2 \sin\theta\, F_{02} \right)_{, 0} &=& 0 
\end{eqnarray}
$$
{F_{02}}_{,3} - \left( \sqrt{\Upsilon}\, F_{03} \right)_{, 2} + 
\left(\frac{r^2}{\sqrt{\Upsilon}}\, F_{23}  \right)_{,0}
$$
\begin{eqnarray}
&=&  \sin\theta \sqrt{\Upsilon}\, F_{01}~Q_{23} + \frac{r^2 \sin\theta}{\sqrt{\Upsilon}}~F_{12}\,Q_{03} - r^2 \sin\theta\, F_{13}\, Q_{02}
 ~\label{maxeven}
\end{eqnarray}
where the partial derivative of a function $f$ is given with the notation, 
\begin{eqnarray}
( f )_{,~\mu} &=&  \frac{\partial f}{\partial x^{\mu}}
\end{eqnarray}
The function $Q_{AB}$ are given by,
\begin{eqnarray}
Q_{AB} &=& \frac{\partial q_A}{\partial x^B} - \frac{\partial q_B}{\partial x^A}
\hspace{0.5cm}
\mbox{and}\hspace{0.5cm}
Q_{A0} = \frac{\partial q_A}{\partial x^0} - \frac{\partial \omega}{\partial x^A}
\hspace{0.5cm} ( A,B = 2,3)
\end{eqnarray}
It should be noted that the first set of equations~(\ref{maxodd}) having the quantities which reverse 
their signs when $\varphi$ is replaced by -$\varphi$. Whereas the second set of equations~(\ref{maxeven}) 
having the quantities which are invariant to the reversal in the sign of $\varphi$. That's why we call them 
odd or axial quantities and even or polar quantities. 

Further it should be noted that $F_{02}=-\frac{M}{r^2}$ is the only non-vanishing components of the Maxwell 
tensor for the unperturbed extremal metric, therefore the linearized version of equations~(\ref{maxodd}) and 
(\ref{maxeven}) which governs the perturbations are 
\begin{eqnarray}
 \left( \sqrt{\Upsilon}~ F_{01} \sin \theta \right)_{,r} +   \frac{r^2}{ \sqrt{\Upsilon}}~ F_{12,0} \sin \theta &=& 0\\
\sqrt{\Upsilon} \left( F_{01} \sin \theta \right)_{, \theta} + r^2\, F_{13,0} \sin \theta &=& 0\\
 \frac{r^2}{ \sqrt{\Upsilon}}\, F_{01,0} + \left( \sqrt{\Upsilon} F_{12} \right)_{,r} + F_{13,\theta} &=& - M ( \omega_{,2} - q_{2,0})\sin \theta \\
 \frac{r^2}{ \sqrt{\Upsilon}} \,F_{03,0} &=& \left( \sqrt{\Upsilon} F_{23} \right)_{,r} \\
\delta F_{02,0} - \frac{M}{r^2} \left( \delta \psi + \delta \mu_3 \right)_{,0} + \frac{e^{\nu}}{r \sin \theta } 
\left( F_{23} \sin \theta \right)_{, \theta} &=& 0\\
\left[ \delta F_{02} - \frac{M}{r^2} \left( \delta \nu + \delta \mu_2 \right) \right]_{,\theta} 
+  ( \sqrt{\Upsilon} F_{ 30} )_{,r} +  \frac{r^2}{ \sqrt{\Upsilon}}\, F_{23, 0} &=& 0  ~\label{maxp}
\end{eqnarray}
Since the perturbed components of Ricci tensor of extremal RN metric can not be zero like 
Schwarzschild BH therefore these components can be computed from the following identity:
\begin{eqnarray}
\delta R_{(a)(b)} &=& -2 \left[\eta^{(n)(m)} \left\{\delta F_{(a)(n)}F_{(b)(m)} +
 F_{(a)(n)}\delta_{(b)(m)}\right\}-\eta_{(a)(b)}\, M\, \frac{\delta F_{(0)(2)}}{r^2} \right]. ~\label{maxxp}
\end{eqnarray}
From this identity, we will get 
\begin{eqnarray}
\delta R_{(0)(0)} &=& \delta R_{(1)(1)}= - \delta R_{(2)(2)} =\delta R_{(3)(3)}=- 2\frac{M}{r^2} \delta F_{(0)(2)}\nonumber\\
\delta R_{(0)(1)} &=& - 2\frac{M}{r^2} \delta F_{(1)(2)},\,\, \delta R_{(0)(3)}= + 2\frac{M}{r^2} \delta F_{(2)(3)}\nonumber\\
\delta R_{(1)(2)} &=& + 2\frac{M}{r^2} \delta F_{(0)(1)},\,\, \delta R_{(2)(3)}= + 2\frac{M}{r^2} \delta F_{(0)(3)}\nonumber\\
\delta R_{(0)(2)} &=& \delta R_{(1)(3)}= 0 
\end{eqnarray}
Here,  $Z_{(\alpha)(\beta)}$ is the $(\alpha)(\beta)$ tetrad component of the tensor $Z$. 
Also, $R$ and $F$ are denoted as the Ricci and Maxwell tensors, and a comma denotes 
partial differentiation.

\section{\label{axx} Axial  perturbations in Extremal Reissner-Nordstr\"{o}m Spacetime}
First we will treat the axial perturbations following the linerization of coupled Einstein-Maxwell equations 
around the extremal RN BH. Now putting the unperturbed metric coefficients in the  $\varphi~r$ and $\varphi~\theta$ 
components of the Ricci tensor, we now get
\begin{eqnarray}
\left[\Upsilon (q_{2,3}-q_{3,2})\sin^{3}\theta\right]_{,3}+r^{4}(\omega_{,2}-q_{2,0})_{,0}\, \sin^{3}\theta
&=& 2r^{2} \sqrt{\Upsilon}\, \sin^{2}\theta \,\delta R_{(1)(2)}\nonumber\\
&=& 4M \sqrt{\Upsilon} F_{(0)(1)}\,\sin^{2}\theta, ~\label{ric}
\end{eqnarray}
and 
\begin{eqnarray}
\left[\Upsilon \,(q_{2,3}-q_{3,2})\sin^{3}\theta\right]_{,2}-\frac{r^4}{\Upsilon}\,(\omega_{,3}-q_{3,0})_{,0}\sin^{3}\theta 
&=& -2r^{2} \sin^{2}\theta \,\delta R_{(1)(3)}\nonumber\\
&=& 0, 
\label{ric1}
\end{eqnarray}
Now the linearized components of the Maxwell tensors are
\begin{eqnarray}
\left[\sqrt{\Upsilon}  \, F_{(0)(1)} \right]_{,2}+ \frac{r^2}{ \sqrt{\Upsilon}} \,F_{(1)(2),0} &=& 0, \label{xmaxw1}\\
\sqrt{\Upsilon}\, \left[F_{(0)(1)}\sin\theta\right]_{,3}+r^{2} F_{(1)(3),0}\, \sin\theta &=& 0, \label{xmaxw2} \\
 \frac{r^2}{ \sqrt{\Upsilon}} \,F_{(0)(1),0} + \left[\sqrt{\Upsilon}\, F_{(1)(2)}\right]_{,2}+F_{(1)(3),3}  
&=& -M (\omega_{2}-q_{2,0})\sin\theta.~\label{xmaxw3}
\end{eqnarray}
Now introducing the functions 
\begin{eqnarray}
{\cal B}(r, \theta) &=& F_{(0)(1)}\, \sin\theta,~\label{xb}\\
{\cal Q}(r, \theta) &=& \Upsilon (q_{2,3}-q_{3,2})\sin^{3}\theta.~\label{xq}
\end{eqnarray}
Putting these functions in Eq.~(\ref{ric}) and Eq.~(\ref{ric1}) and using the linearized Maxwell 
equations~[\ref{xmaxw1},\ref{xmaxw2}, \ref{xmaxw3}], we get the differential equations 
\begin{eqnarray}
\frac{1}{r^{4}}\frac{1}{\sin^{3}\theta}\frac{\,\partial {\cal Q}(r, \theta)}{\,\partial \theta} &=& 
-(\omega_{,2}-q_{2,0})_{,0}+4M \, \frac{\sqrt{\Upsilon}}{r^{4}}  \frac{{\cal B}(r, \theta)}{\sin^{2}\theta} , ~ \label{xdif1}
\end{eqnarray} 
and 
\begin{eqnarray}
\frac{\Upsilon}{r^{4}}\frac{1}{\sin^{3}\theta}\frac{\,\partial {\cal Q}(r, \theta)}{\,\partial r} &=& 
(\omega_{,3}-q_{3,0})_{,0}, ~\label{xdif2}
\end{eqnarray}
Assuming the perturbations have the time dependence $e^{i\sigma t}$ therefore we obtain the 
following equations 

\begin{eqnarray}
\frac{1}{r^{4}} \frac{1}{\sin^{3}\theta}\frac{\,\partial {\cal Q}(r, \theta)}{\,\partial \theta} &=&
-i\sigma\omega_{,2}-\sigma^{2}q_{2}+4M \,\frac{\sqrt{\Upsilon}}{r^{4}} \frac{{\cal B}(r, \theta)}{\sin^{2}\theta}, ~\label{xdif3}
\end{eqnarray}
and
\begin{eqnarray}
\frac{\Upsilon}{r^{4}} \frac{1}{\sin^{3}\theta} \frac{\,\partial {\cal Q}(r, \theta)}{\,\partial r} &=& 
i \sigma\omega_{,3}+\sigma^{2} q_{3}, ~\label{xdif4}
\end{eqnarray}
Differentiating Eq.~(\ref{xdif3}) with respect to $\theta$ and Eq.~(\ref{xdif4}) with respect to 
$r$, and summing the equations we get 
\begin{eqnarray}
r^{4}\frac{\partial}{\,\partial r}\left(\frac{\Upsilon}{r^{2}}
\frac{\,\partial {\cal Q}(r, \theta)}{\,\partial r}\right)&+&\sin^{3}\theta\,
\frac{\partial}{\,\partial\theta}\left(\frac{1}{\sin^{3}\theta}
\frac{\,\partial {\cal Q}(r, \theta)}{\,\partial\theta}\right)+\sigma^{2}\frac{r^{4}}
{\Upsilon}\, {\cal Q}(r, \theta) \nonumber\\
&=& 4M\, \sin^{3}\theta\, \sqrt{\Upsilon}\, \frac{\partial}{\,\partial\theta}
\left(\frac{{\cal B}(r, \theta)}{\sin^{2}\theta}\right). ~\label{xeq1}
\end{eqnarray} 
Again, differentiate Eq.~(\ref{xmaxw1}) with respect to $r$, Eq.~(\ref{xmaxw2}) with respect to $\theta$, 
and Eq.~(\ref{xmaxw3}) with respect to $t$. Putting these  in the latter  equation and using Eq. (\ref{xb}),
one obtains
\begin{eqnarray}
\left[\frac{\Upsilon}{r^2}\,\left\{\sqrt{\Upsilon} \,{\cal B}(r, \theta) \right\}_{,2}\right]_{,2}+\frac{\sqrt{\Upsilon}}{r^2}
\left(\frac{{\cal B}(r, \theta)_{,3}}{\sin\theta}\right)_{,3}\sin\theta-\frac{r^2}{\sqrt{\Upsilon}}\, {\cal B}(r, \theta)_{,0,0}
&=& M \left(\omega_{,2,0}-q_{2,0,0} \right)\, \sin^{2}\theta.~\label{xeq2}
\end{eqnarray}
Thus Eqs.~(\ref{xeq1}) and (\ref{xeq2}) govern  the axial perturbations. 

Next, the variables $r$ and $\theta$ in Eqs.~(\ref{xeq1}) and (\ref{xeq2}) may be separated 
by the following substituations~\cite{sc} 
\begin{eqnarray}
{\cal B}(r, \theta) &=& 3 {\cal B}(r) C^{-1/2}_{l+1}(\cos\theta), ~\label{brx}
\end{eqnarray}
and 
\begin{eqnarray}
{\cal Q}(r, \theta) &=& {\cal Q}(r) C^{-3/2}_{l+2}(\cos\theta),~\label{qrx}
\end{eqnarray} 
where $C^{j}_{i}$ denotes the Gegenbauer function of order $i$ and index 
$j$~\cite{stegun}. It should be noted  that the relations given in Ref.~ 
\cite{sc} between the Gegenbauer functions and the Legendre functions 
are valid only for $l\ge 2$~\cite{bur99}.  The valid expression for the 
$i=3$ and $j=-3/2$ Gegenbauer function is 
$$
C^{-3/2}_{3}(\cos\theta)=\frac{1}{2}(\cos^{3}\theta-3\cos\theta)
$$
The Gegenbauer functions may be computed directly for all orders and indices from the Ref~\cite{stegun}
\begin{eqnarray}
C^{j}_{i}(\cos\theta) &=& \sum_{k=0}^{i}\frac{\Gamma(j+k)\Gamma
(j+i-k)}{k!(i-k)![\Gamma(j)]^{2}}\cos(i-2k)\theta.
\end{eqnarray}
Putting Eqs.~(\ref{brx}) and (\ref{qrx}) in Eqs.~(\ref{xeq1}) and 
(\ref{xeq2}) we get the radial equations
\begin{eqnarray}
\Upsilon\, \frac{d}{\,dr}\left(\frac{\Upsilon}{r^{4}}\frac{\,d{\cal Q}(r)}{\,dr}\right)
-2n \frac{\Upsilon}{r^{4}}\,{\cal Q}(r)+\sigma^{2}\,{\cal Q}(r) &=& -8nM\, \frac{\Upsilon^{\frac{3}{2}}}{r^{3}}\, {\cal B}(r)
\label{xeq3}
\end{eqnarray}
and 
\begin{eqnarray}
\frac{d}{\,dr}\left[\frac{\Upsilon}{r^2}\, \frac{d}{\,dr}(\sqrt{\Upsilon}\,{\cal B}(r))\right]
-2(n+1)\frac{\sqrt{\Upsilon}}{r^2} \,{\cal B}(r)+\left(\sigma^{2}\frac{r^2}{\sqrt{\Upsilon}}
-4M^{2}\frac{\sqrt{\Upsilon}}{r^{4}}\right) {\cal B}(r) &=&
-M \frac{{\cal Q}(r)}{r^{4}} .
\label{xeq4}
\end{eqnarray}
where 
$$
n=\frac{(l-1)(l+2)}{2}
$$
Here, $n$ is the principal quantum number, and $l$ is an orbital quantum number. 

With the further substituations 
\begin{eqnarray}
H_{1}^{(-)} &=& -2 \sqrt{2n}\,\sqrt{\Upsilon}{\cal B}(r), ~\label{xh1}
\end{eqnarray}
and
\begin{eqnarray}
H_{2}^{(-)} &=& \frac{1}{r}{\cal Q}(r), ~\label{xh2}
\end{eqnarray}
and changing  to the Regge-Wheeler ``tortoise'' coordinate $r_{*}$ 
defined by 
$$
\frac{d}{dr_{*}} = \frac{\Upsilon}{r^2} \frac{d}{dr}
$$
with these definitions, we get a pair of coupled  differential equations 
of second order
\begin{eqnarray}
\Lambda^{2} {H}_{1}^{(-)} &=& \frac{\Upsilon}{r^{5}}\left[\left\{2(n+1)r-3M+4\frac{M^2}{r}\right\}{H}_{1}^{(-)}+
3M {H}_{1}^{(-)}+2M \sqrt{2n} {H}_{2}^{(-)}\right], ~\label{xeq5}
\end{eqnarray} 
and
\begin{eqnarray}
\Lambda^{2} {H}_{2}^{(-)} &=& \frac{\Upsilon}{r^{5}}\left[\left\{2(n+1)r
-3M+4\frac{M^2}{r}\right\}{H}_{2}^{(-)}-3M {H}_{2}^{(-)}+2M\sqrt{2n} \bar{H}_{1}^{(-)}\right].
\label{xeq6}
\end{eqnarray}
where, $\Lambda^{2}=\frac{d^{2}}{dr_{*}^{2}}+\sigma^{2}$. 

\begin{figure}
\begin{center}
\subfigure[]{
\includegraphics[width=2in,angle=0]{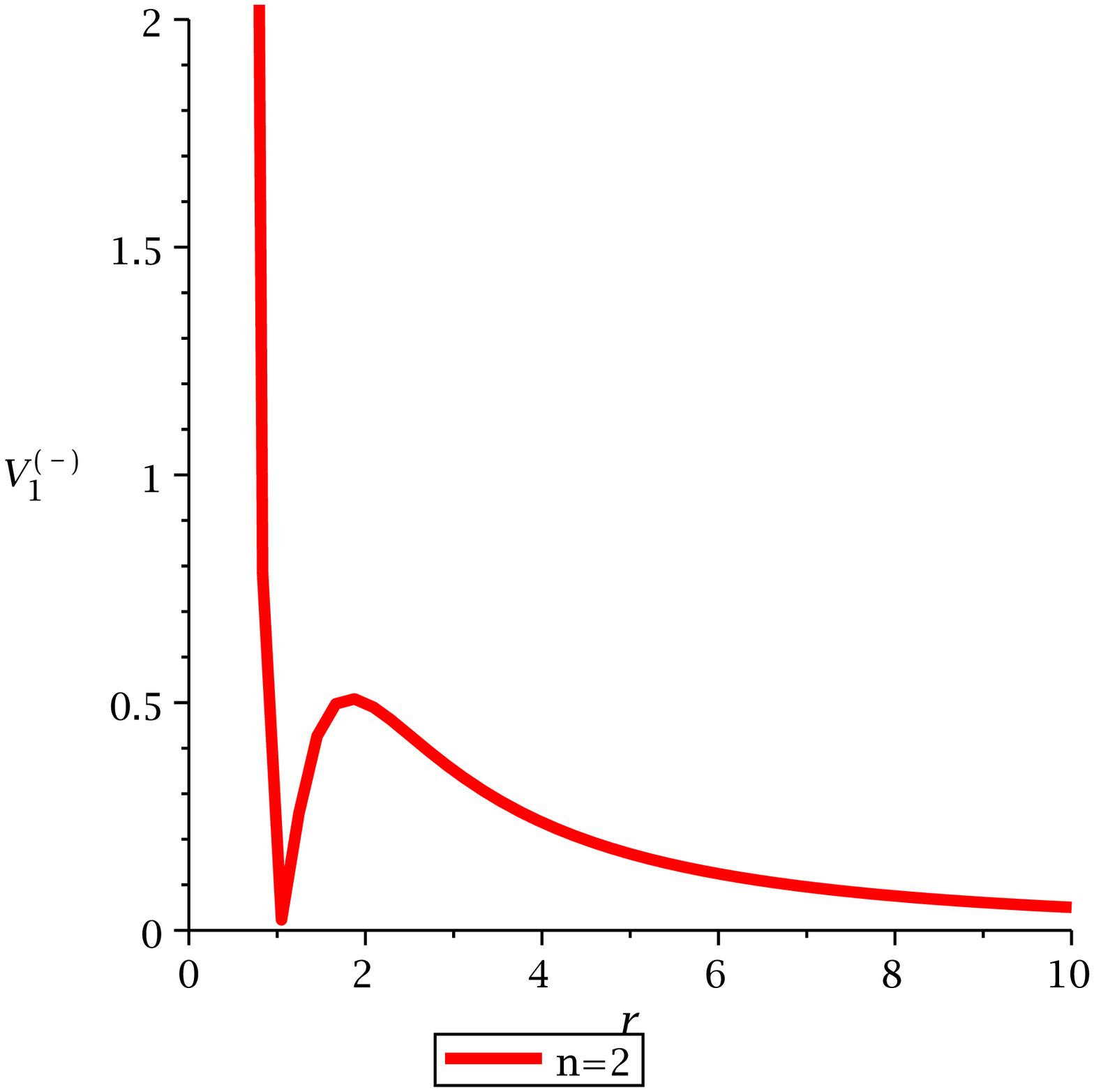}} 
\subfigure[]{
\includegraphics[width=2in,angle=0]{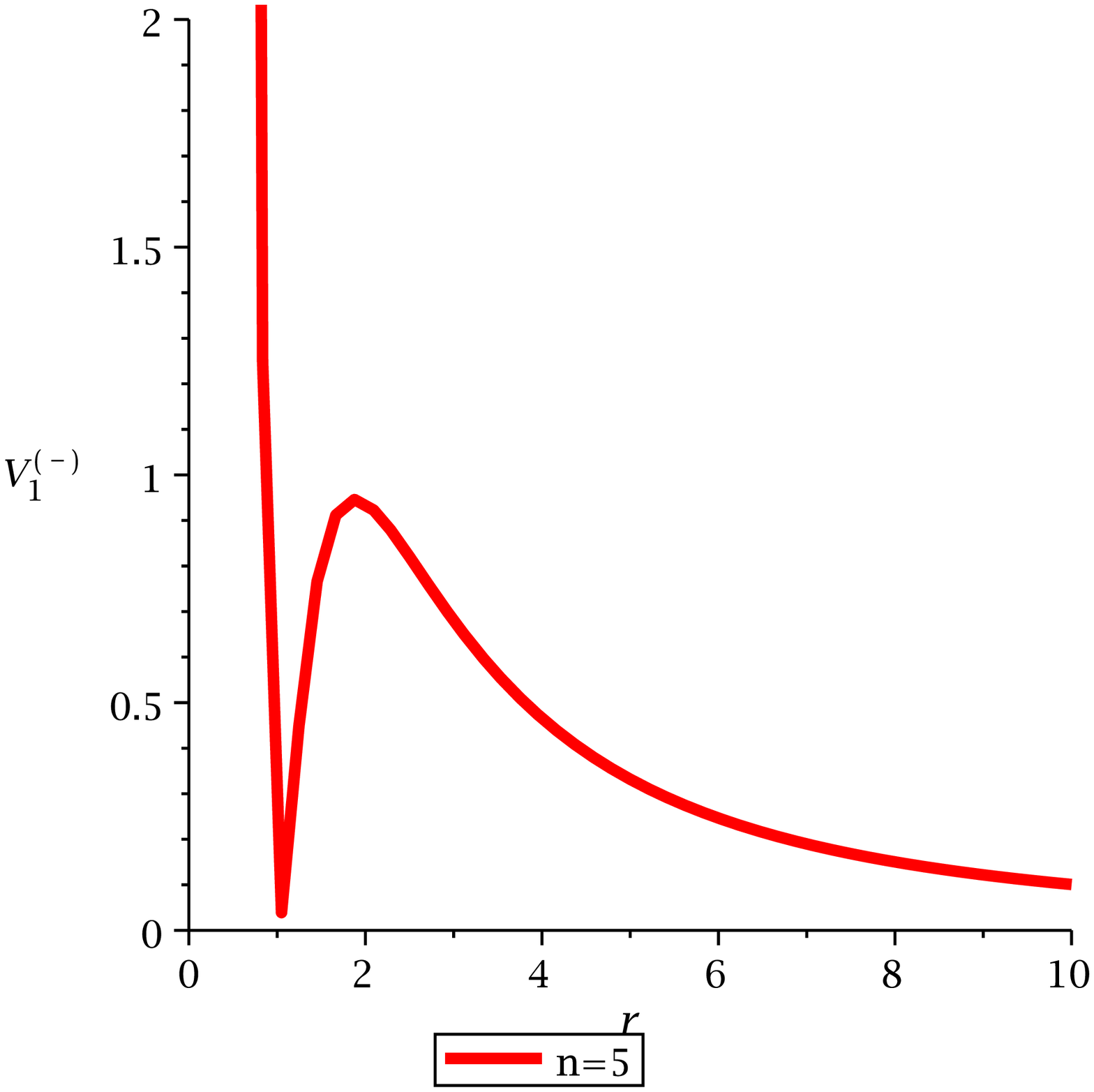}} 
\subfigure[]{
\includegraphics[width=2in,angle=0]{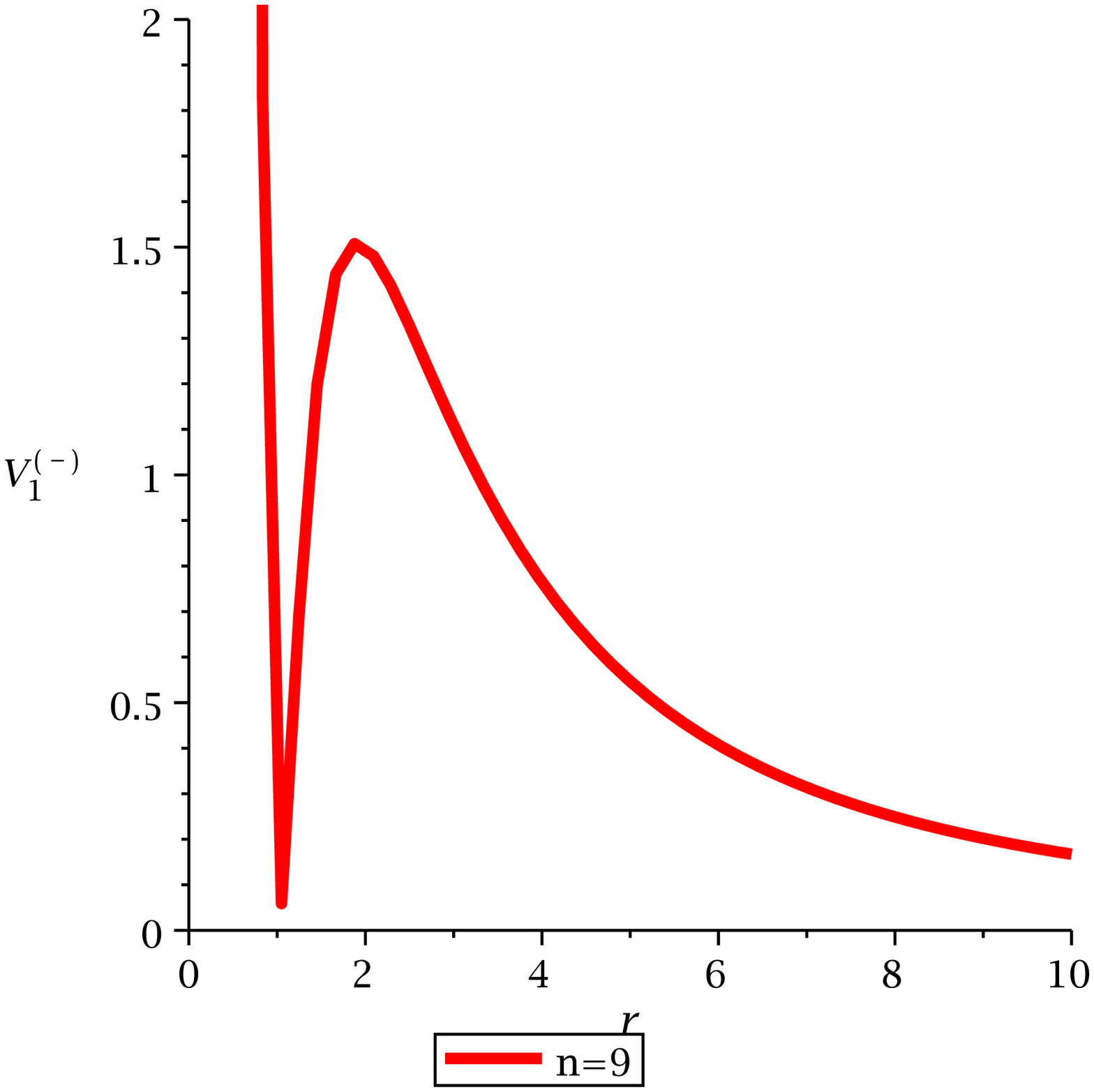}} 
\subfigure[]{
\includegraphics[width=2in,angle=0]{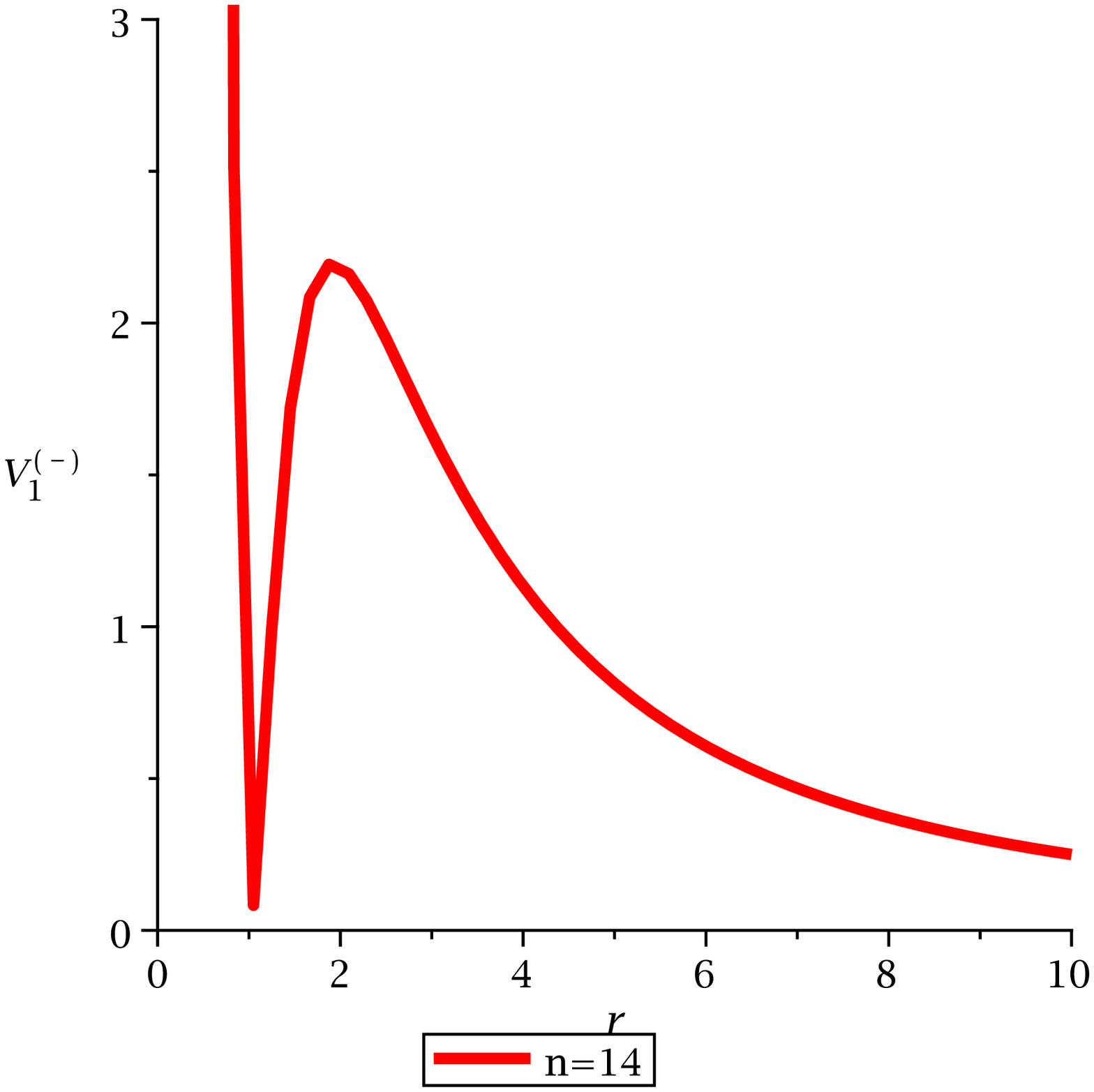}} 
\subfigure[]{
\includegraphics[width=2in,angle=0]{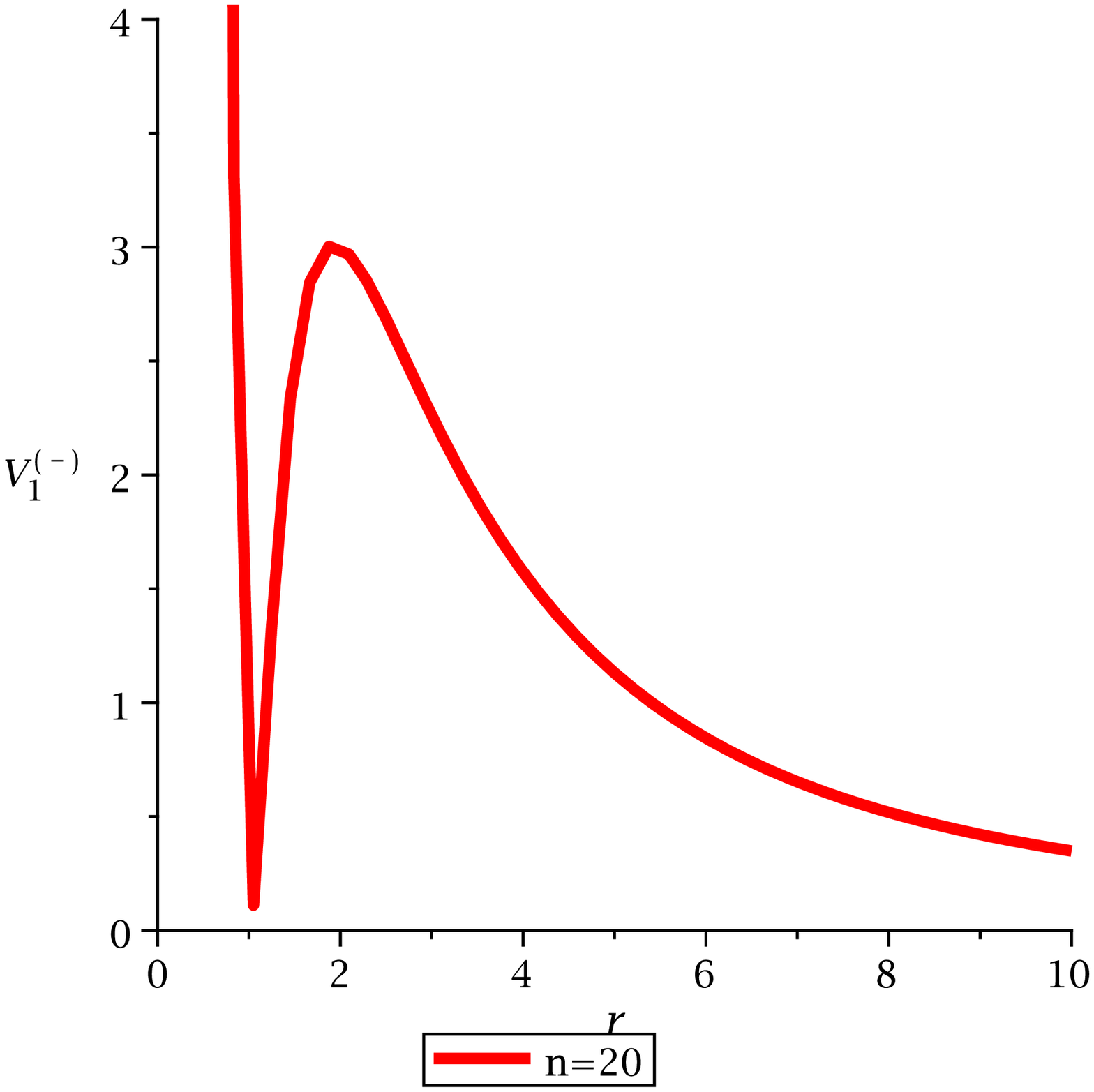}} 
\subfigure[]{
\includegraphics[width=2in,angle=0]{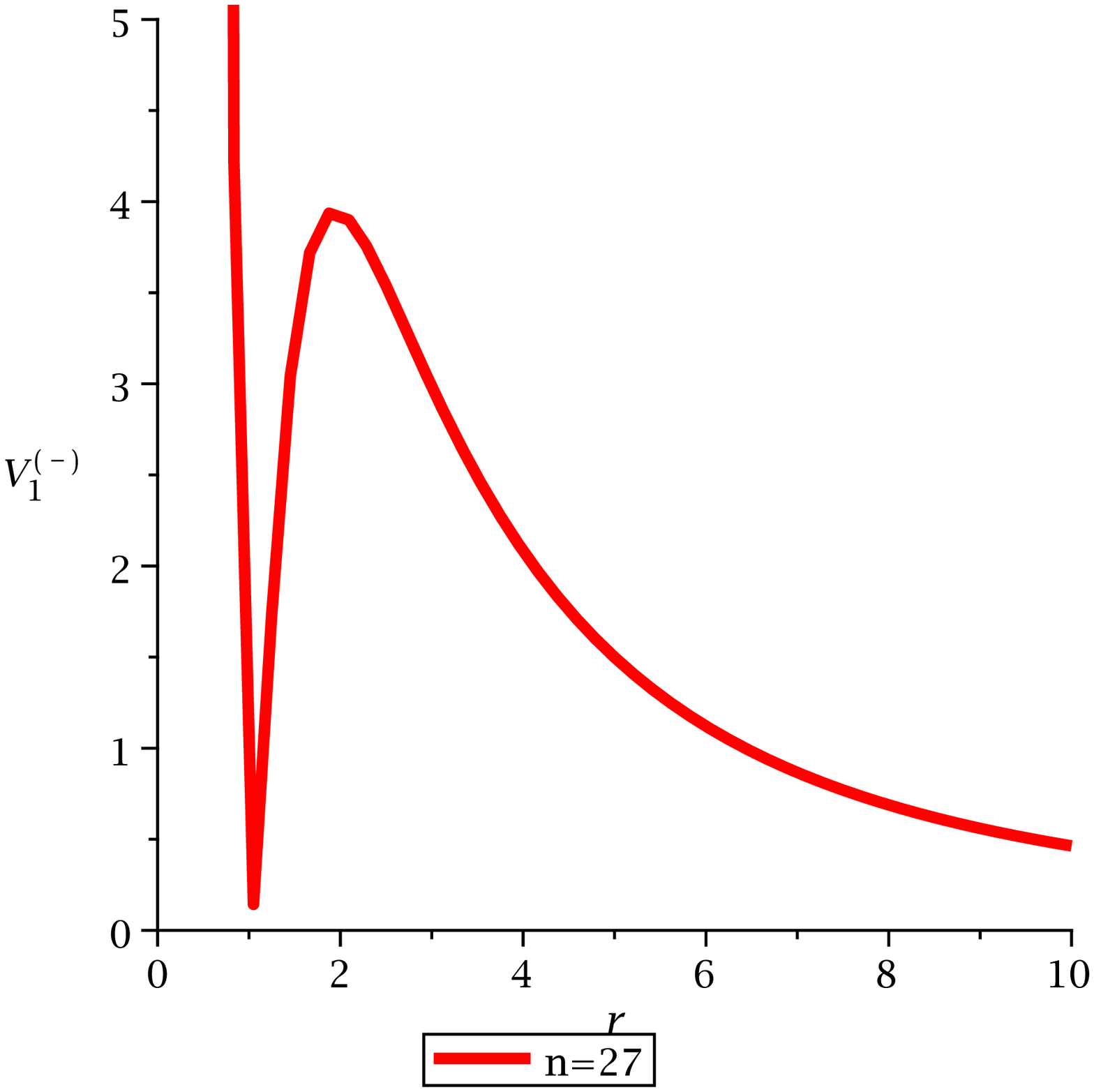}} 
\subfigure[]{
\includegraphics[width=2in,angle=0]{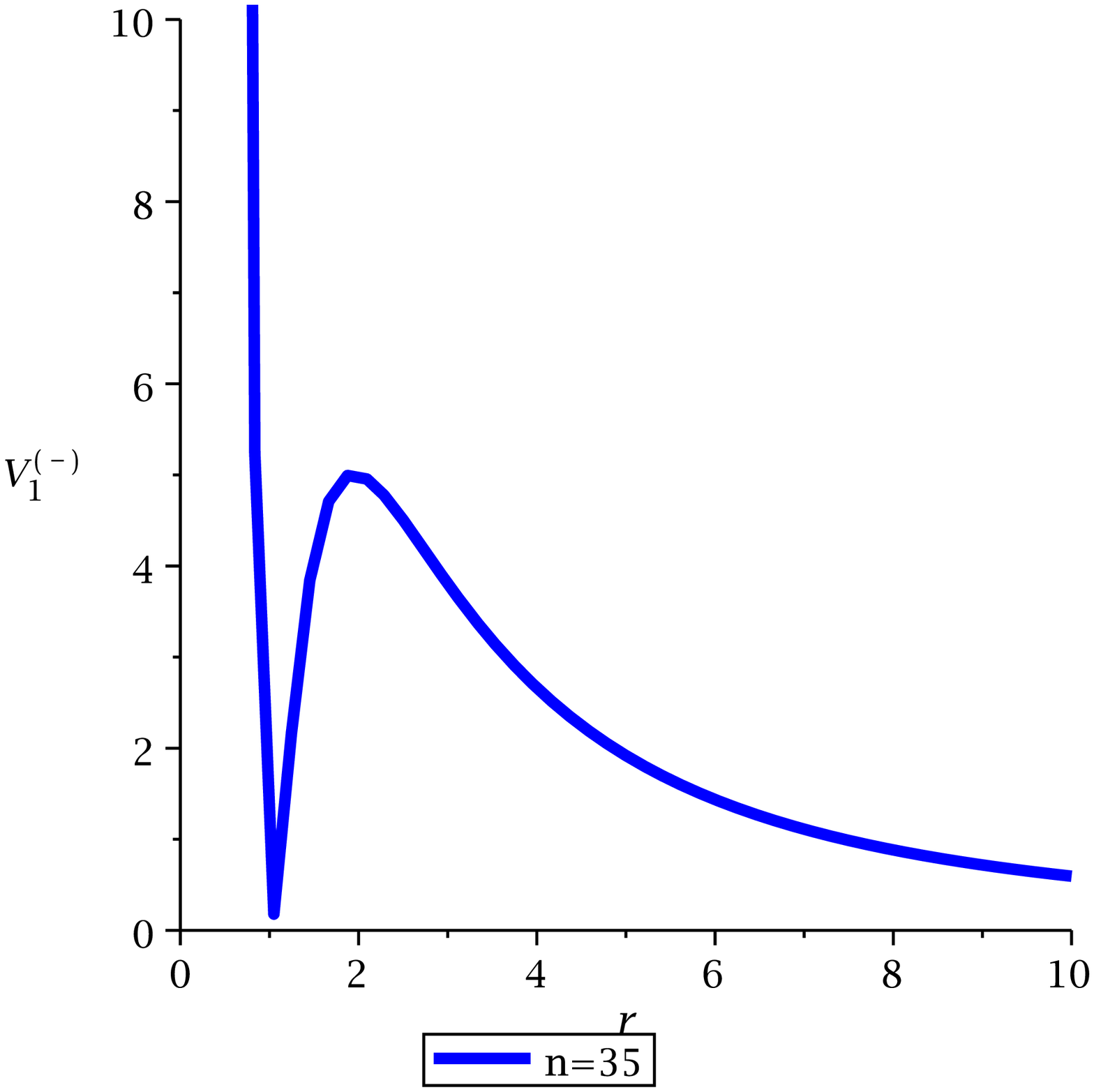}} 
\subfigure[]{
\includegraphics[width=2in,angle=0]{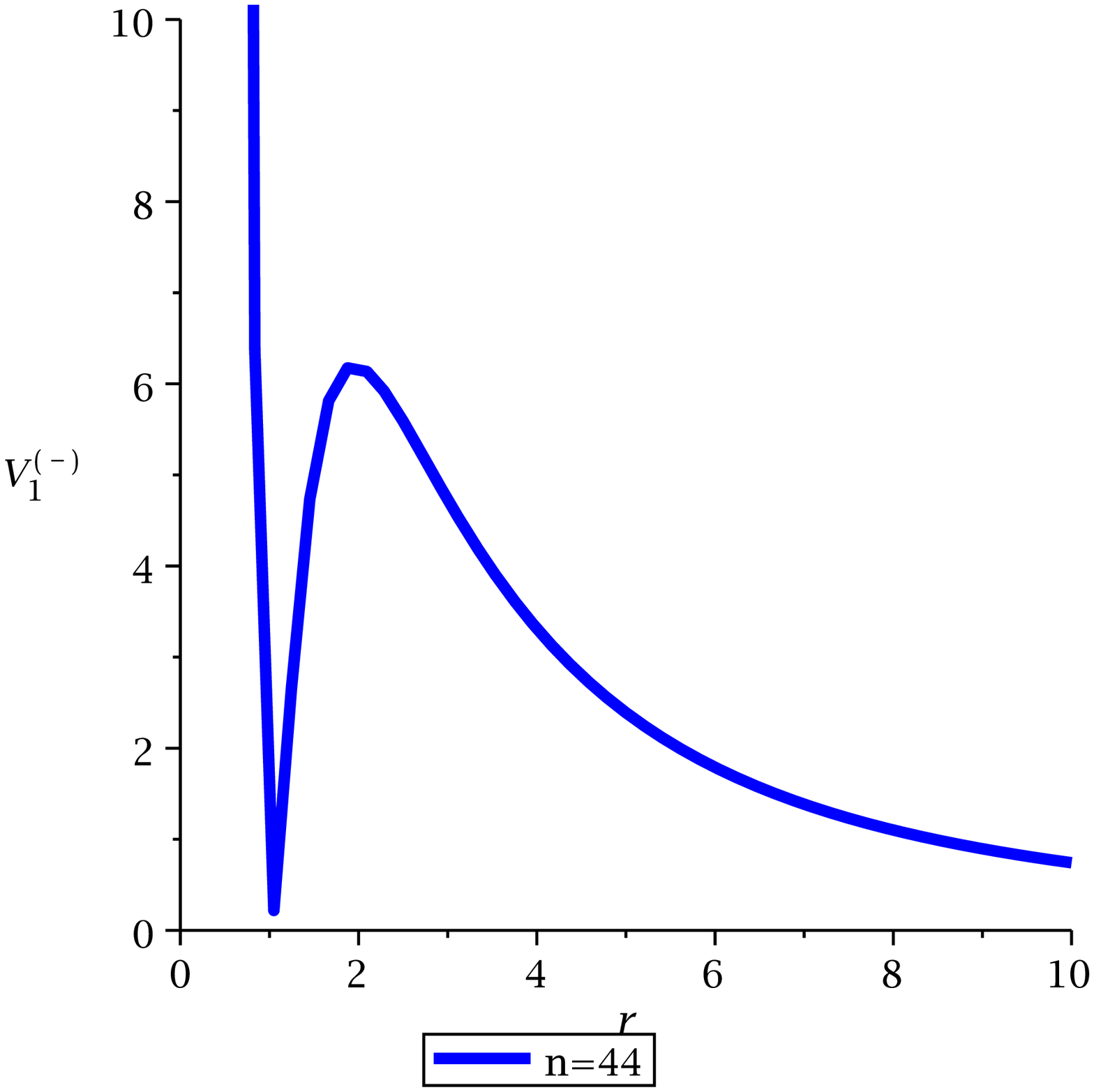}} 
\subfigure[]{
\includegraphics[width=2in,angle=0]{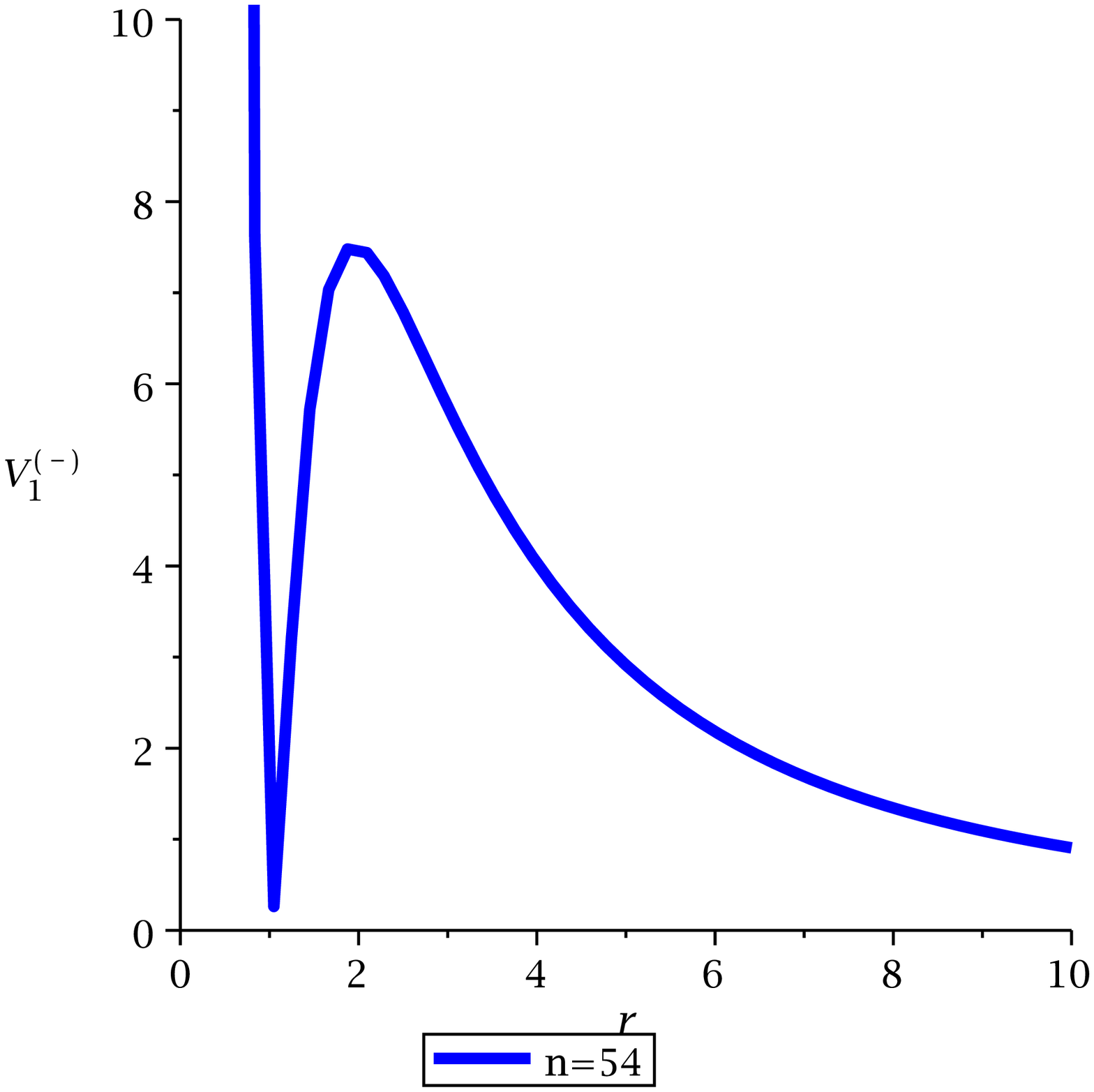}} 
\subfigure[]{
\includegraphics[width=2in,angle=0]{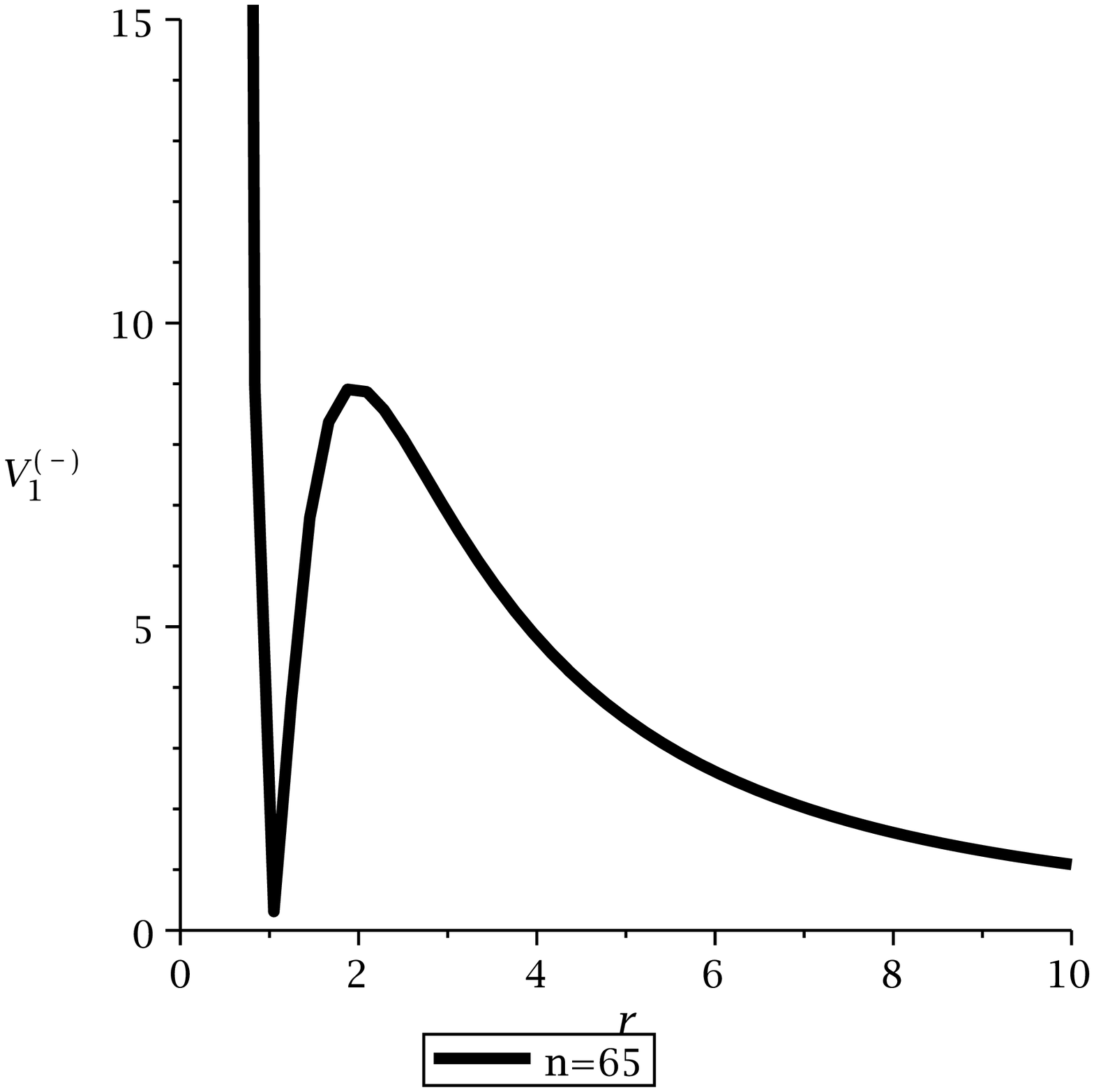}} 
\subfigure[]{
\includegraphics[width=2in,angle=0]{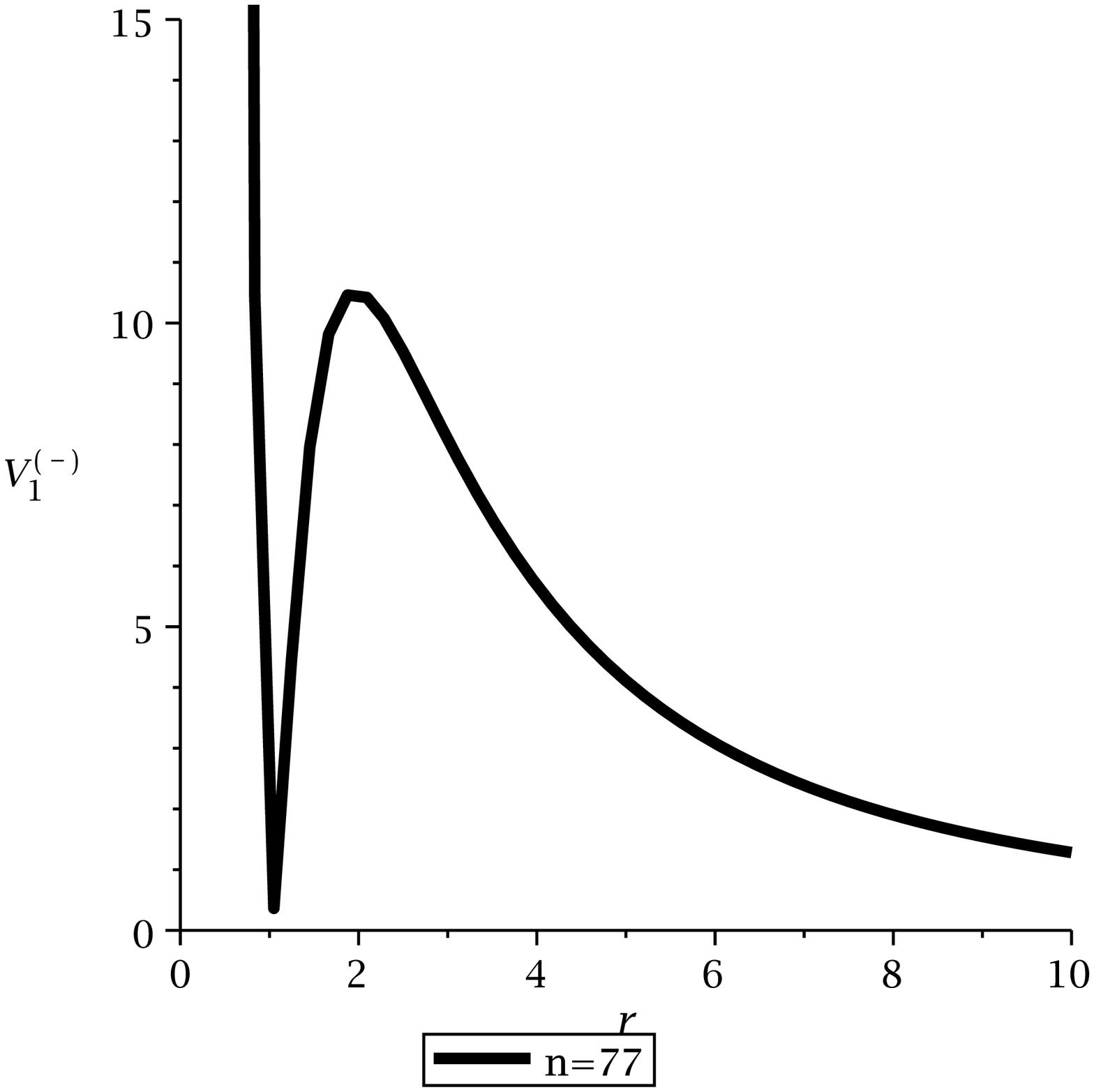}} 
\subfigure[]{
\includegraphics[width=2in,angle=0]{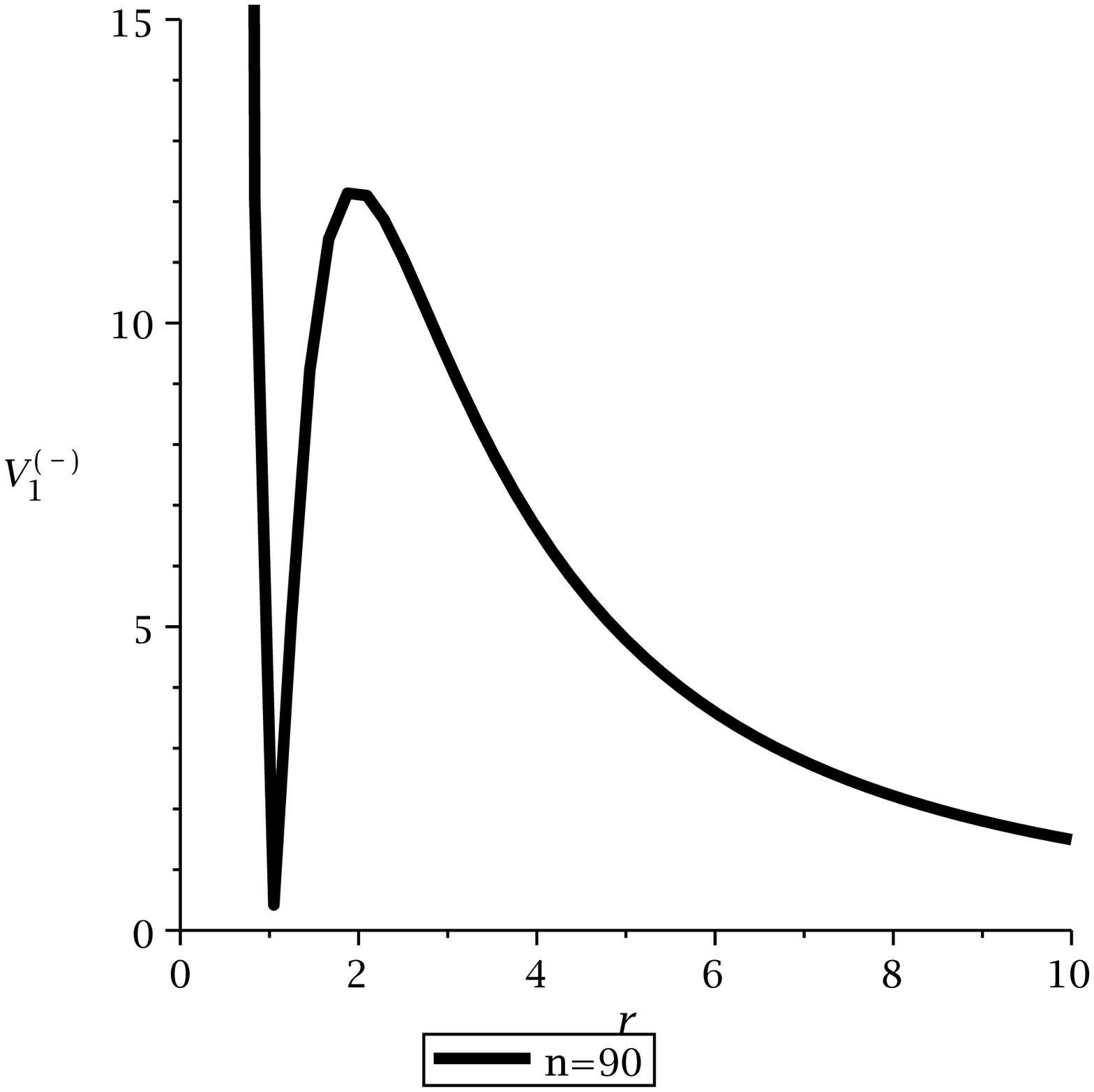}} 
\caption{The structure of  effective potential~($V_{1}^{(-)}$) barriers 
surrounding the extremal Reissner Nordstr\"{o}m  BH for axial perturbations.}
\label{axialx} 
\end{center}
\end{figure}

\begin{figure}
\begin{center}
\subfigure[]{
\includegraphics[width=2in,angle=0]{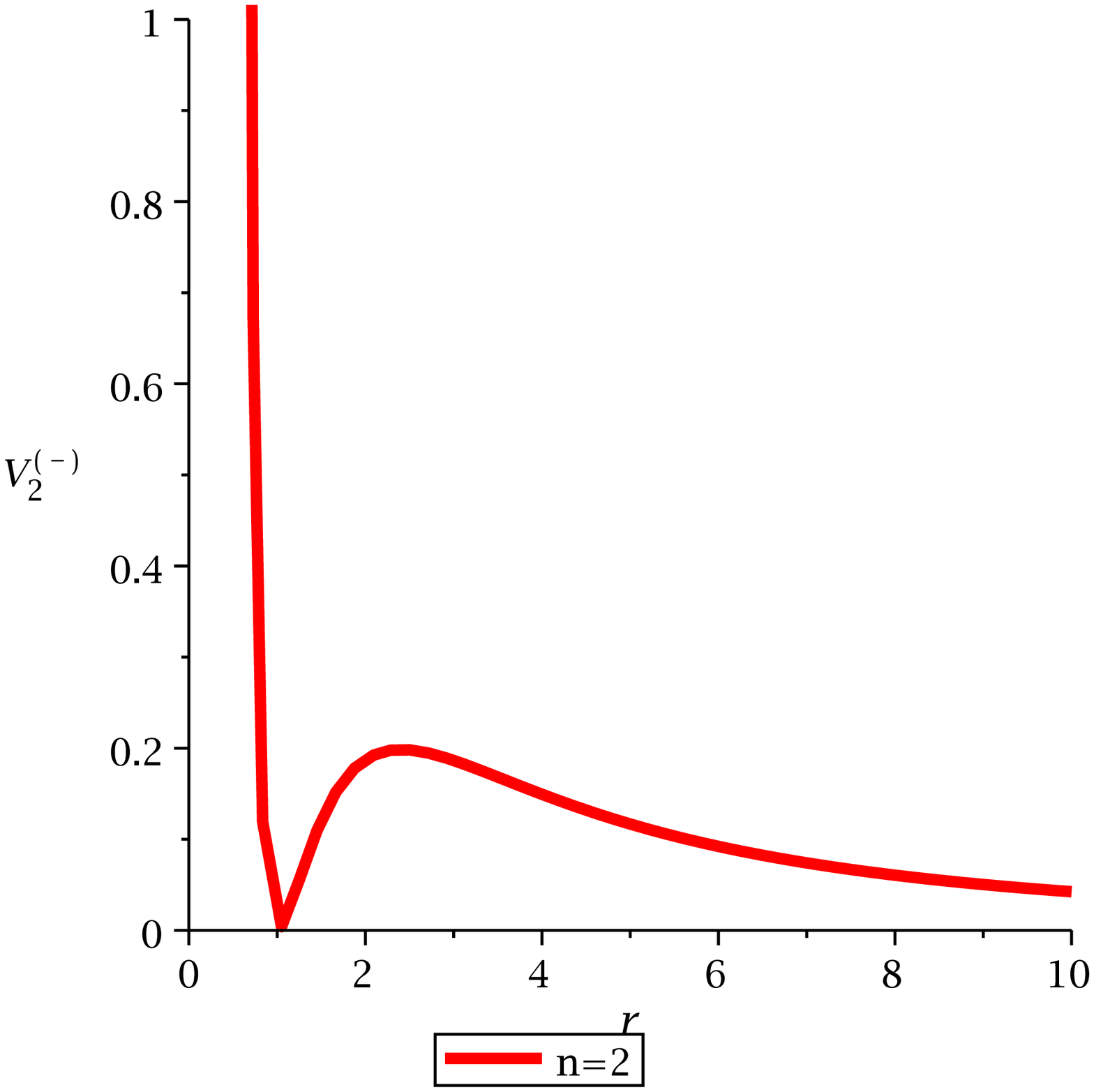}} 
\subfigure[]{
\includegraphics[width=2in,angle=0]{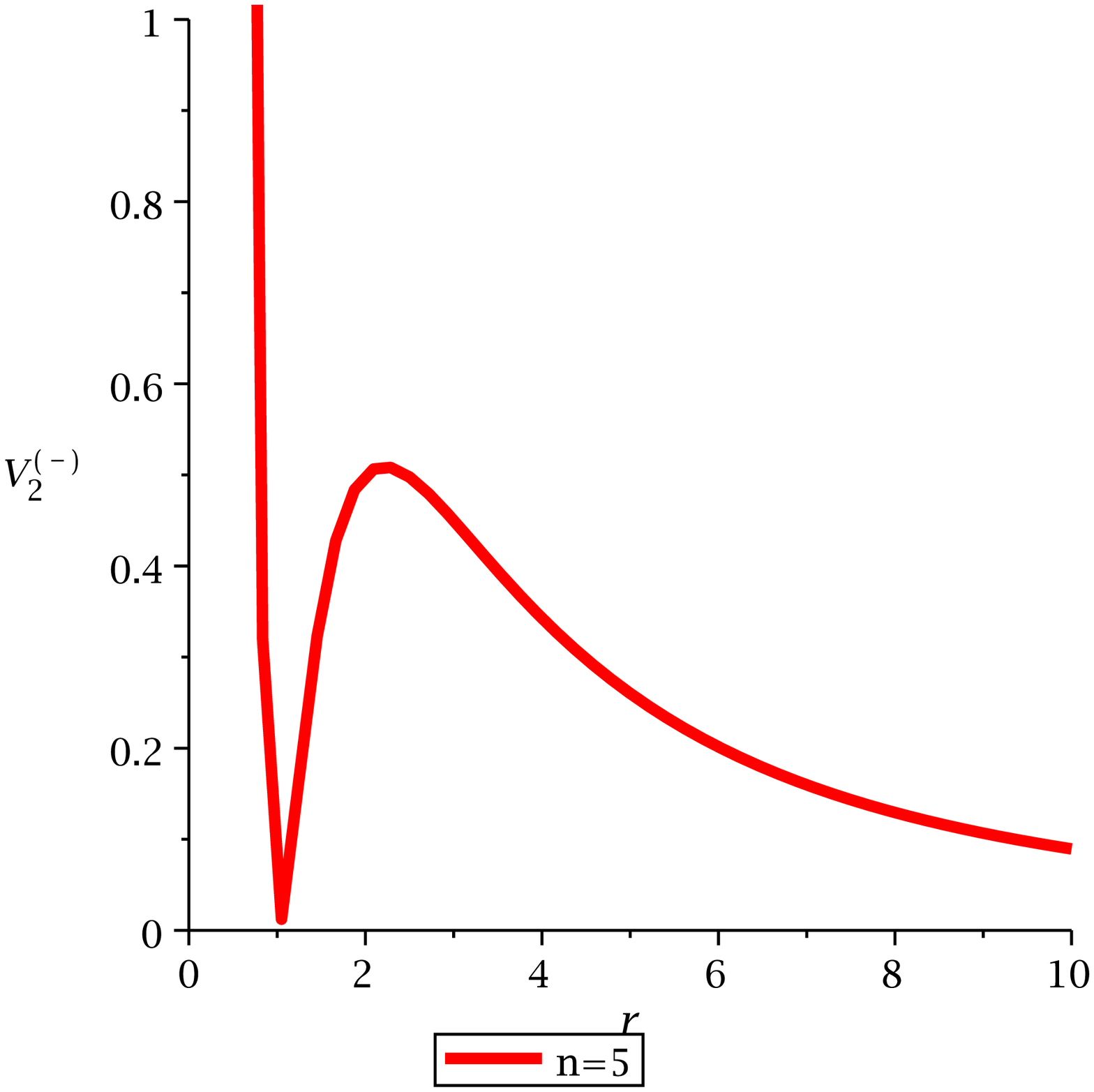}} 
\subfigure[]{
\includegraphics[width=2in,angle=0]{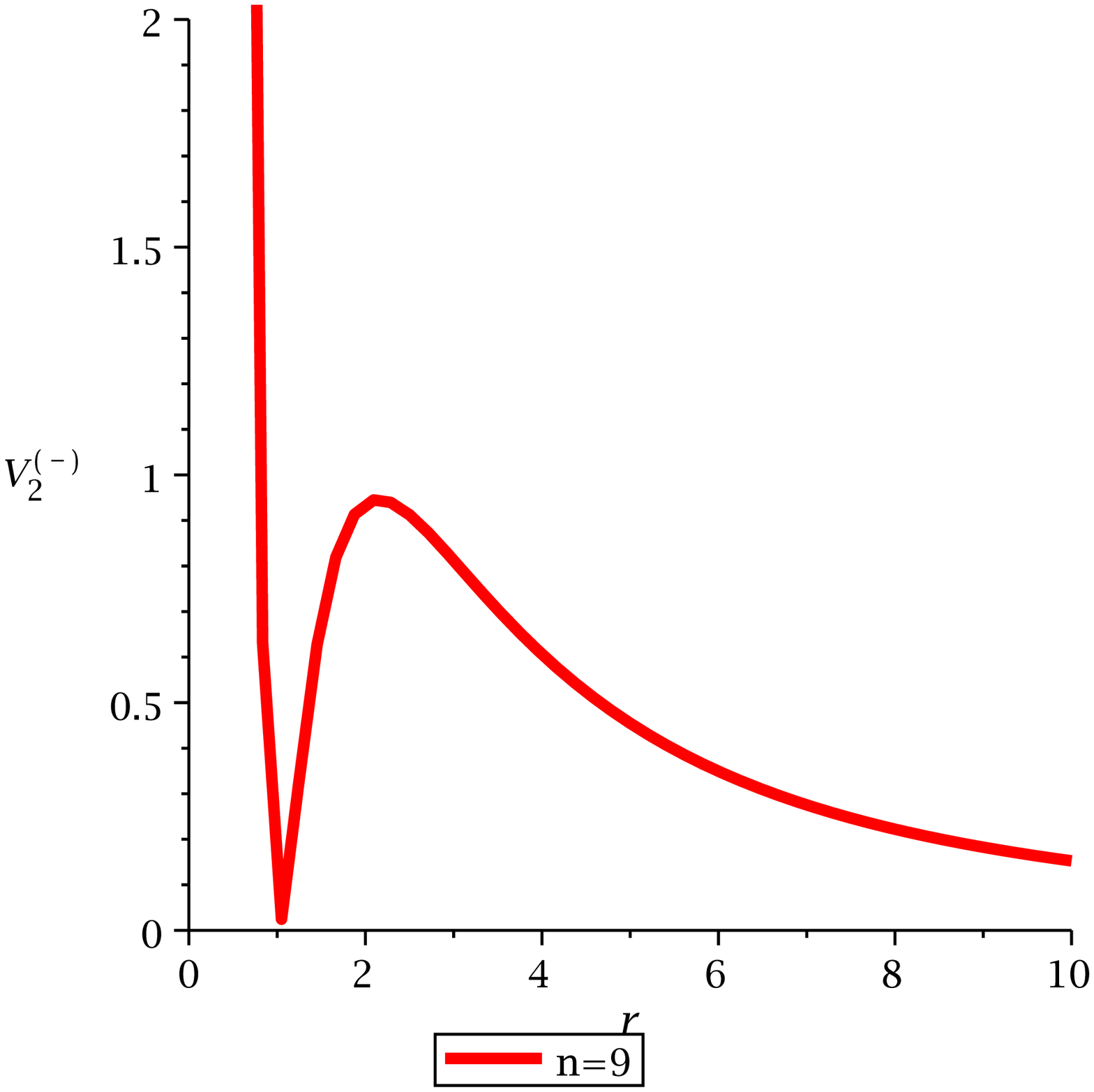}} 
\subfigure[]{
\includegraphics[width=2in,angle=0]{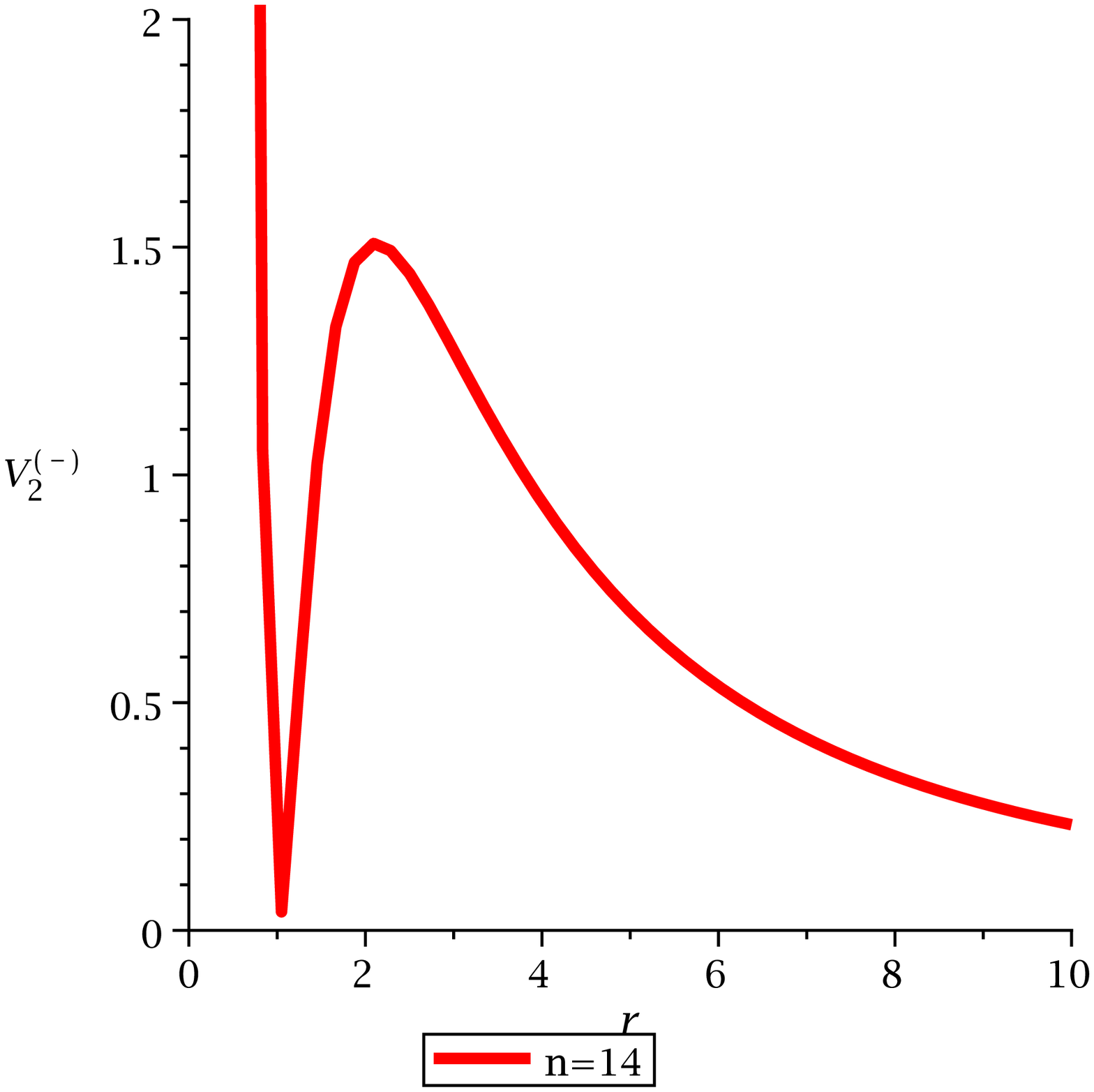}} 
\subfigure[]{
\includegraphics[width=2in,angle=0]{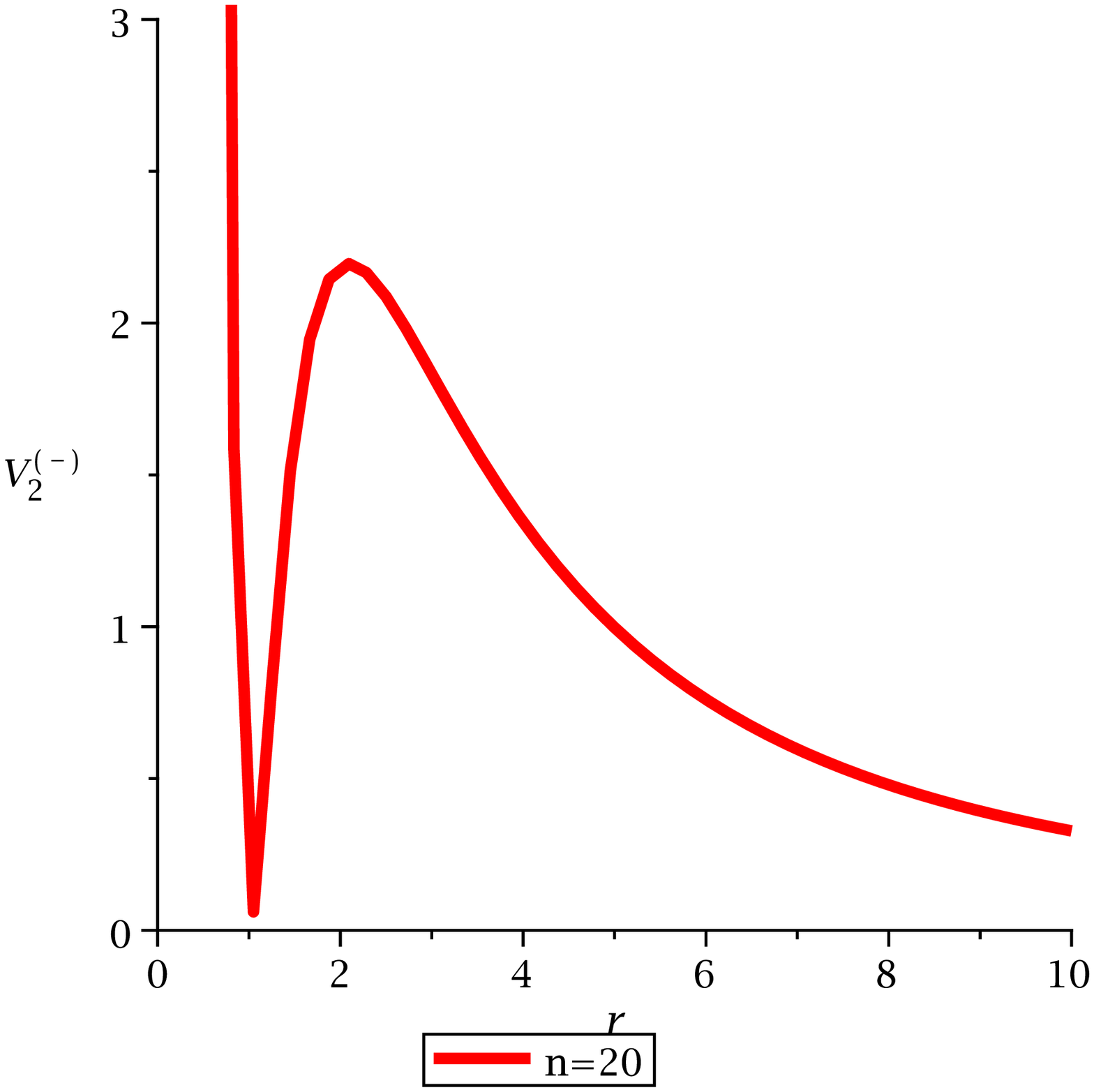}} 
\subfigure[]{
\includegraphics[width=2in,angle=0]{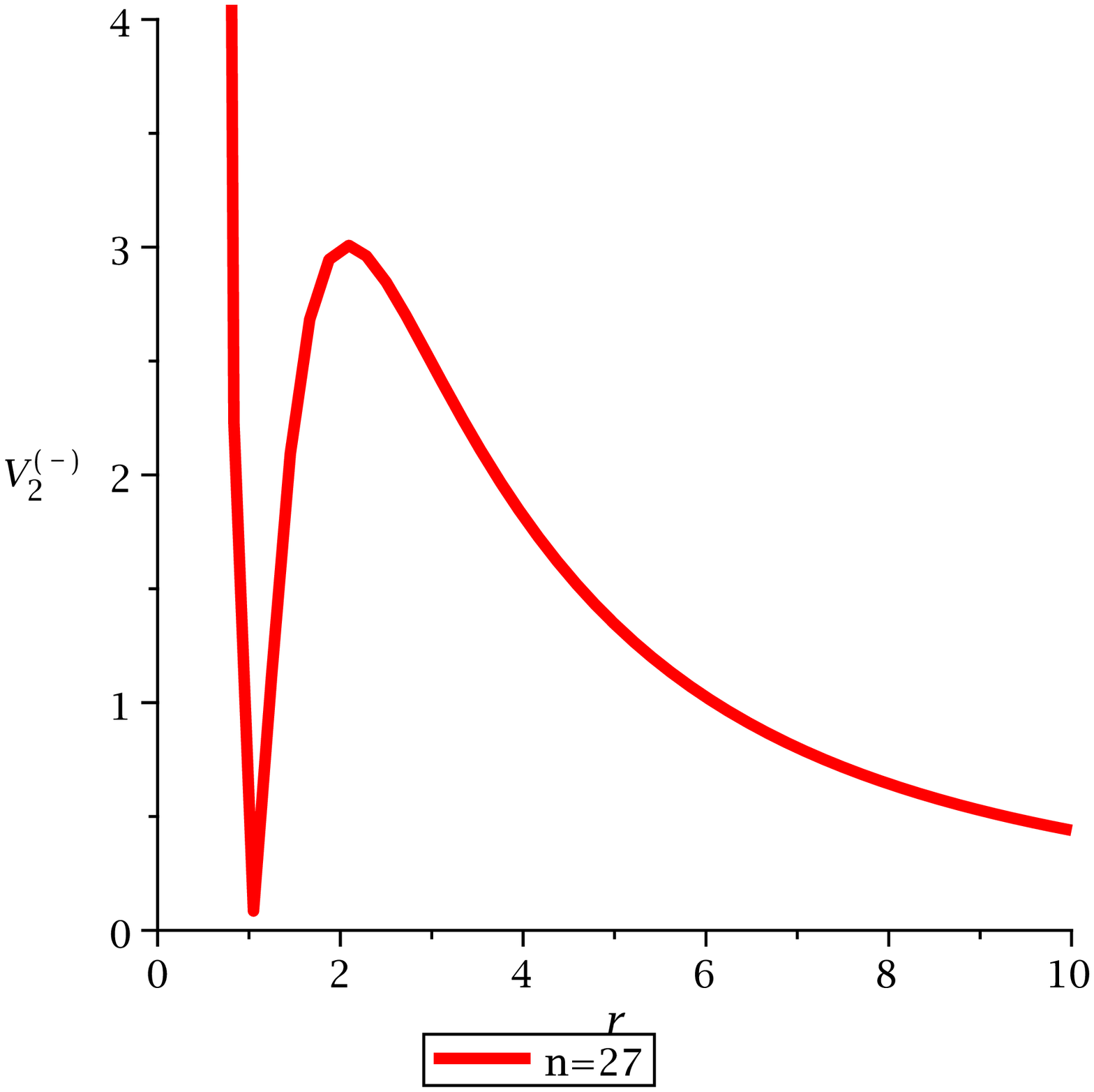}}
\subfigure[]{
\includegraphics[width=2in,angle=0]{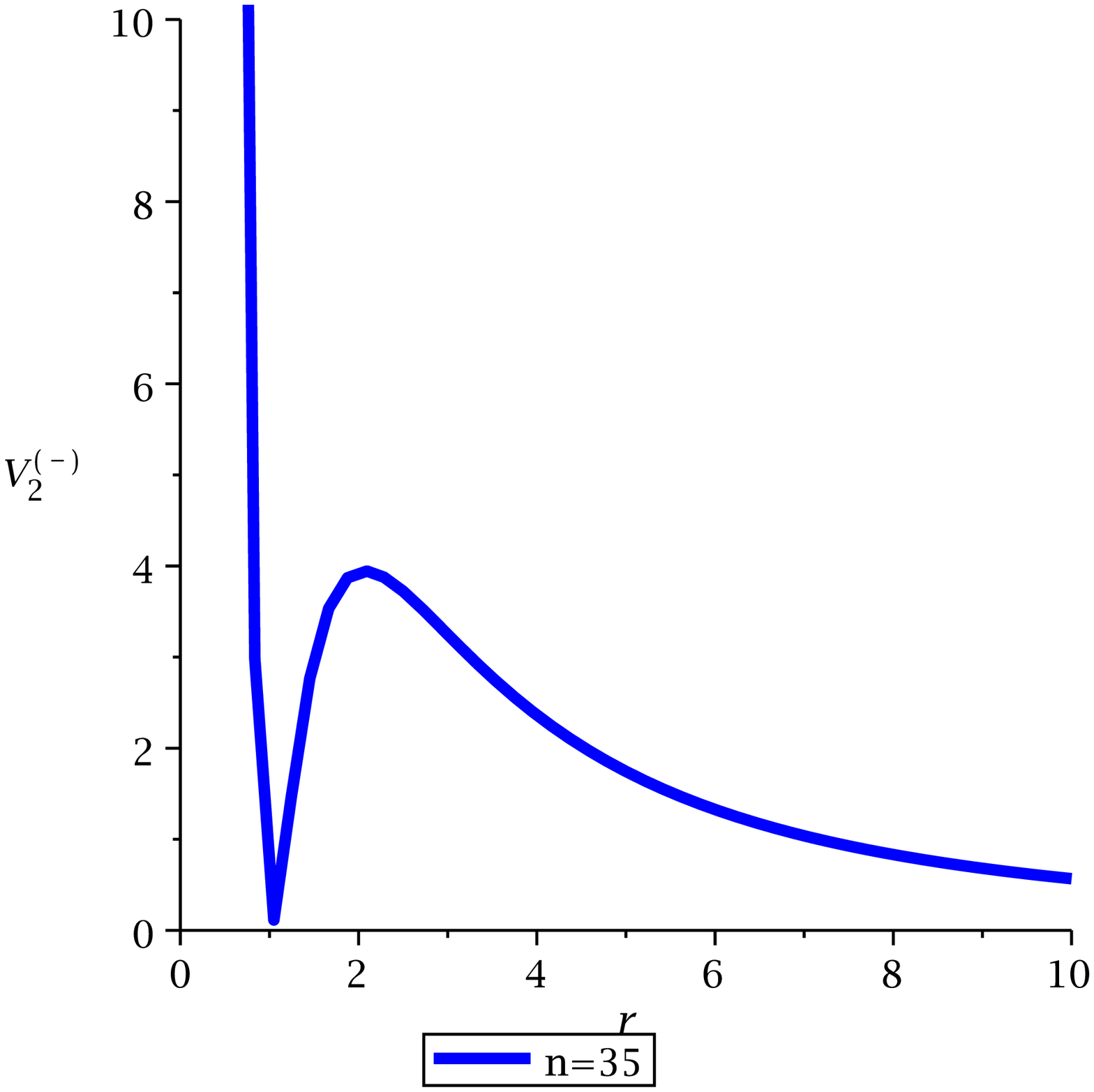}} 
\subfigure[]{
\includegraphics[width=2in,angle=0]{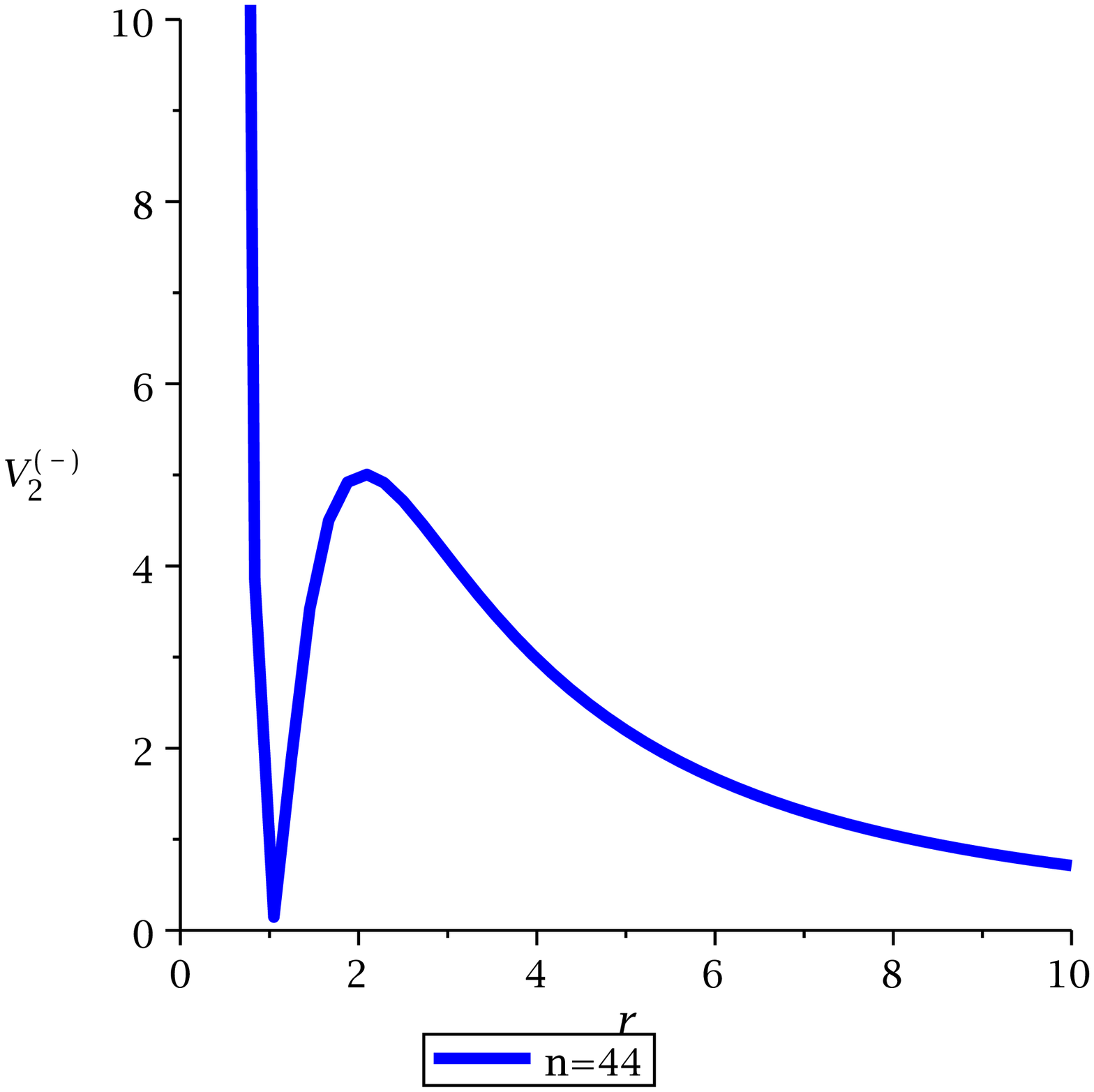}} 
\subfigure[]{
\includegraphics[width=2in,angle=0]{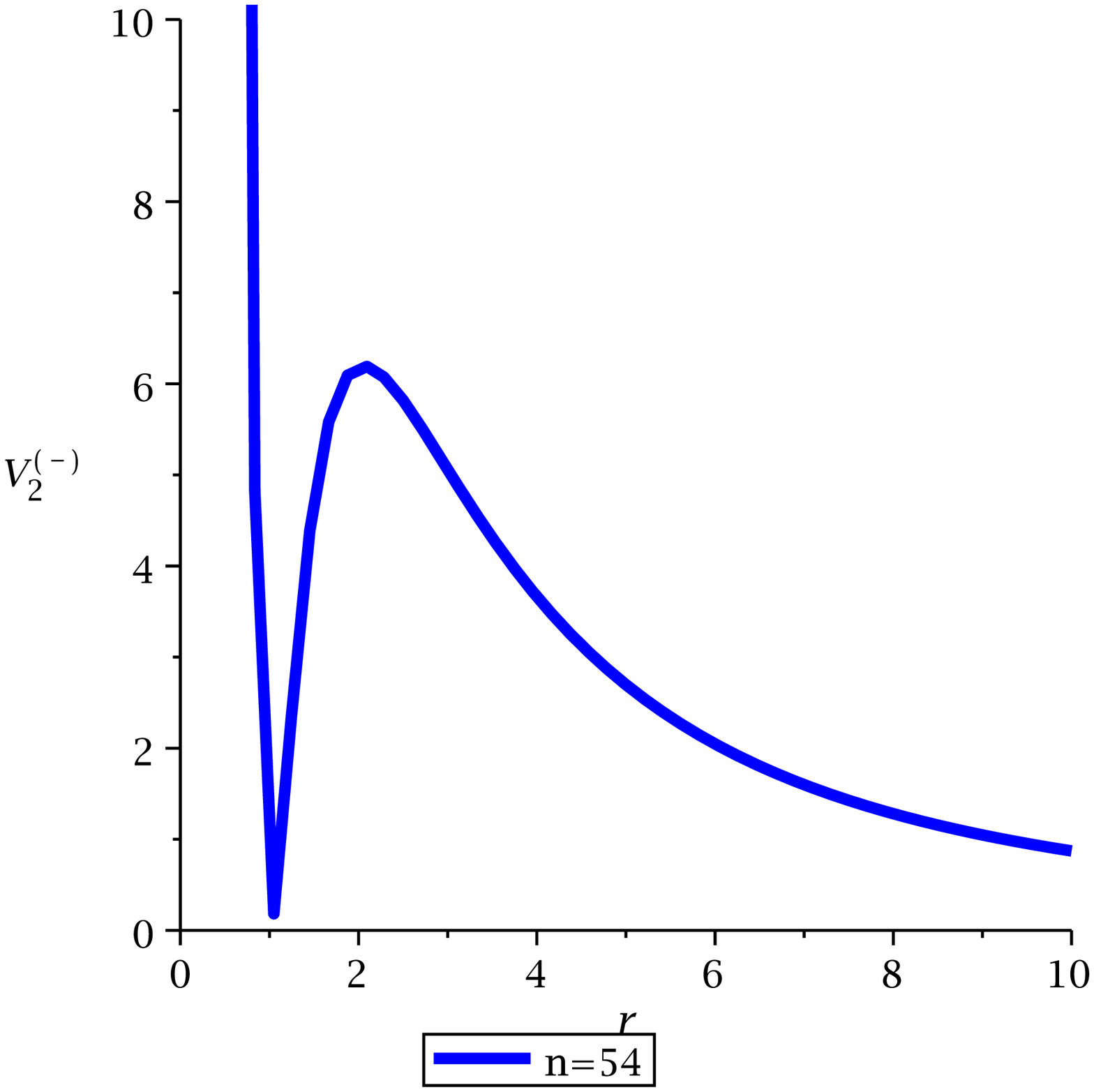}} 
\subfigure[]{
\includegraphics[width=2in,angle=0]{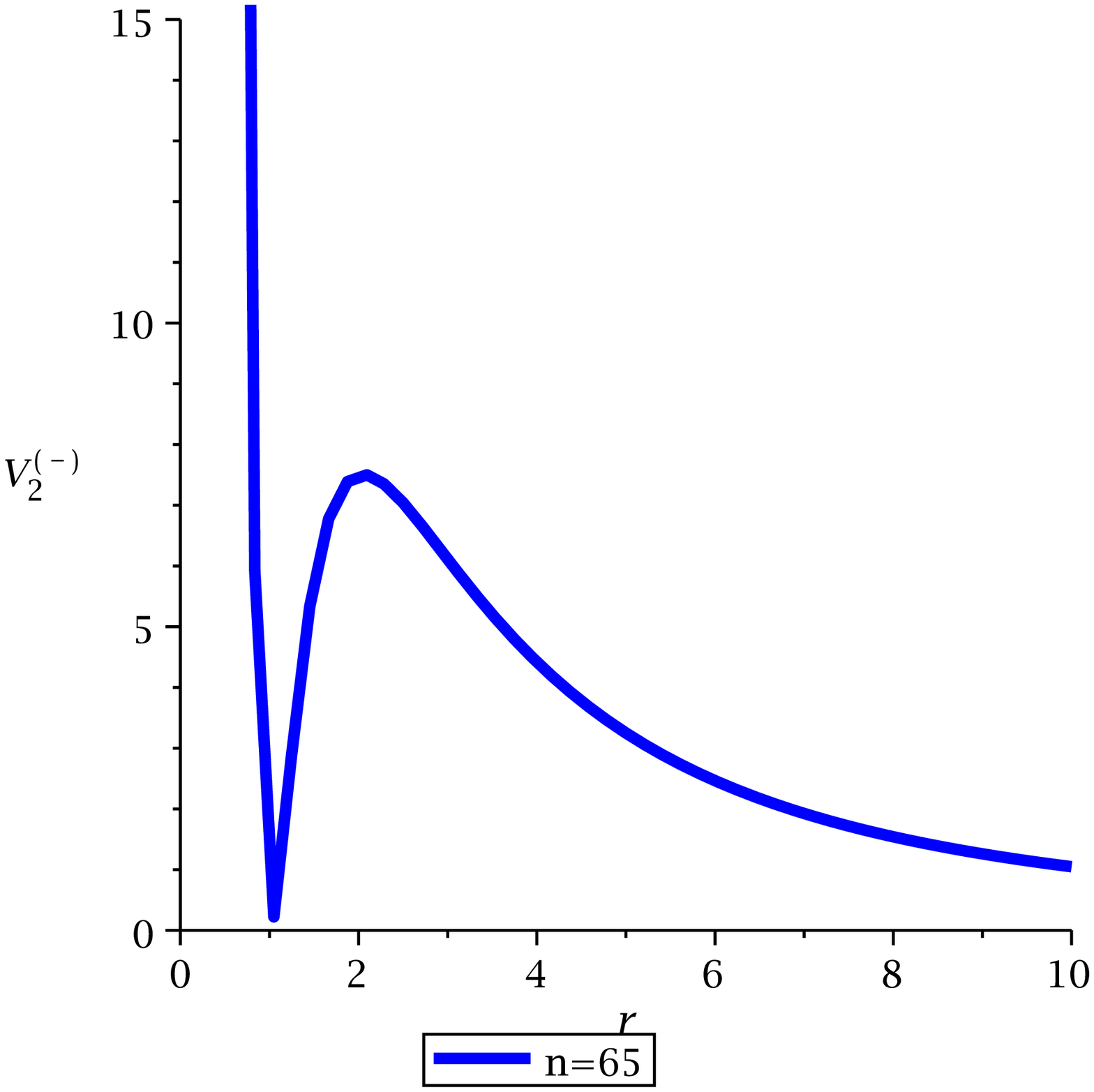}} 
\subfigure[]{
\includegraphics[width=2in,angle=0]{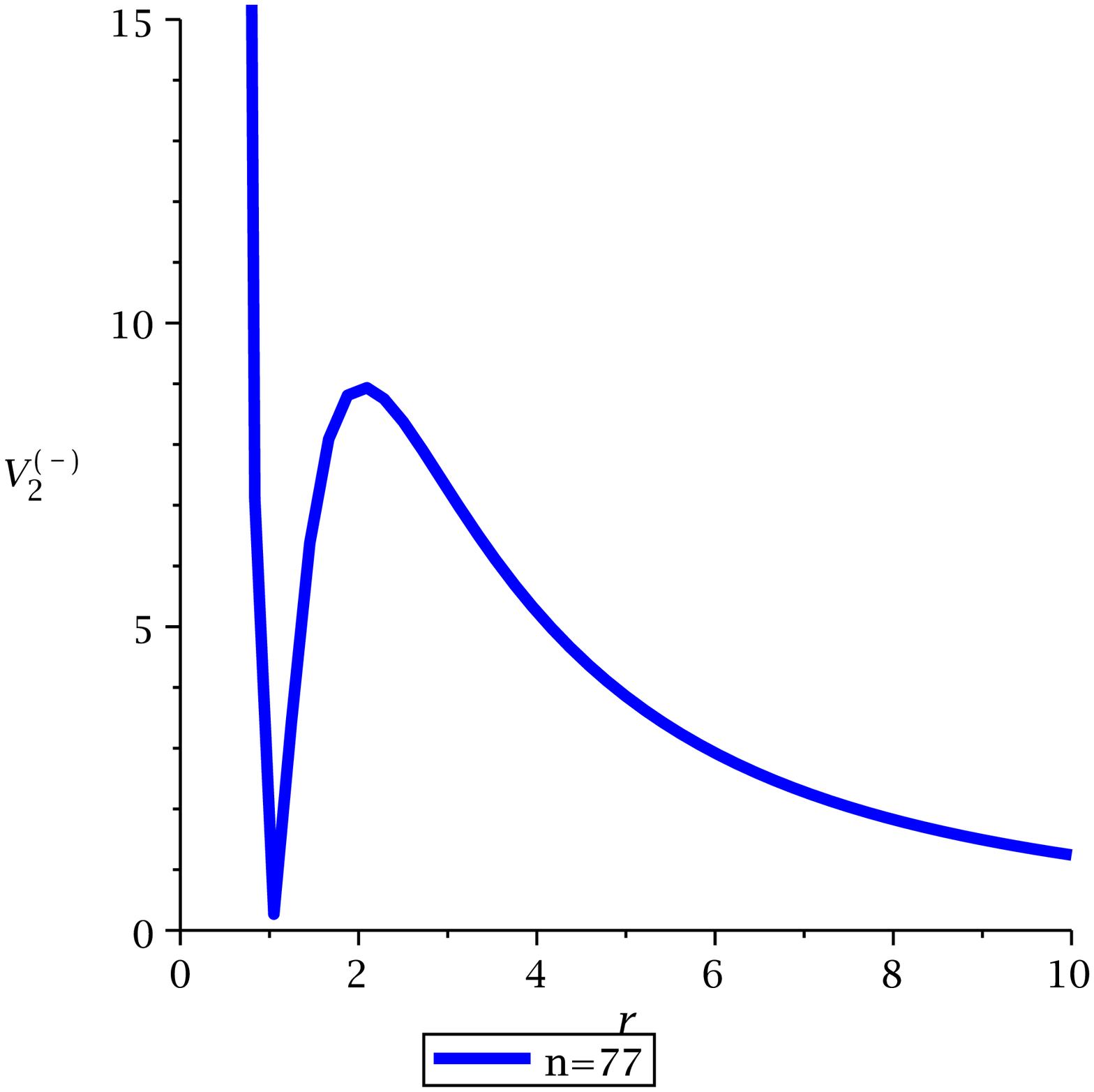}} 
\subfigure[]{
\includegraphics[width=2in,angle=0]{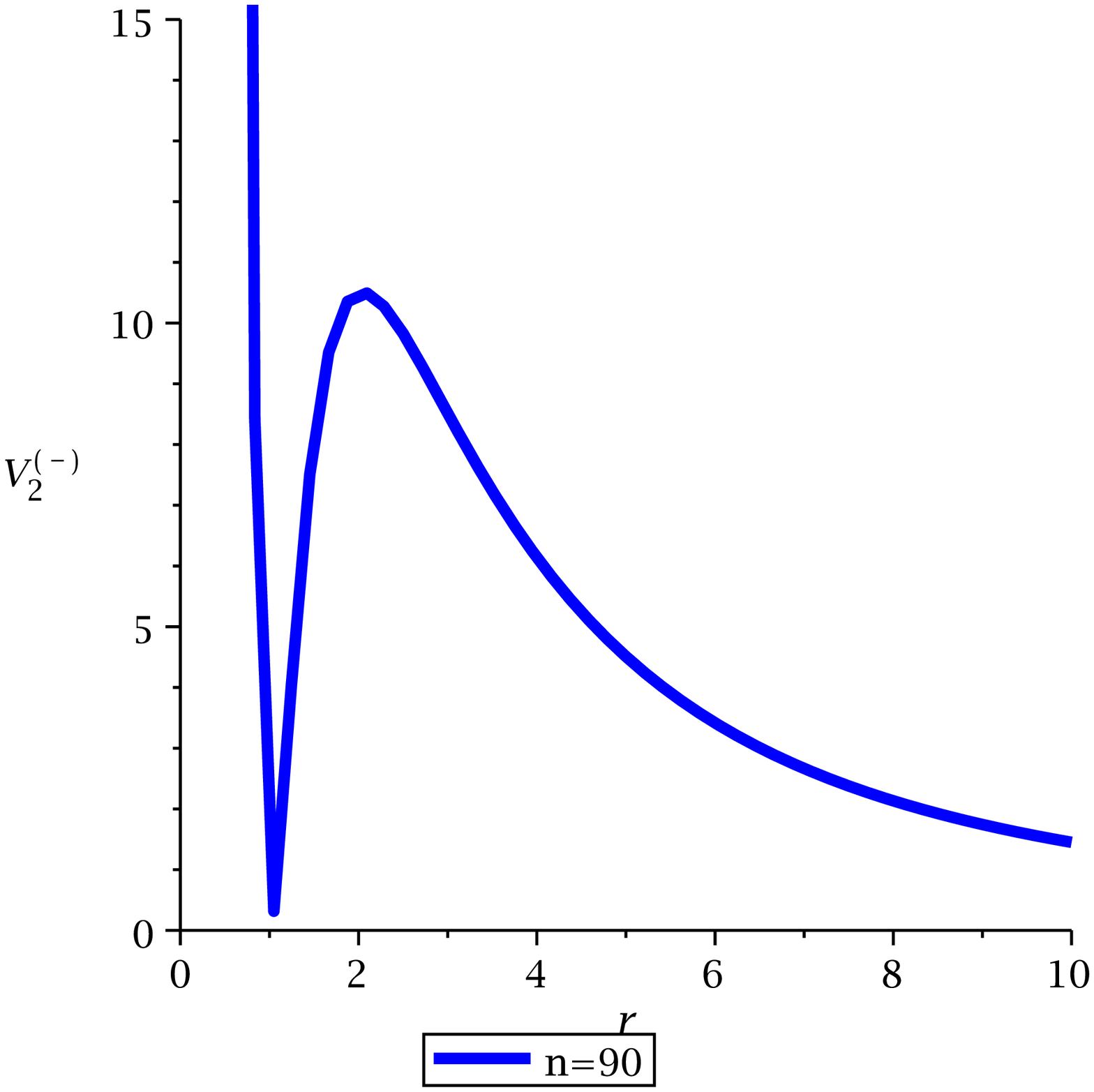}} 
\caption{The structure of effective potential~($V_{2}^{(-)}$) barriers 
surrounding the extremal Reissner Nordstr\"{o}m  BH for axial 
perturbations.}
\label{axialx1} 
\end{center}
\end{figure}

The above pair of coupled second order differential equations can be decoupled by further substituations 
\begin{eqnarray}
Z_{1}^{(-)} &=& +q_{1} H_{1}^{(-)}+\sqrt{-q_{1}q_{2}}{H}_{2}^{(-)} \nonumber\\
Z_{2}^{(-)} &=& -\sqrt{-q_{1}q_{2}}{H}_{1}^{(-)} +q_{1} H_{2}^{(-)},~ ~\label{xeq7}
\end{eqnarray}
where 
\begin{eqnarray}
q_{1} &=& M \left(3+\sqrt{9+8n}\right) \\ 
q_{2} &=& M \left(3-\sqrt{9+8n}\right).~ \label{xeq8}
\end{eqnarray}
It is proved that $Z_{1}^{(-)}$ and $Z_{2}^{(-)}$ satisfy one-dimensional Schro\"{o}dinger-type 
wave-equations, 
\begin{eqnarray}
\Lambda^{2} Z_{1}^{(-)} &=& V_{1}^{(-)} Z_{1}^{(-)}, ~\label{xz1}
\end{eqnarray}
and
\begin{eqnarray}
\Lambda^{2} Z_{2}^{(-)} &=& V_{2}^{(-)} Z_{2}^{(-)}, ~\label{xz2}
\end{eqnarray}
where the effective potential for axial perturbations of extremal RN BH after substituting the value of 
$\Upsilon$ are 
\begin{eqnarray}
V_{1}^{(-)} & \equiv & V_{1}^{(-)}(r)= \frac{(r-M)^2}{r^{6}}\left[2(n+1)r^2-\left(3-\sqrt{9+8n}\right)Mr+4 M^{2}\right], 
~\label{v3}
\end{eqnarray}
and 
\begin{eqnarray}
V_{2}^{(-)} & \equiv & V_{2}^{(-)}(r)= \frac{(r-M)^2}{r^{6}}\left[2(n+1)r^2-\left(3+\sqrt{9+8n}\right)Mr+4 M^{2}\right].
~\label{v4}
\end{eqnarray}
These equations governing the axial perturbations to the pair of one dimensional wave equations for the 
extremal Reissner Nordstr\"{o}m BH. Now we will see the behavior  of these potentials graphically~
[See Fig.~(\ref{axialx})) and Fig.~(\ref{axialx1}))]. It is evident from the plot that the height of 
the potential barrier increases when $n$ increases.  

\subsection{Stability Analysis of Extremal RN BH via the Effective potentials $V_{1}^{(-)}$ and $V_{2}^{(-)}$}
The shape of the effective potentials depends on $n$. The barrier-height is increasing when the value of $n$ is increasing. First, we will analyze the effective potential 
$V_{1}^{(-)}$. 

\subsubsection{Properties of the Effective potential $V_{1}^{(-)}$}
The extrema of the effective potential is determined by 
solving the following equation
\begin{eqnarray}
\frac{d\,V_{1}^{(-)}}{dr} &=& \frac{\left(r-M \right)\left(a\,r^3+b\,r^2+c\,r+d\right)}{r^7} = 0,  ~\label{sv}
\end{eqnarray}
where 
\begin{eqnarray}
a &=& -4\left(n+1 \right) \\
b &=& \left(17+8n-3\sqrt{9+8n}\right)M\\
c &=& - \left(31-5\sqrt{9+8n}\right)M^2\\
d &=& 24 M^3. ~\label{sv1}
\end{eqnarray}
It follows from the above equation that the only solution of Eq.~(\ref{sv}) is   
\begin{eqnarray}
r=M
\end{eqnarray}
or
\begin{eqnarray}
a\,r^3+b\,r^2+c\,r+d = 0 
\end{eqnarray}
The first solution i.e. $r=r_{x}=M$ is the horizon of extremal RN BH. Let us check the stability of this 
extremal horizon. To do so, we have to calculate the second derivative of effective potential. Thus 
one find 
\begin{eqnarray}
\frac{d^2\,V_{1}^{(-)}}{dr^2} &=& \frac{2}{r^8}\left[a_1\,r^4+b_1\,r^3+c_1\,r^2+d_1\,r+e_1\right]. ~\label{sv2}
\end{eqnarray} 
where 
\begin{eqnarray}
a_1 &=& 6\left(n+1 \right) \\
b_1 &=& -6\left(7+4n-\sqrt{9+8n} \right)M\\
c_1 &=& 20\left(6+n-\sqrt{9+8n}\right)M^2 \\
d_1 &=& -15\left(11-\sqrt{9+8n}\right)M^3\\
e_1 &=& 84M^4. ~\label{sv3}
\end{eqnarray}
Now 
\begin{eqnarray}
\left(\frac{d^2\,V_{1}^{(-)}}{dr^2}\right)_{r_{x}=M} &=& \frac{2}{M^4} \left(2n+3+\sqrt{9+8n}\right)> 0 \,\, 
\mbox{for}\,\, \forall n ~\label{sv4}
\end{eqnarray}
Thus the radius of the orbit $r=r_{x}=M$ is a stable orbit for $\forall n$\footnote{It should be emphasized that $r_{x}=M$ is not only the horizon of extremal RN BH. It is also the radius of the null photon sphere of extremal RN BH. 
As it was already proved in \cite{pla11} that there exists stable photon orbit at radius $r_{x}=M$. We here strongly suggest that indeed there exists a stable photon orbit at $r_{x}=M$ which is reverified via axial gravitational perturbations.}. Hence \emph{the extremal horizon of extremal RN BH under 
gravitational axial perturbation is stable}. 

Now we analyze the second solution i. e. $a\,r^3+b\,r^2+c\,r+d = 0$. To determine the nature of the roots 
we have to compute the discriminate 
\begin{eqnarray}
\Delta = 1728 \left(n+1 \right) \left(17+8n-3\sqrt{9+8n}\right) \left(31-5\sqrt{9+8n}\right)M^6
-96\left(17+8n-3\sqrt{9+8n}\right)M^6 \nonumber\\
-\left(17+8n-3\sqrt{9+8n}\right)^2 \left(31-5\sqrt{9+8n}\right)^2 M^6
-16\left(n+1 \right) \left(31-5\sqrt{9+8n}\right)^3 M^6 \nonumber\\
-248832 \left(n+1 \right)M^6 ~\label{sv5}
\end{eqnarray}
It implies that $\Delta<0$ for $\forall n$. So the cubic equation has only one real root 
and two complex conjugate roots. 
For simplicity, we consider the value of $n=2$. In this case we find $a=-12$, $b=18M$, $c=-6M^2$ and 
$d=24M^3$ therefore the value of $\Delta=-1703376<0$. Then the radial equation becomes 
\begin{eqnarray}
2r^3-3Mr^2+M^2r-4M^3 &=& \left(r-r_{1}\right) \left(r-r_{2}\right) \left(r-r_{3}\right)= 0 ~\label{sv6}
\end{eqnarray}
The only real root of this equation is 
\begin{eqnarray}
r_{1} &=& \frac{1}{6} \left[3+\left(216-3\sqrt{5181} \right)^{\frac{1}{3}}
+3^{\frac{1}{3}}\left(72+\sqrt{5181} \right)^{\frac{1}{3}} \right]M ~\label{sv7}
\end{eqnarray}
Taking the value of $\sqrt{5181}=71.979\approx 71.98$, we get the value of real root 
\begin{eqnarray}
r_{1} & \approx & 1.82 M ~\label{sv8}
\end{eqnarray}
and other roots $r_{2}$ and $r_{3}$ are imaginary. Note that $r_{1}>r_{x}$. We can 
check now whether this radius $r_{1}$ which is located outside the horizon is stable 
or unstable.  

We see that
\begin{eqnarray}
\left(\frac{d^2\,V_{1}^{(-)}}{dr^2}\right)_{r=r_{1}} &\approx & -\frac{261}{M^4} <0  ~\label{sv9}
\end{eqnarray}
Thus the radius $r=r_{1}$ which is located at $r>M$  has a local maximum.  Now we will see  
the more generalized case. 
To do so we consider the horizon location away from extremality i.e. $r=\left(1+\chi \right)M$, 
where $\chi<<1$. In this limit 
\begin{eqnarray}
\left(\frac{d^2\,V_{1}^{(-)}}{dr^2}\right)_{r=\left(1+\chi \right)M} &=& 
\frac{2}{M^4} \left[2n+3+\sqrt{9+8n}-\left(24n+51+15\sqrt{9+8n}\right)\chi+{\cal O}(\chi^2) \right] ~\label{sv11}
\end{eqnarray}
This immediately implies that 
\begin{eqnarray}
\left(\frac{d^2\,V_{1}^{(-)}}{dr^2}\right)_{r=\left(1+\chi \right)M}< 0 \,\, \mbox{for}\,\, \forall n 
~\label{sv12}
\end{eqnarray}
This means that outside the horizon the effective potential barrier has a local maximum. 

\subsubsection{Properties of the Effective potential $V_{2}^{(-)}$}
Proceeding analogously the extrema of the effective potential is determined by  
\begin{eqnarray}
\frac{d\,V_{2}^{(-)}}{dr} &=& \frac{\left(r-M \right)\left(a\,r^3+b\,r^2+c\,r+d\right)}{r^7} = 0,  ~\label{sv1.1}
\end{eqnarray}
where 
\begin{eqnarray}
a &=& -4\left(n+1 \right) \\
b &=& \left(17+8n-3\sqrt{9+8n}\right)M\\
c &=& - \left(31-5\sqrt{9+8n}\right)M^2\\
d &=& 24 M^3. ~\label{sv1.2}
\end{eqnarray}
There are two possible solutions. Either $r=M$ or  $a\,r^3+b\,r^2+c\,r+d = 0$. The first one is 
extremal horizon. To determine the stability of it we have to compute 
\begin{eqnarray}
\frac{d^2\,V_{2}^{(-)}}{dr^2} &=& \frac{2}{r^8}\left[a_1\,r^4+b_1\,r^3+c_1\,r^2+d_1\,r+e_1\right]. ~\label{sv1.3}
\end{eqnarray} 
where 
\begin{eqnarray}
a_1 &=& 6\left(n+1 \right) \\
b_1 &=& -6\left(7+4n+\sqrt{9+8n}\right)M\\
c_1 &=& 20\left(6+n+\sqrt{9+8n}\right)M^2 \\
d_1 &=&-15\left(11+\sqrt{9+8n}\right)M^3\\
e_1 &=& 84 M^4. ~\label{sv1.4}
\end{eqnarray}
Now
\begin{eqnarray}
\left(\frac{d^2\,V_{2}^{(-)}}{dr^2}\right)_{r=M} &=& \frac{2}{M^4} \left(2n+3-\sqrt{9+8n}\right) > 0 \,\, 
\mbox{for}\,\, \forall n ~\label{sv1.3}
\end{eqnarray}
Again it proves that \emph{the extremal horizon of extremal RN BH under axial gravitational perturbation is stable}. 

Now we would like to study the near-extremal situations i.e.  at  $r=\left(1+\chi \right)M$. In this limit 
\begin{eqnarray}
\left(\frac{d^2\,V_{2}^{(-)}}{dr^2}\right)_{r=\left(1+\chi \right)M} &=& 
\frac{2}{M^4} \left[2n+3-\sqrt{9+8n}-\left(24n+51-15\sqrt{9+8n}\right)\chi+{\cal O}(\chi^2) \right] ~\label{sv1.5}
\end{eqnarray}
This also indicates that 
\begin{eqnarray}
\left(\frac{d^2\, V_{2}^{(-)}}{dr^2}\right)_{r=\left(1+\chi \right)M}< 0 \,\, \mbox{for}\,\, \forall n 
~\label{sv1.6}
\end{eqnarray}
Thus outside the horizon the \emph{effective potential barrier has a local maximum}. 
It is evident from the graphical plots of $V_{1}^{(-)}$ and $V_{2}^{(-)}$ that 
\emph{for extremal RN BHs the structure of effective potentials look like a potential 
barrier  outside the horizon i.e. $r>M$. Also the potentials $V_{1}^{(-)}$ and $V_{2}^{(-)}$  
are real and positive everywhere outside the horizon.} Hence following the statement of 
Chandrasekhar~\cite{sc} we could say that \emph{extremal RN BH is stable under 
gravitational axial perturbations.}

\section{\label{axn}Axial Perturbation in Naked Singularity case of Reissner-Nordstr\"{o}m Spacetime}
Since in this work we are interested to study both $M^2=Q_{\ast}^2$ and $M^2<Q_{\ast}^2$ situations. In 
previous section, we have studied the axial perturbations in case of $M^2=Q_{\ast}^2$. In the 
present section, we will study the NS case of RN spacetime. We will also observe how the potential 
barrier vary for non-extremal case, extremal case and NS. In previous section we have already been
calculated the effective potential due to axial perturbations for the extremal situations, here we 
will calculate the effective potential for NS case. Therefore the metric function for Eq.~(\ref{metric}) 
can be written as 
\begin{eqnarray}
 e^{2\nu} &=& e^{-2\mu_{2}}= \left(1-\frac{M}{r} \right)^2+\left(\frac{Q_{\ast}^2-M^2}{r^2}\right)
 =\left(\frac{\Upsilon+\Xi}{r^2}\right).~\label{nsf}
\end{eqnarray}
where $\Xi=Q_{\ast}^2-M^2$. Clearly for $Q_{\ast}^2>M^2$, $\Xi>0$ there  
is no horizon. This is called NS situations. For $Q_{\ast}^2=M^2$, $\Xi=0$ that means we get 
extremal BH solution.  For $Q_{\ast}^2<M^2$, $\Xi<0$ that means we get non-extremal BH 
solution having two horizons namely the event horizon, $r_{+}=M+(M^2-Q_{\ast}^2)^{\frac{1}{2}}$ and the 
Cauchy horizon, $r_{+}=M-(M^2-Q_{\ast}^2)^{\frac{1}{2}}$. 

The fundamental difference between non-extremal BH, extremal BH, and NS is as follows. Since the 
spacetime metric admits Killing vector field $\xi^{\mu}_{(t)}=(1,0,0,0)$. Thus the norm of the 
Killing vector field is defined by the quantity $\xi^{\mu}\xi_{\mu}$ and for 

(i) non-extremal BH it is given by 
\begin{eqnarray}
\xi^{\mu}\xi_{\mu} &=& \left(r-r_{+}\right)\left(r-r_{-}\right)r^{-2} 
\end{eqnarray}
and
\begin{eqnarray}
\xi^{\mu}\xi_{\mu} &>& 0~(\mbox{time-like})\, \mbox{for}\, r_{-}<r<r_{+}\\
                   &=& 0~(\mbox{null})\,  \mbox{for}\, r=r_{\pm} \\
                   &<& 0~(\mbox{space-like})\, \mbox{for}\, 0<r<r_{-}
\end{eqnarray}
(ii) for extremal BH it is defined by 
\begin{eqnarray}
\xi^{\mu}\xi_{\mu} &=& \left(1-\frac{M}{r}\right)^{2} > 0~(\mbox{time-like})\,\, \forall r\\
                    &=& 0~(\mbox{null})\,  \mbox{for}\, r=M 
\end{eqnarray}

(iii) for NS it is
\begin{eqnarray}
\xi^{\mu}\xi_{\mu} &=& \left(1-\frac{M}{r} \right)^2+\left(\frac{Q_{\ast}^2-M^2}{r^2}\right) 
> 0~(\mbox{time-like})\, \forall r
\end{eqnarray}
Analogously, for $Q_{\ast}^2>M^2$, the pair of  coupled second order differential equations 
become
\begin{eqnarray}
\Lambda^{2} {H}_{1}^{(-)} &=& \left(\frac{\Upsilon+\Xi}{r^{5}}\right)
\left[\left\{2(n+1)r-3M+4\frac{Q_{\ast}^2}{r}\right\}{H}_{1}^{(-)}+3M {H}_{1}^{(-)}
+2Q_{\ast}^2 \sqrt{2n} {H}_{2}^{(-)}\right], ~\label{nseq1}
\end{eqnarray} 
and
\begin{eqnarray}
\Lambda^{2} {H}_{2}^{(-)} &=& \left(\frac{\Upsilon+\Xi}{r^{5}}\right)\left[\left\{2(n+1)r
-3M+4\frac{Q_{\ast}^2}{r}\right\}{H}_{2}^{(-)}-
 3M {H}_{2}^{(-)}+2Q_{\ast}^2\sqrt{2n} \bar{H}_{1}^{(-)}\right].
\label{nseq2}
\end{eqnarray}
where, $\Lambda^{2}=\frac{d^{2}}{dr_{*}^{2}}+\sigma^{2}$. 

These equations  can be decoupled by further substituations 
\begin{eqnarray}
Z_{1}^{(-)} &=& +q_{1} H_{1}^{(-)}+\sqrt{-q_{1}q_{2}}{H}_{2}^{(-)} \nonumber\\
Z_{2}^{(-)} &=& -\sqrt{-q_{1}q_{2}}{H}_{1}^{(-)} +q_{1} H_{2}^{(-)},~ ~\label{nseq3}
\end{eqnarray}
where 
\begin{eqnarray}
q_{1} &=&  \left(3M+\sqrt{9M^2+8n Q_{\ast}^2}\right) \\ 
q_{2} &=&  \left(3M-\sqrt{9M^2+8n Q_{\ast}^2}\right).~ \label{nseq4}
\end{eqnarray}
We have already proved that  $Z_{1}^{(-)}$ and $Z_{2}^{(-)}$ satisfy one-dimensional 
Schro\"{o}dinger-type wave-equations, 
\begin{eqnarray}
\Lambda^{2} Z_{1}^{(-)} &=& V_{1}^{(-)} Z_{1}^{(-)}, ~\label{nsxz1}
\end{eqnarray}
and
\begin{eqnarray}
\Lambda^{2} Z_{2}^{(-)} &=& V_{2}^{(-)} Z_{2}^{(-)}, ~\label{nsxz2}
\end{eqnarray}
where the effective potential for axial perturbations of NS case
\begin{eqnarray}
V_{1}^{(-)} & \equiv & V_{1}^{(-)}(r)= \left(\frac{\Upsilon+\Xi}{r^{6}}\right)\left[2(n+1)r^2-
\left(3M-\sqrt{9M^2+8n Q_{\ast}^2}\right)r+4 Q_{\ast}^2\right],~~\label{nseq5}
\end{eqnarray}
and 
\begin{eqnarray}
V_{2}^{(-)} & \equiv & V_{2}^{(-)}(r)= \left(\frac{\Upsilon+\Xi}{r^{6}}\right)\left[2(n+1)r^2-
\left(3M+\sqrt{9M^2+8n Q_{\ast}^2}\right)r+4 Q_{\ast}^2 \right].~\label{nseq6}
\end{eqnarray}
These equations governing the axial perturbations for $Q_{\ast}^2>M^2$ cases
of Reissner Nordstr\"{o}m BH and for  $M^2>Q_{\ast}^2$, it was given in S. Chandrasekhar's 
book~\cite{sc}.
The simple way we can distinguish two geometric structures, namely $M^2>Q_{\ast}^2$ and $M^2<Q_{\ast}^2$ via graphical plot. It is evident from the graphical 
plots of $V_{1}^{(-)}$ and $V_{2}^{(-)}$ that \emph{for non-extremal BH~($M^2>Q_{\ast}^2$) the structure of the
effective potentials look like a potential well rather than a potential barrier and potentials are negative in the 
region $r_{-}<r<r_{+}$}. \emph{But for NS~($Q_{\ast}^2>M^2$), the structure of the effective potentials is neither a potential barrier nor a potential well. Rather it looks like an exponential decay function.} 
This is the prime difference between the non-extremal BH and the NS. 

Therefore we have found three different types of effective potential structures for non-extremal BH, extremal BH, and NS. 
Hence these three different types of potentials might allow us to distinguish between non-extremal BH, 
extremal BH, and NS. Hence,  we can conclude that BH and NS are quite distinguished objects. 

\begin{figure}
\begin{center}
\subfigure[]{
\includegraphics[width=1.5in,angle=0]{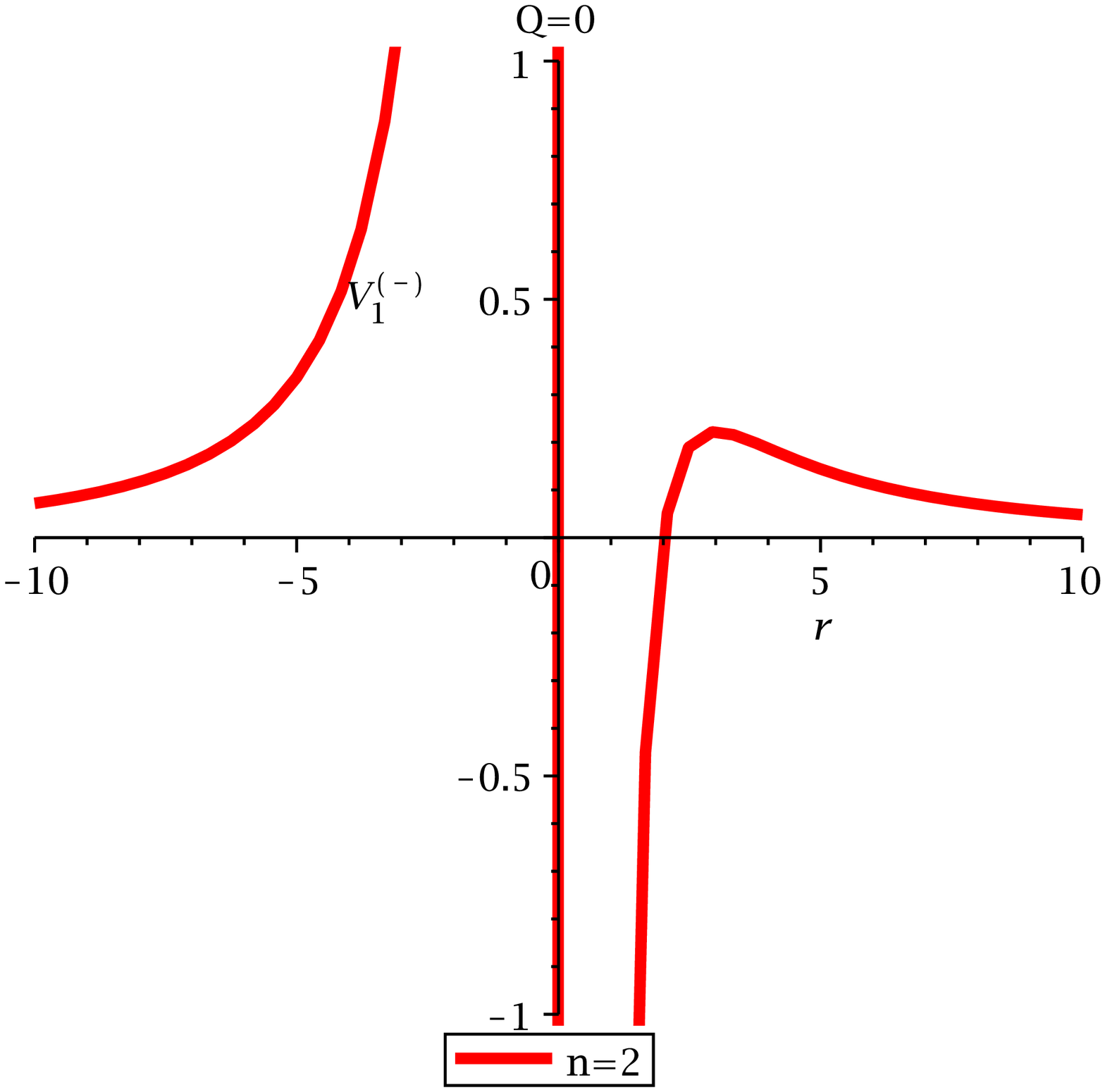}} 
\subfigure[]{
\includegraphics[width=1.5in,angle=0]{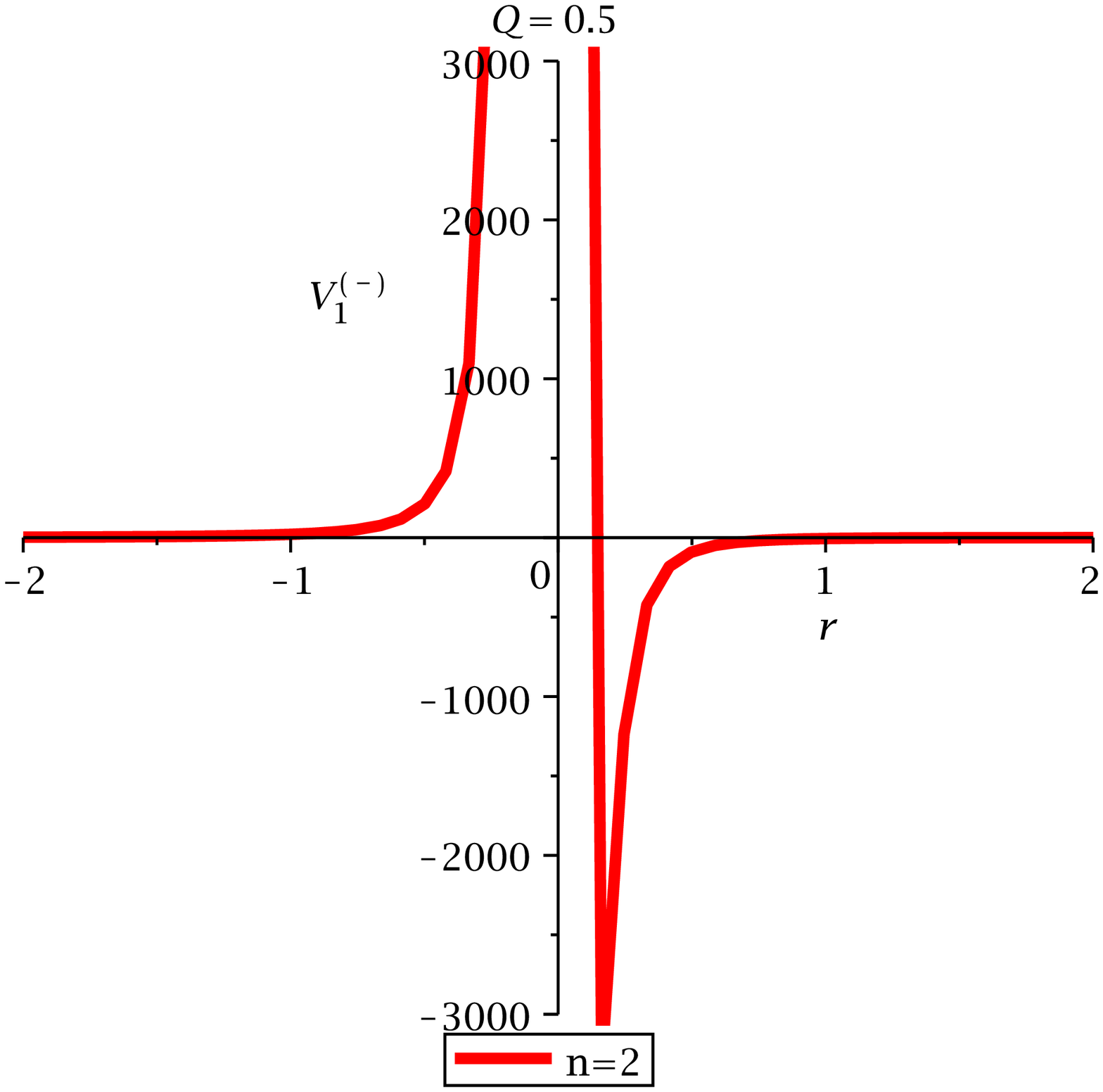}} 
\subfigure[]{
\includegraphics[width=1.5in,angle=0]{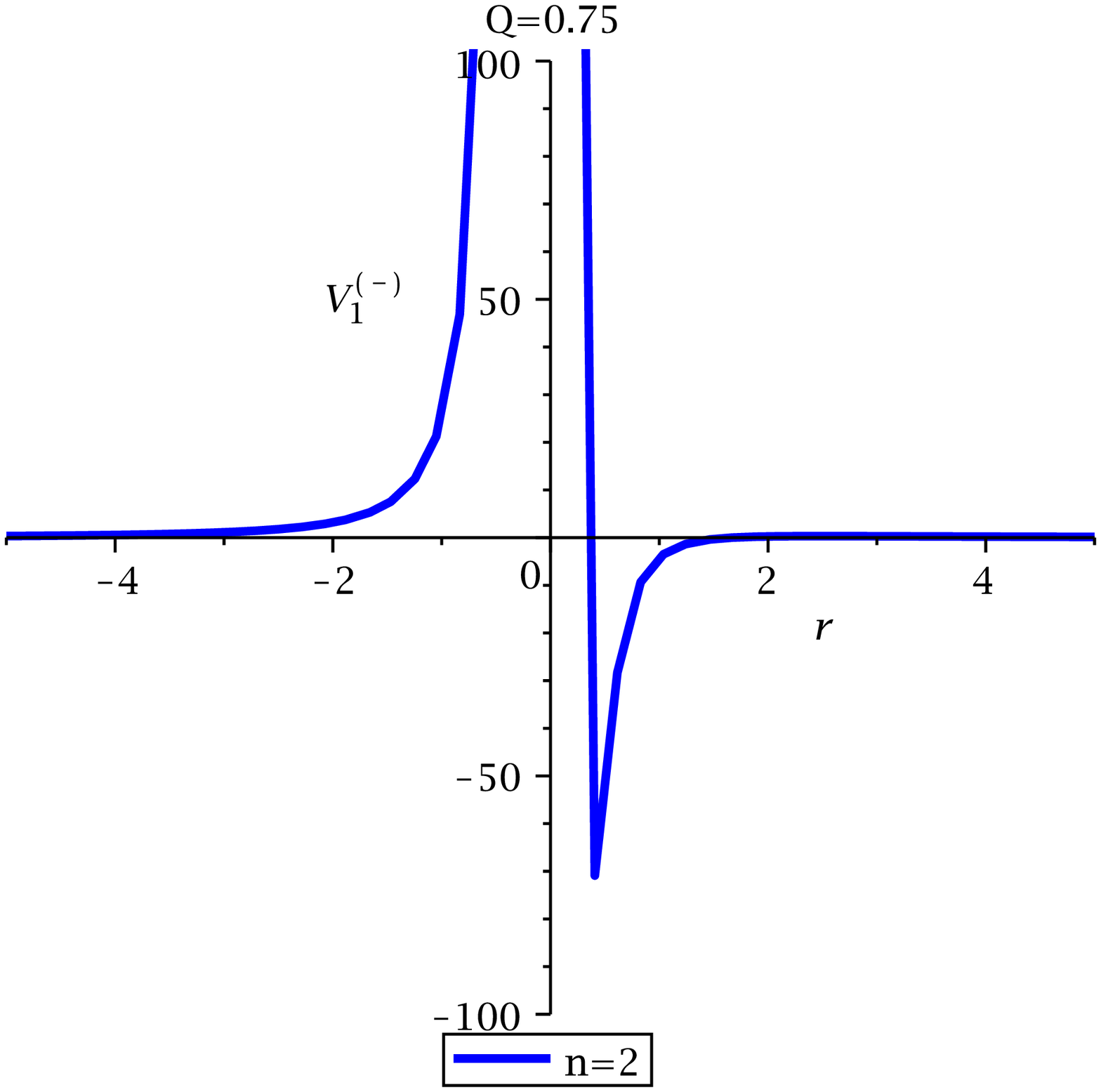}} 
\subfigure[]{
\includegraphics[width=1.5in,angle=0]{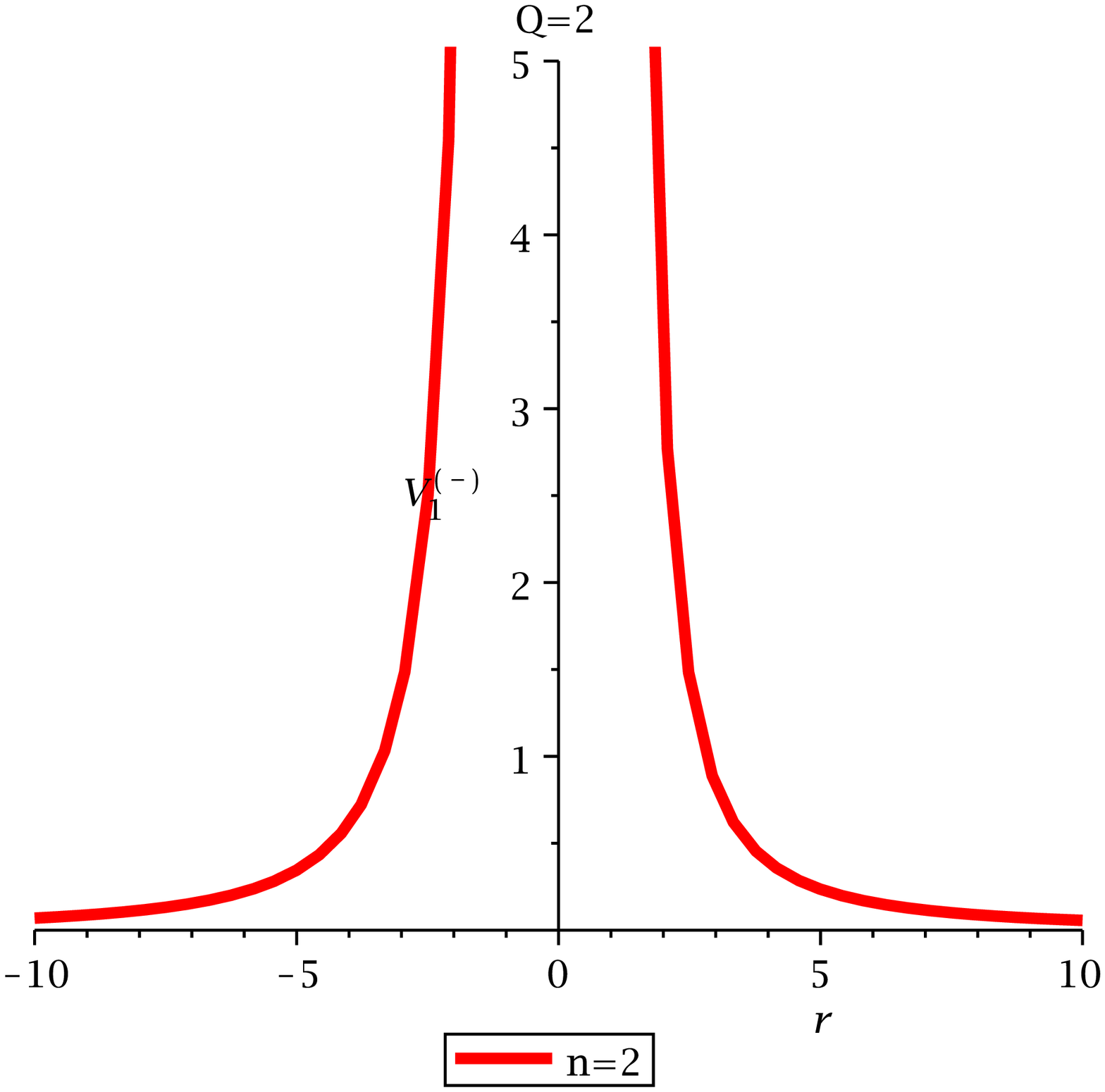}} 
\subfigure[]{
\includegraphics[width=1.5in,angle=0]{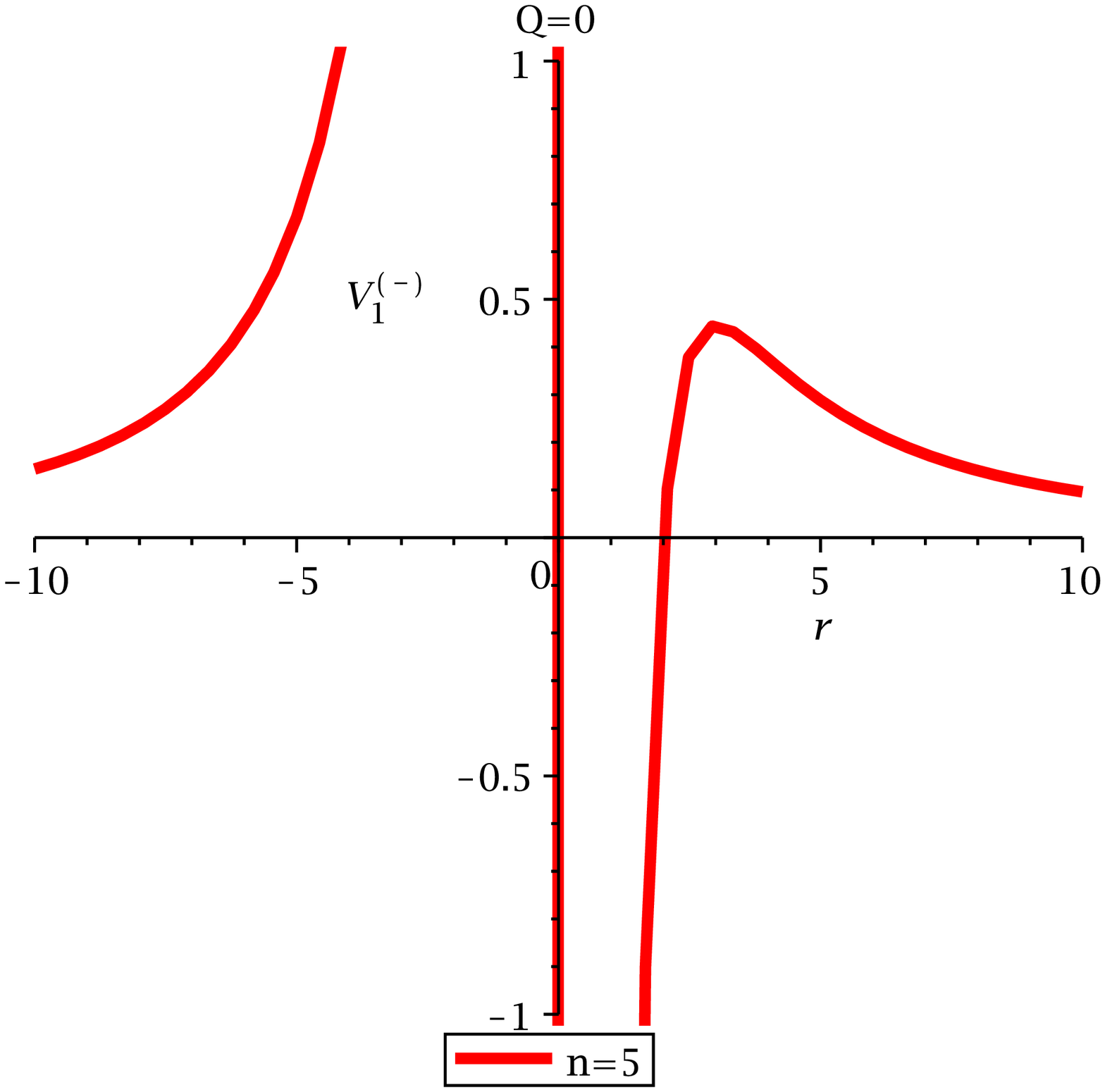}} 
\subfigure[]{
\includegraphics[width=1.5in,angle=0]{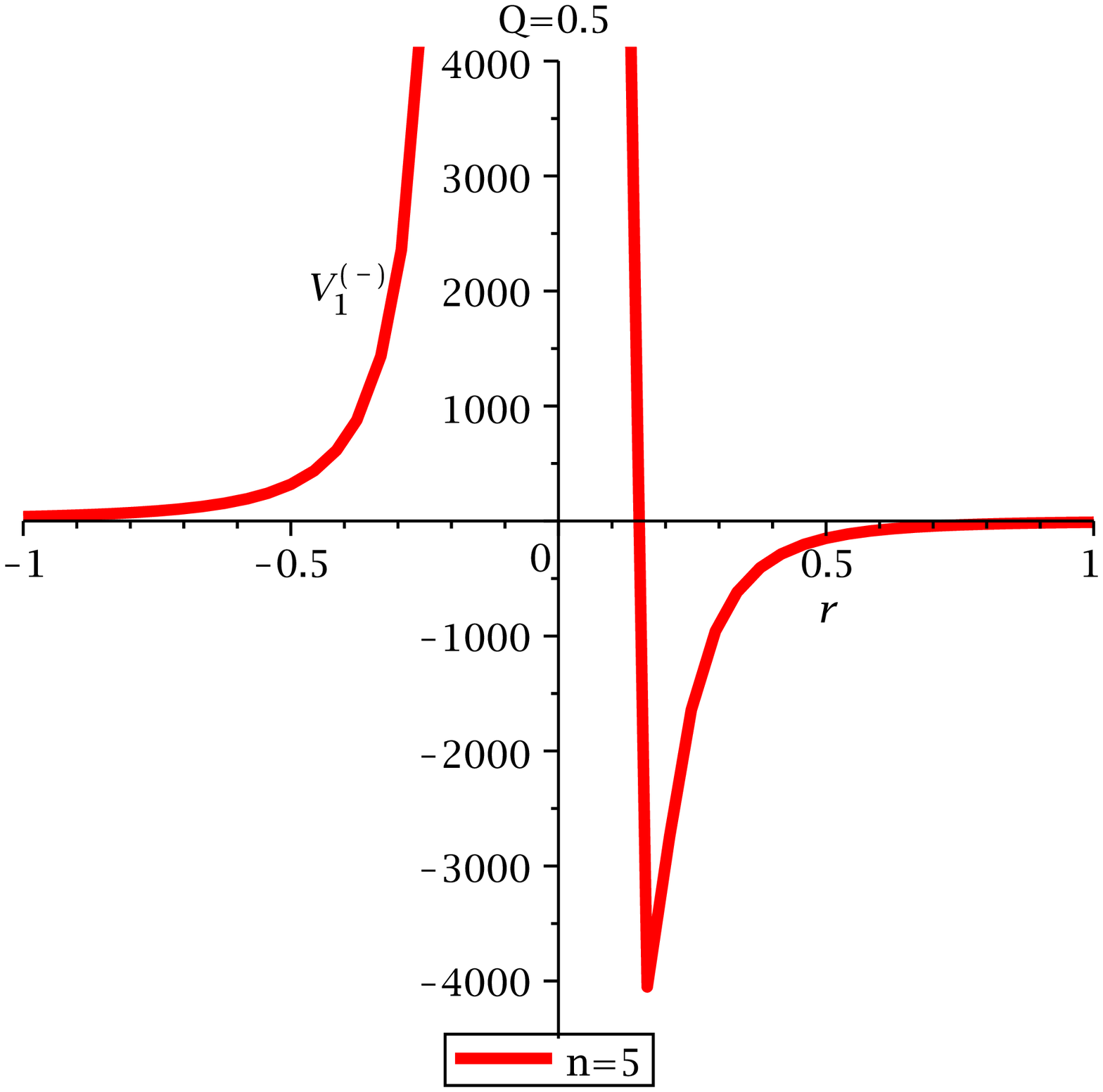}} 
\subfigure[]{
\includegraphics[width=1.5in,angle=0]{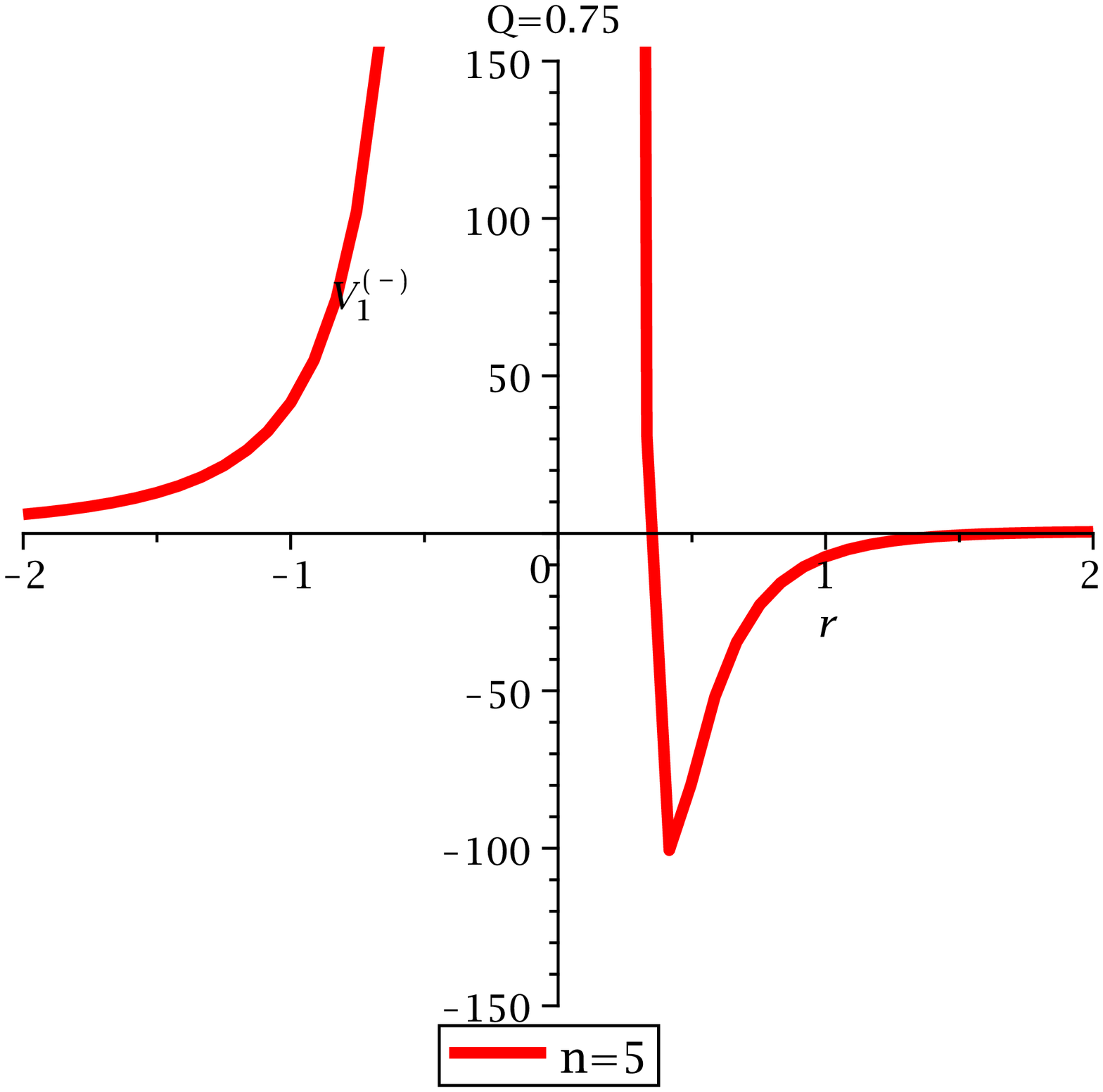}} 
\subfigure[]{
\includegraphics[width=1.5in,angle=0]{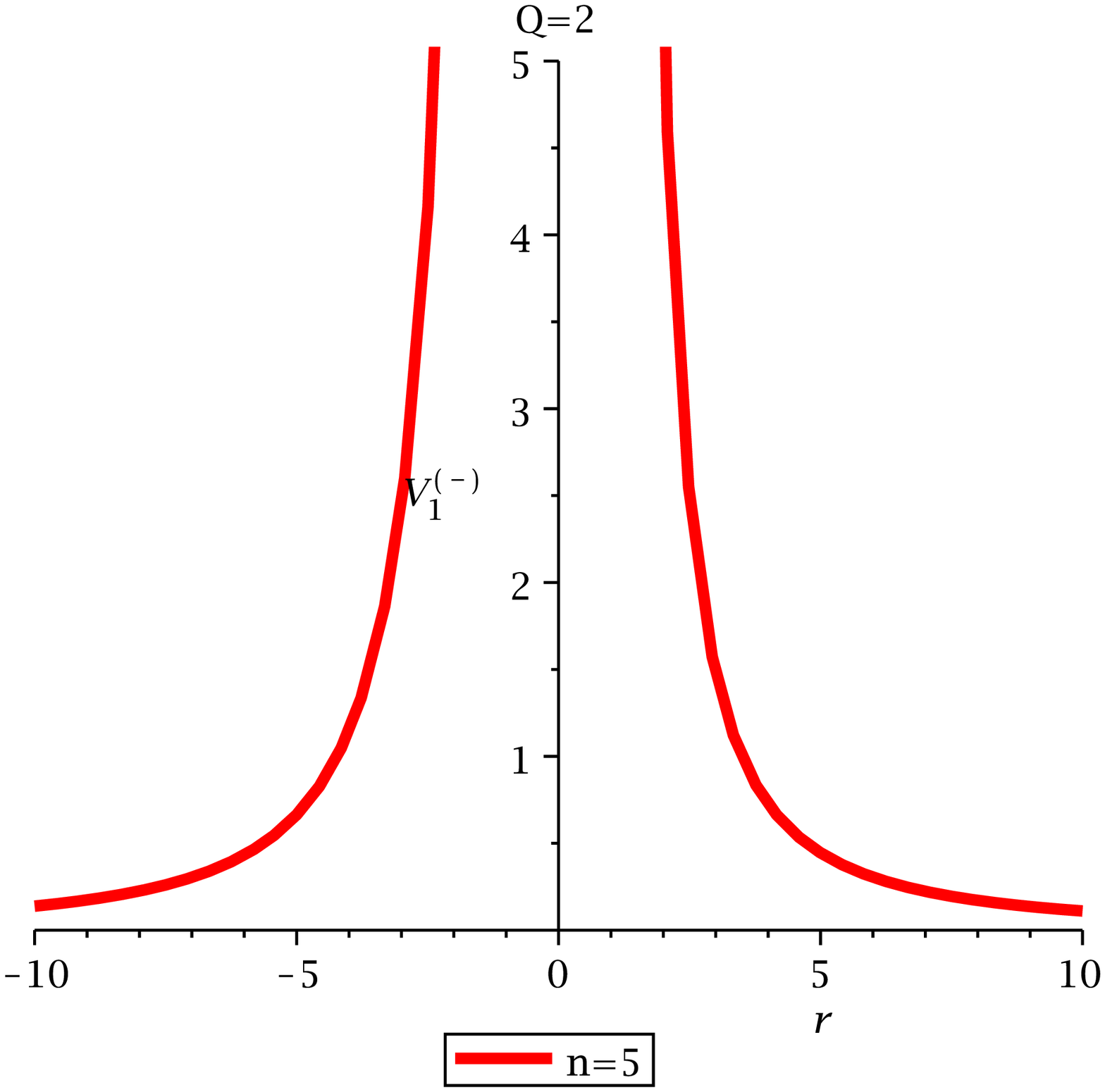}} 
\subfigure[]{
\includegraphics[width=1.5in,angle=0]{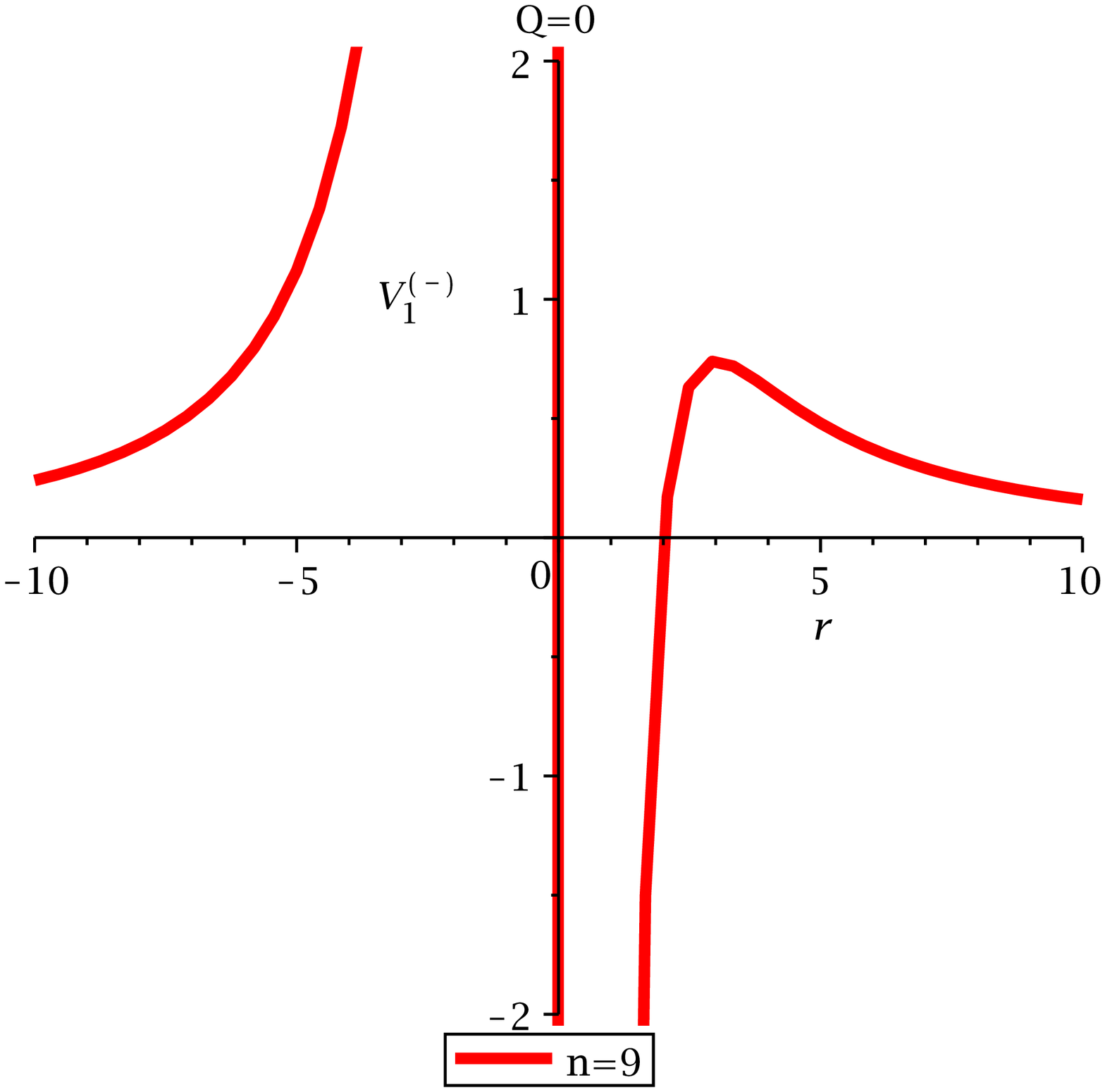}} 
\subfigure[]{
\includegraphics[width=1.5in,angle=0]{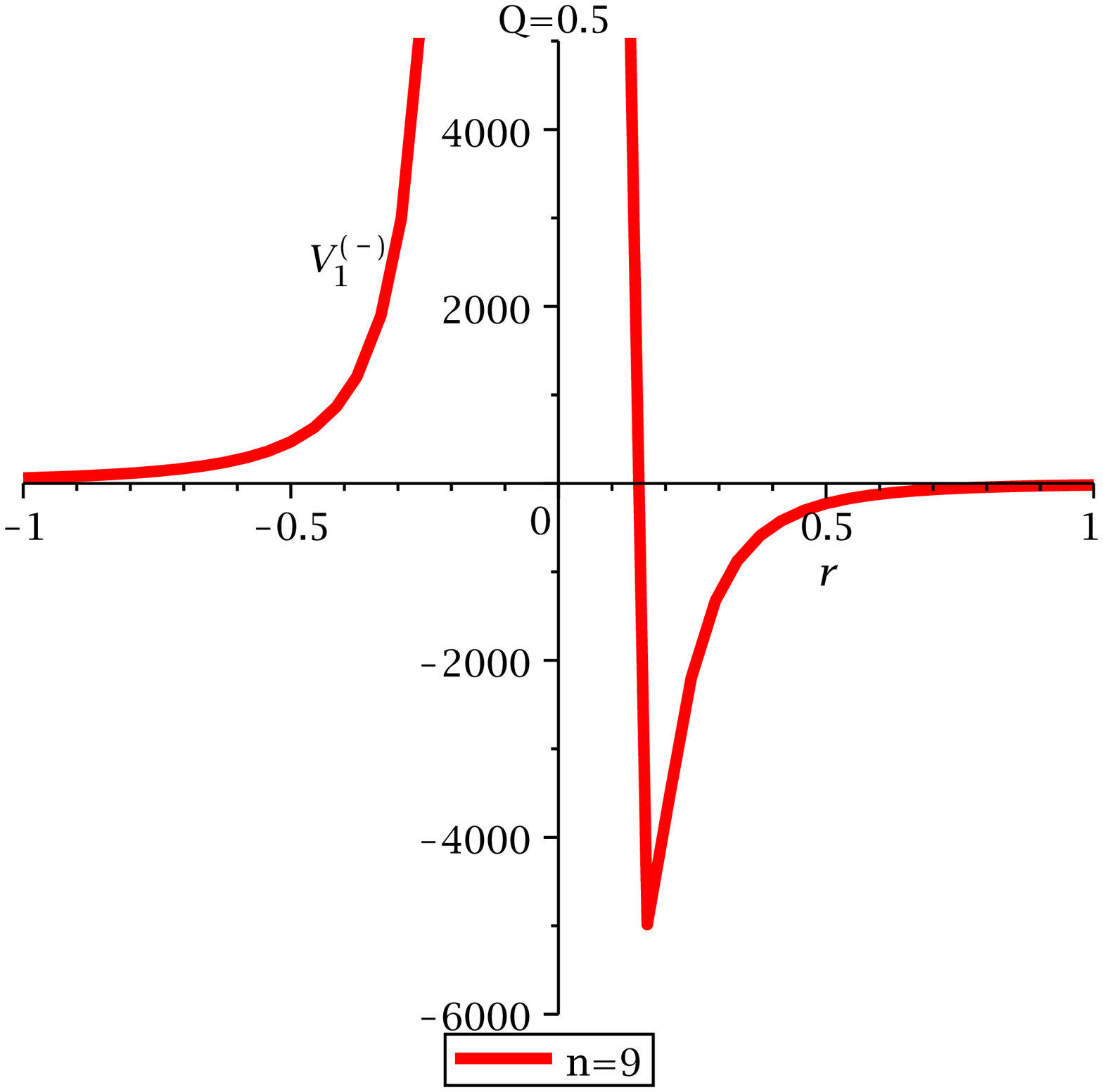}} 
\subfigure[]{
\includegraphics[width=1.5in,angle=0]{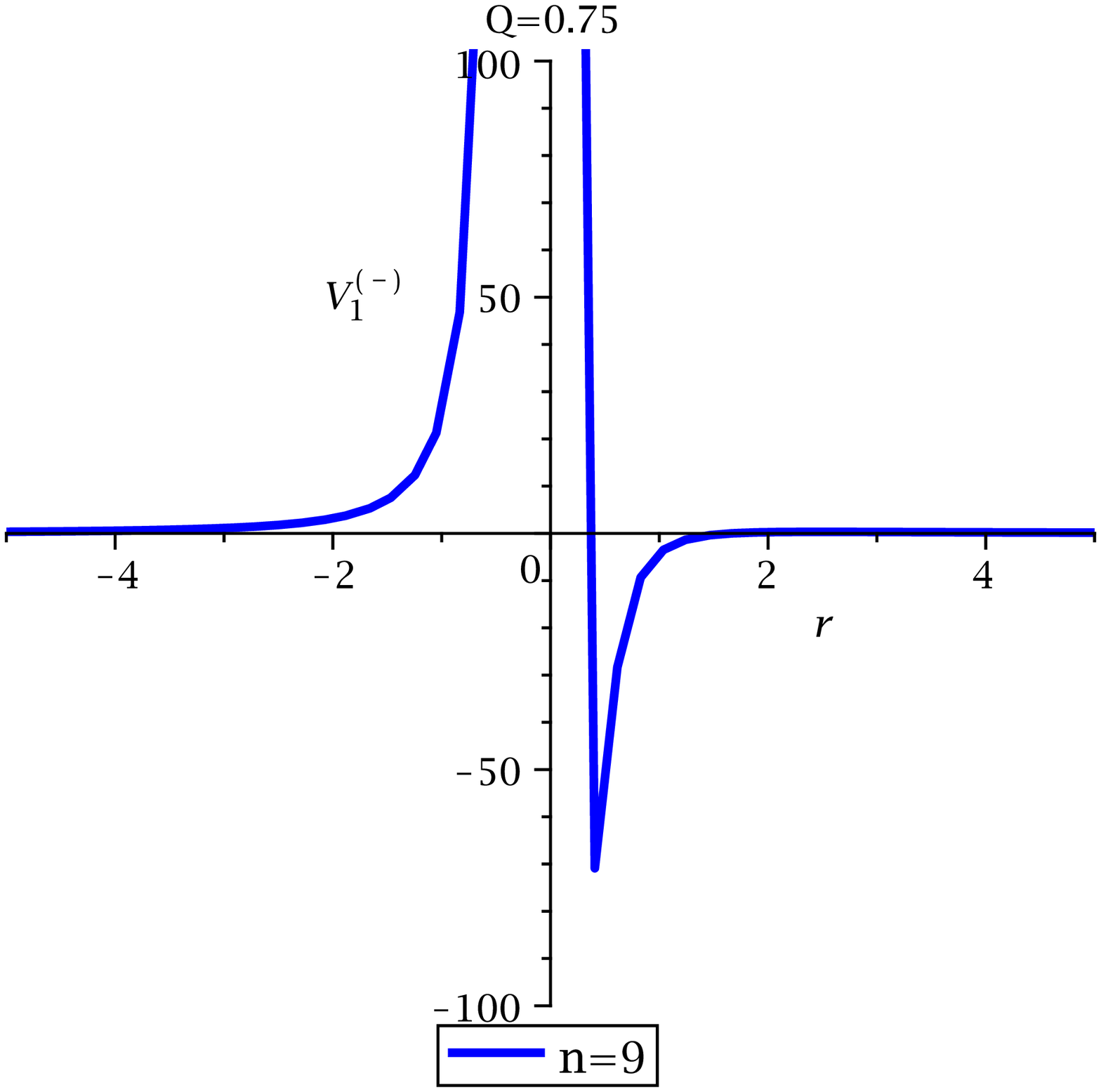}} 
\subfigure[]{
\includegraphics[width=1.5in,angle=0]{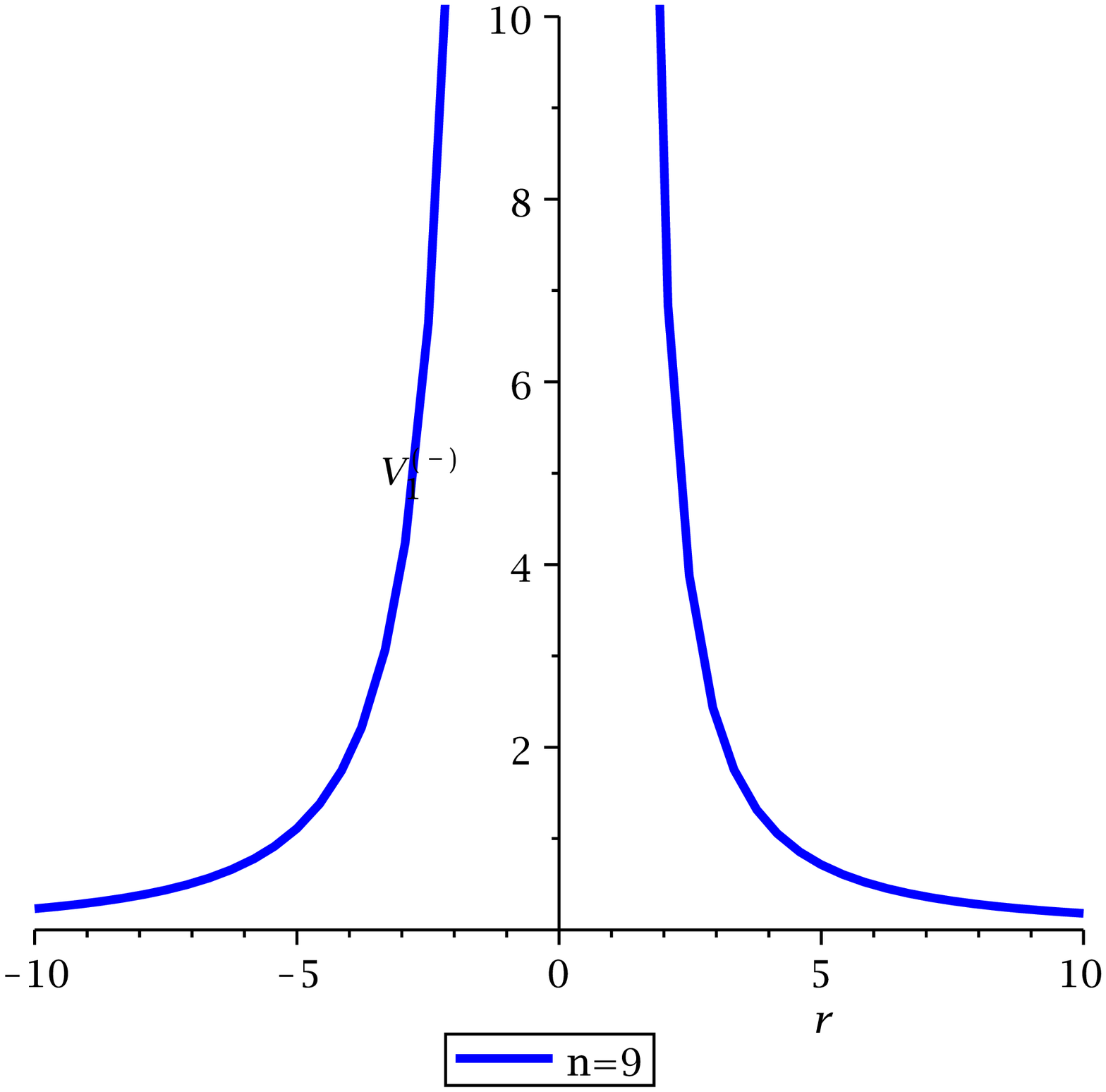}} 
\subfigure[]{
\includegraphics[width=1.5in,angle=0]{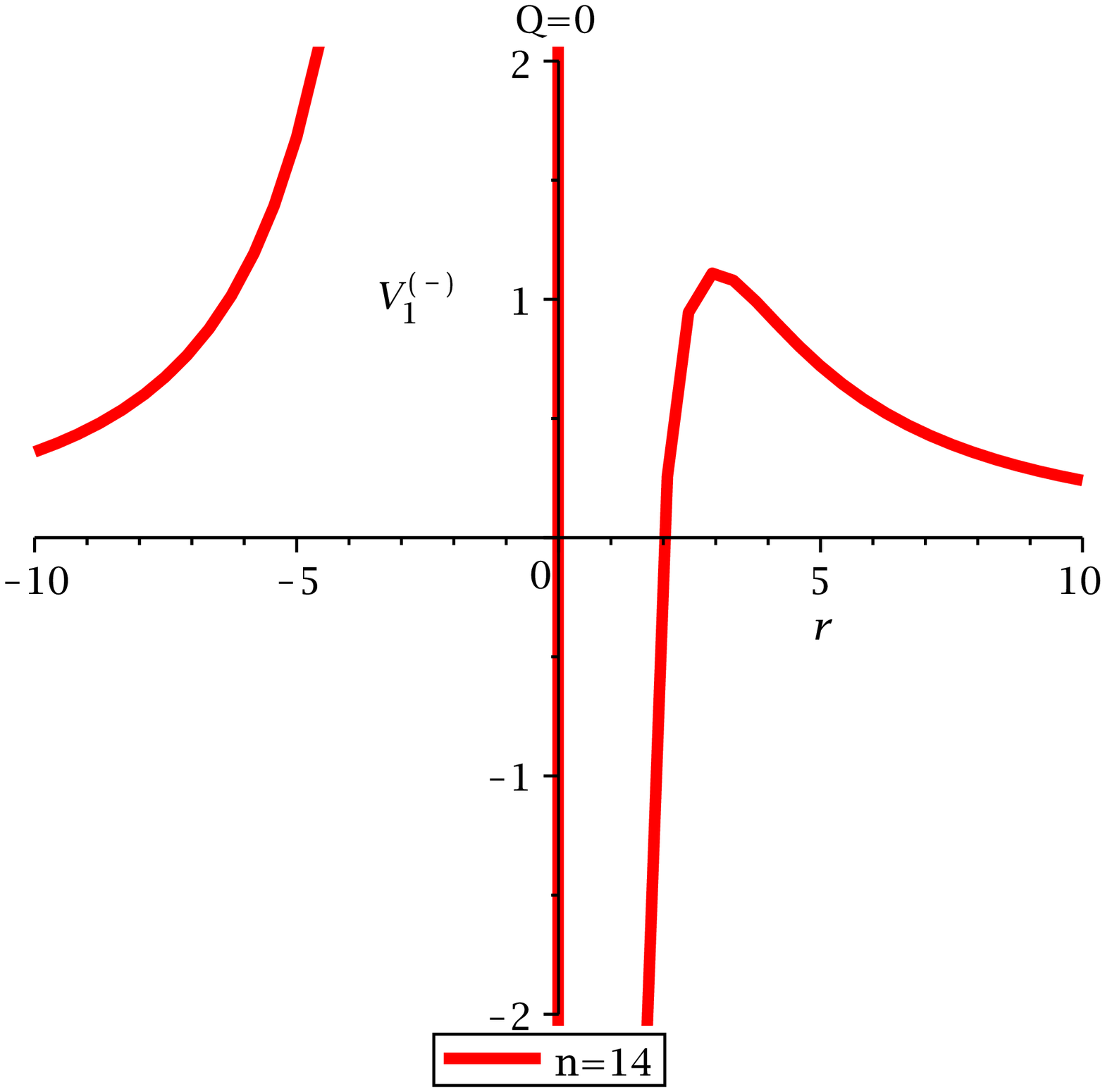}} 
\subfigure[]{
\includegraphics[width=1.5in,angle=0]{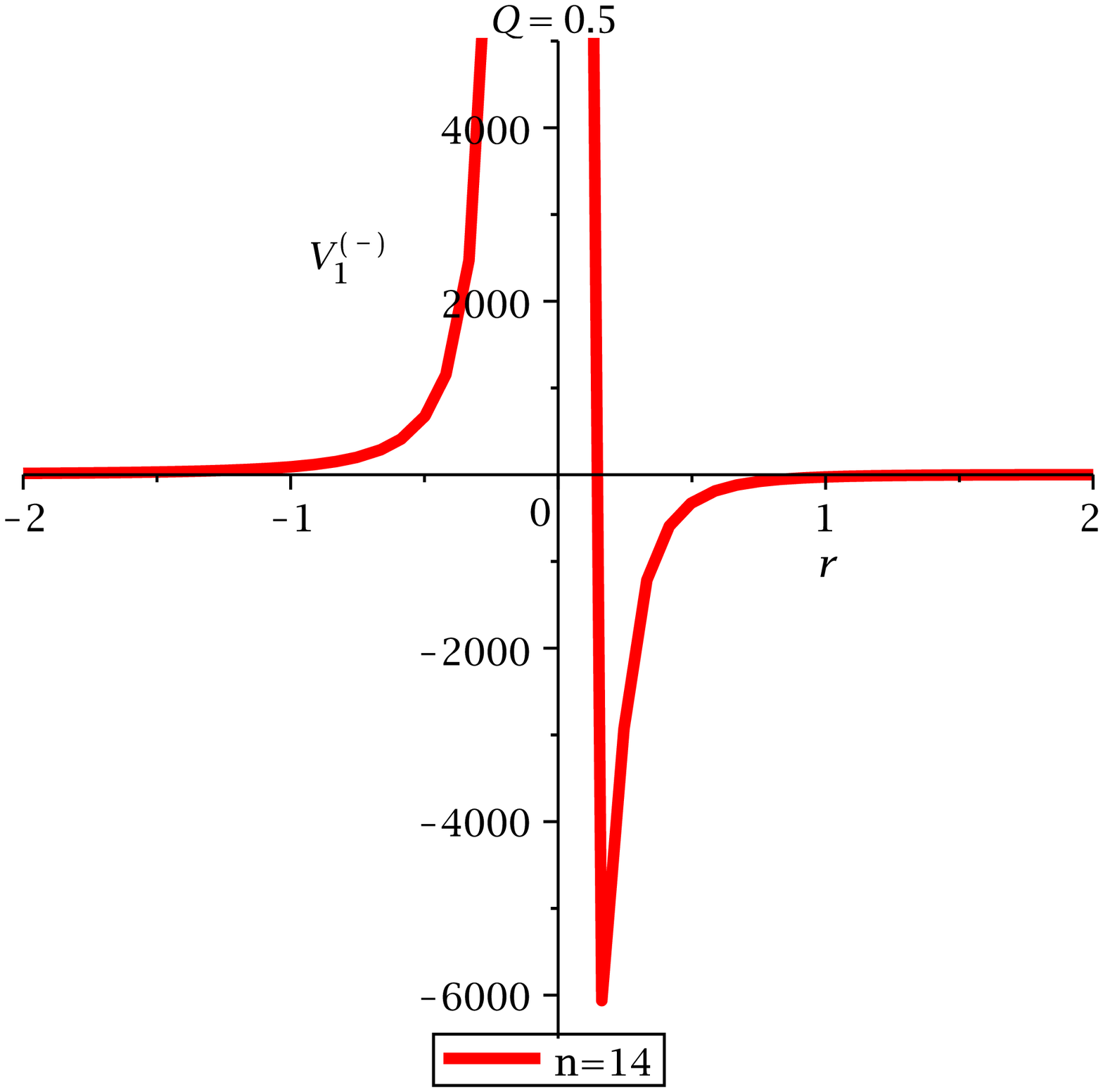}} 
\subfigure[]{
\includegraphics[width=1.5in,angle=0]{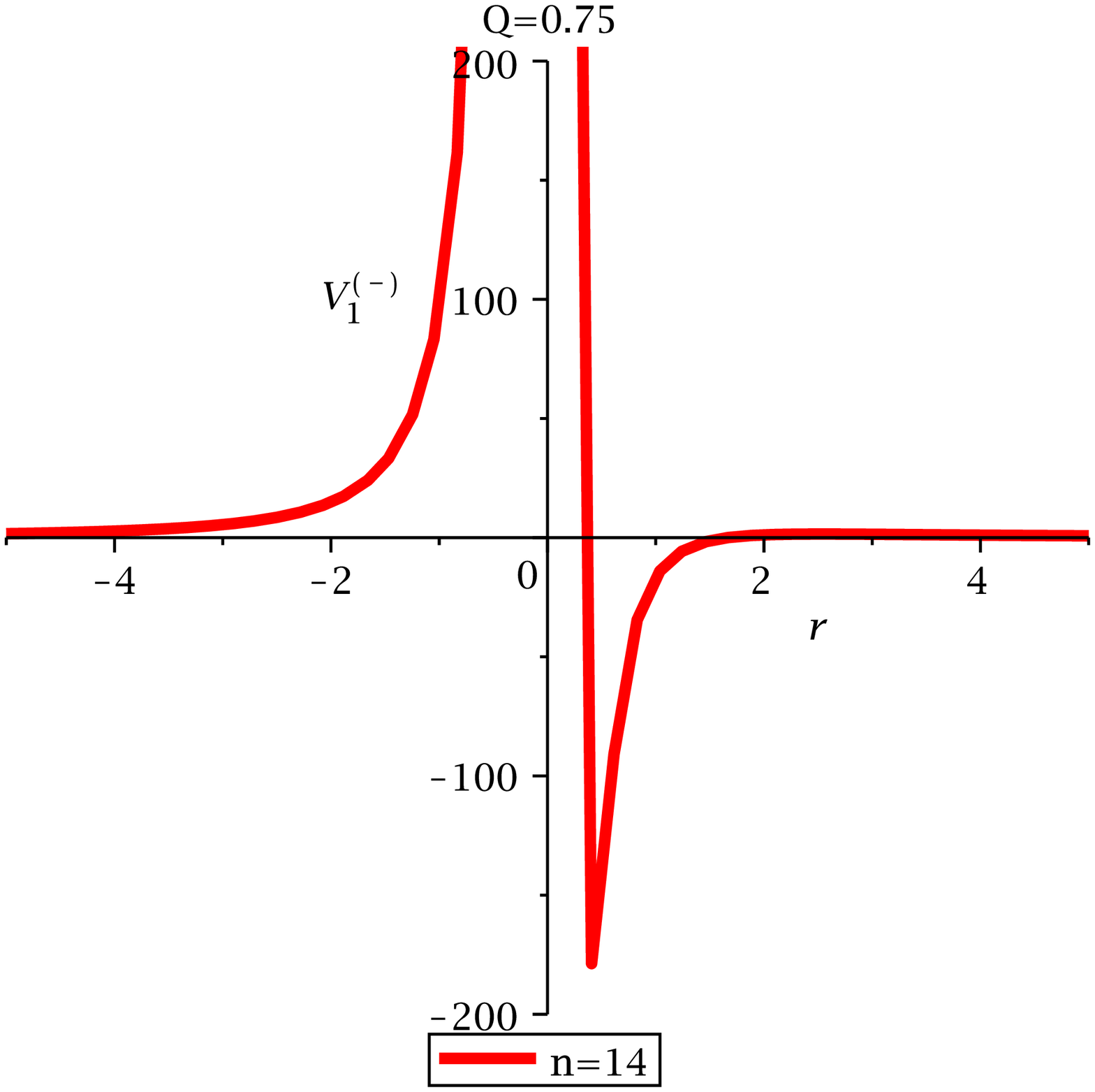}} 
\subfigure[]{
\includegraphics[width=1.5in,angle=0]{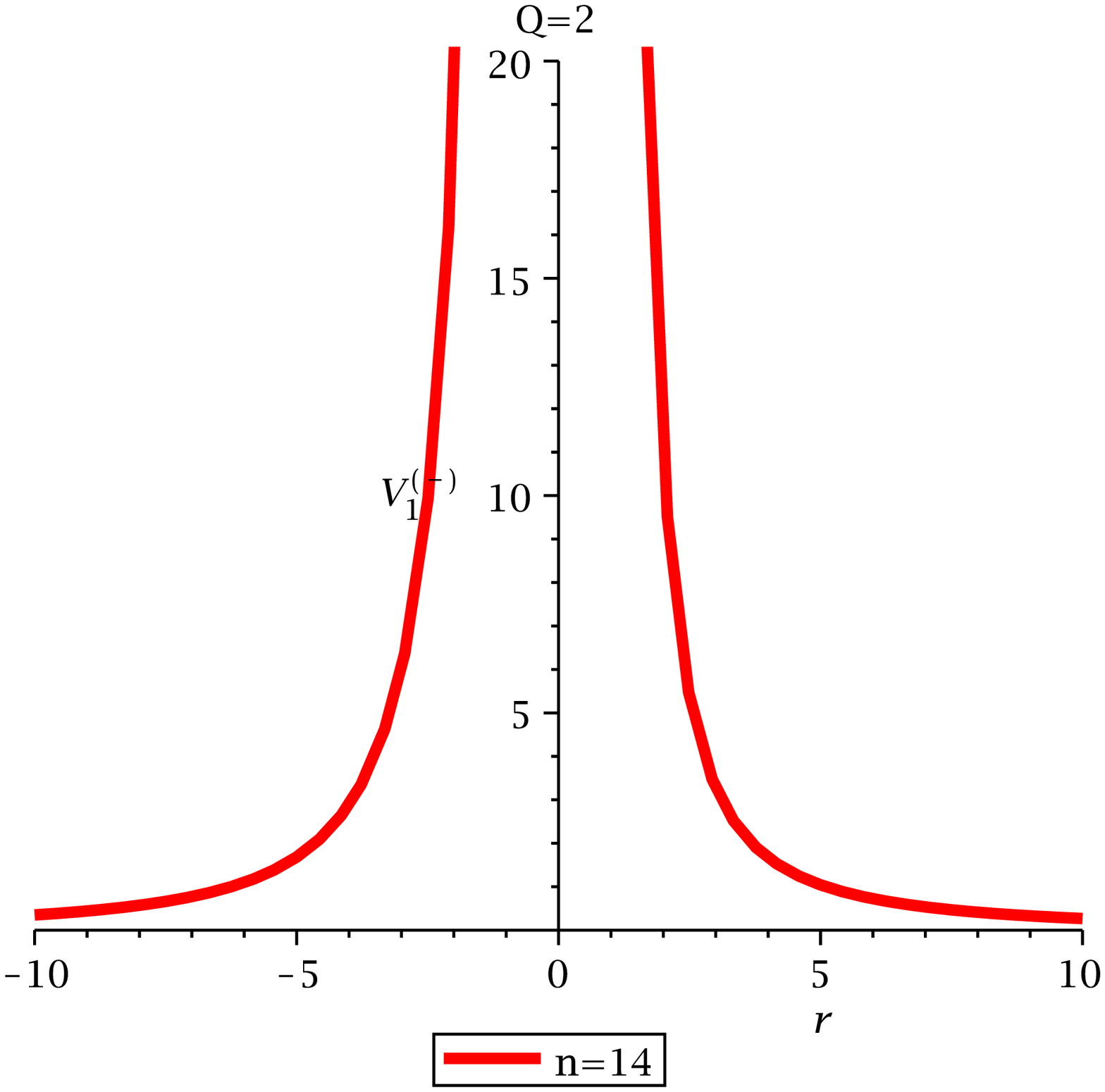}} 
\caption{The shape effective potential~($V_{1}^{(-)}$) barriers surrounding the NS and non-extremal 
Reissner Nordstr\"{o}m  BH for axial perturbations. In the plot $Q_{\ast}=0$ corresponds to Schwarzschild 
BH. While $Q_{\ast}=0.5$ and $Q_{\ast}=0.75$ correspond to non-extremal BH. Finally 
$Q_{\ast}=2$ corresponds to NS. }
\label{nsaxial} 
\end{center}
\end{figure}
%%%%%%%%%%%%%%%%%%%%%%%%%%%%%%
\begin{figure}
\begin{center}
\subfigure[]{
\includegraphics[width=1.5in,angle=0]{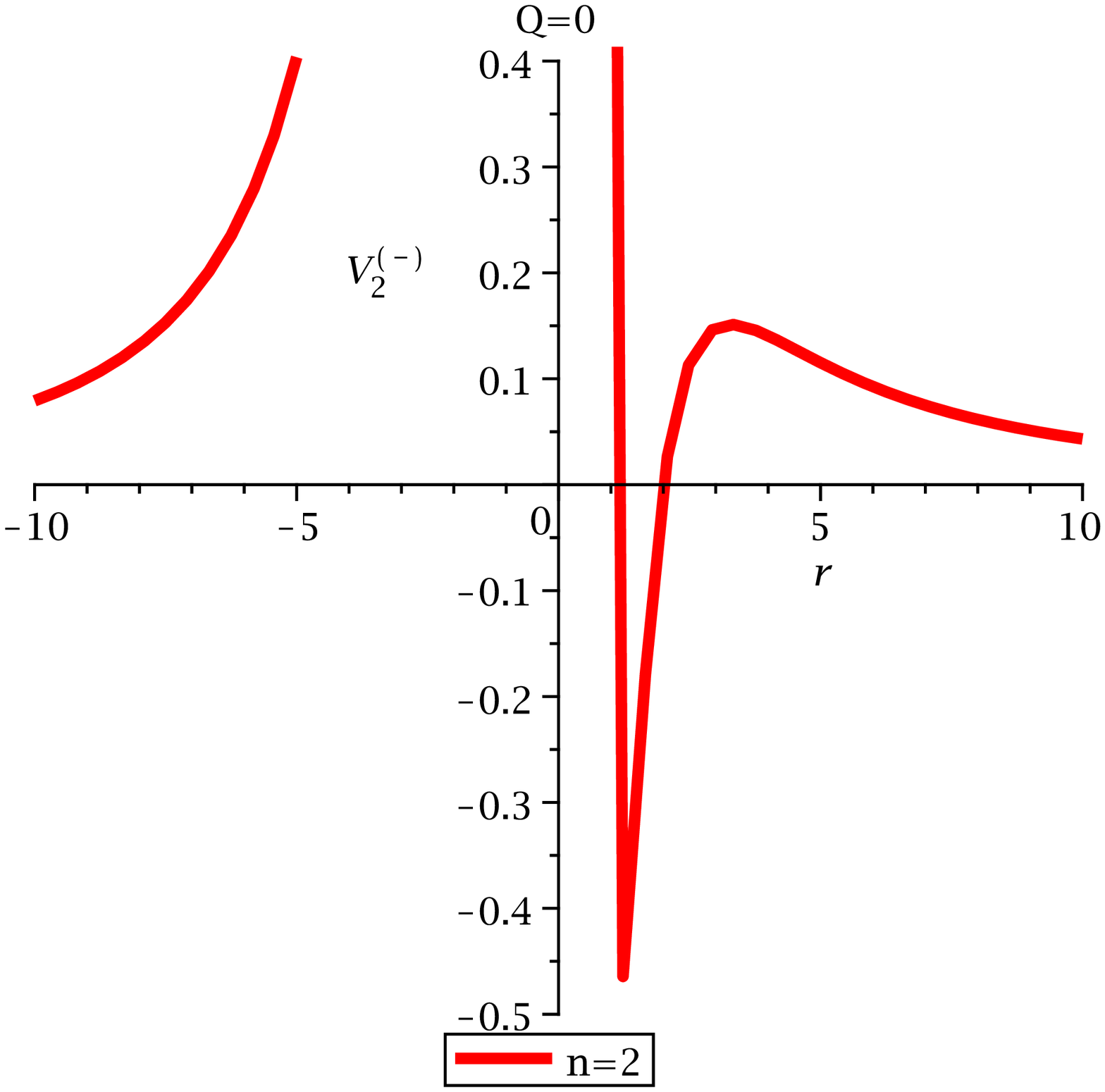}} 
\subfigure[]{
\includegraphics[width=1.5in,angle=0]{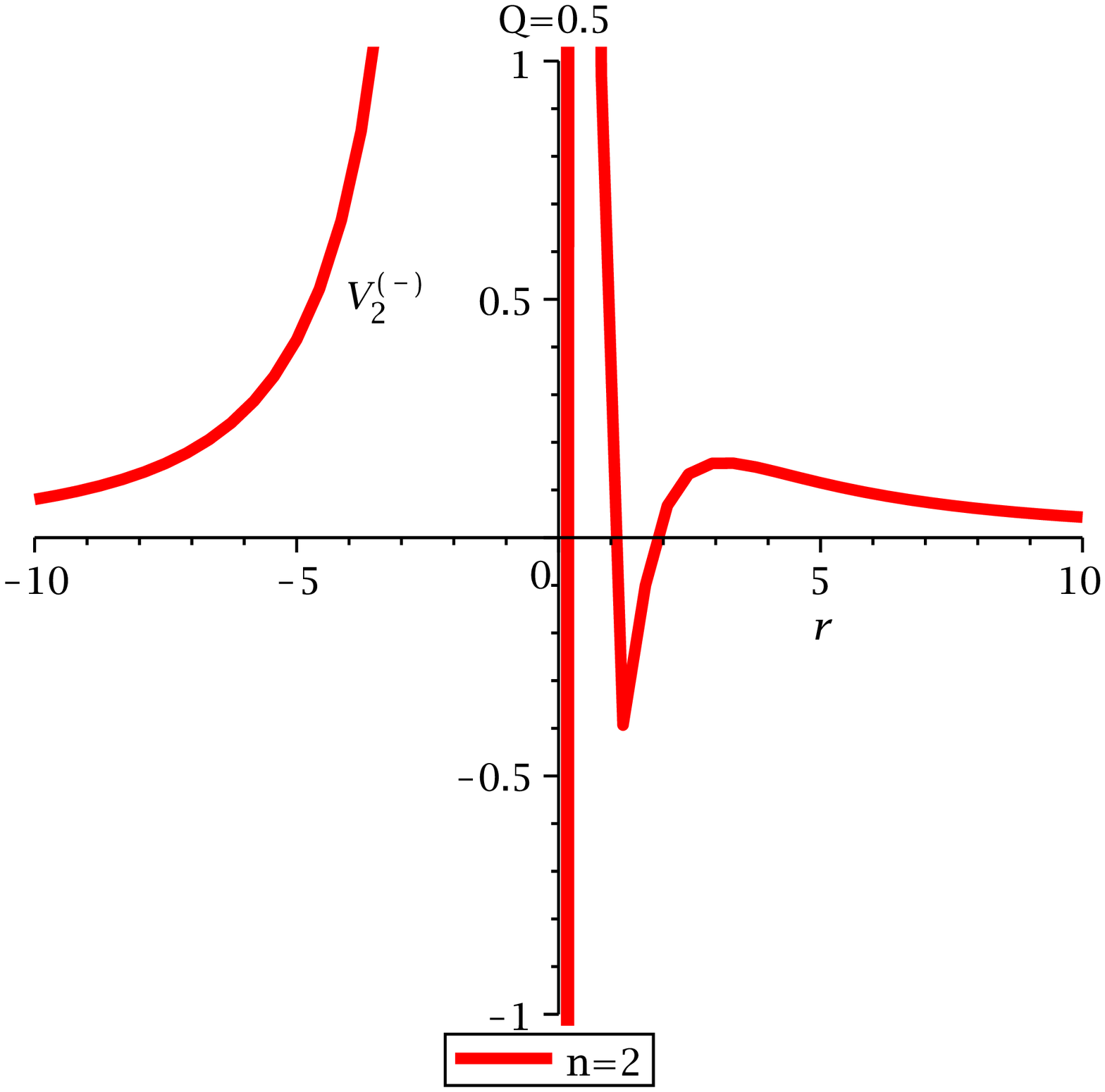}} 
\subfigure[]{
\includegraphics[width=1.5in,angle=0]{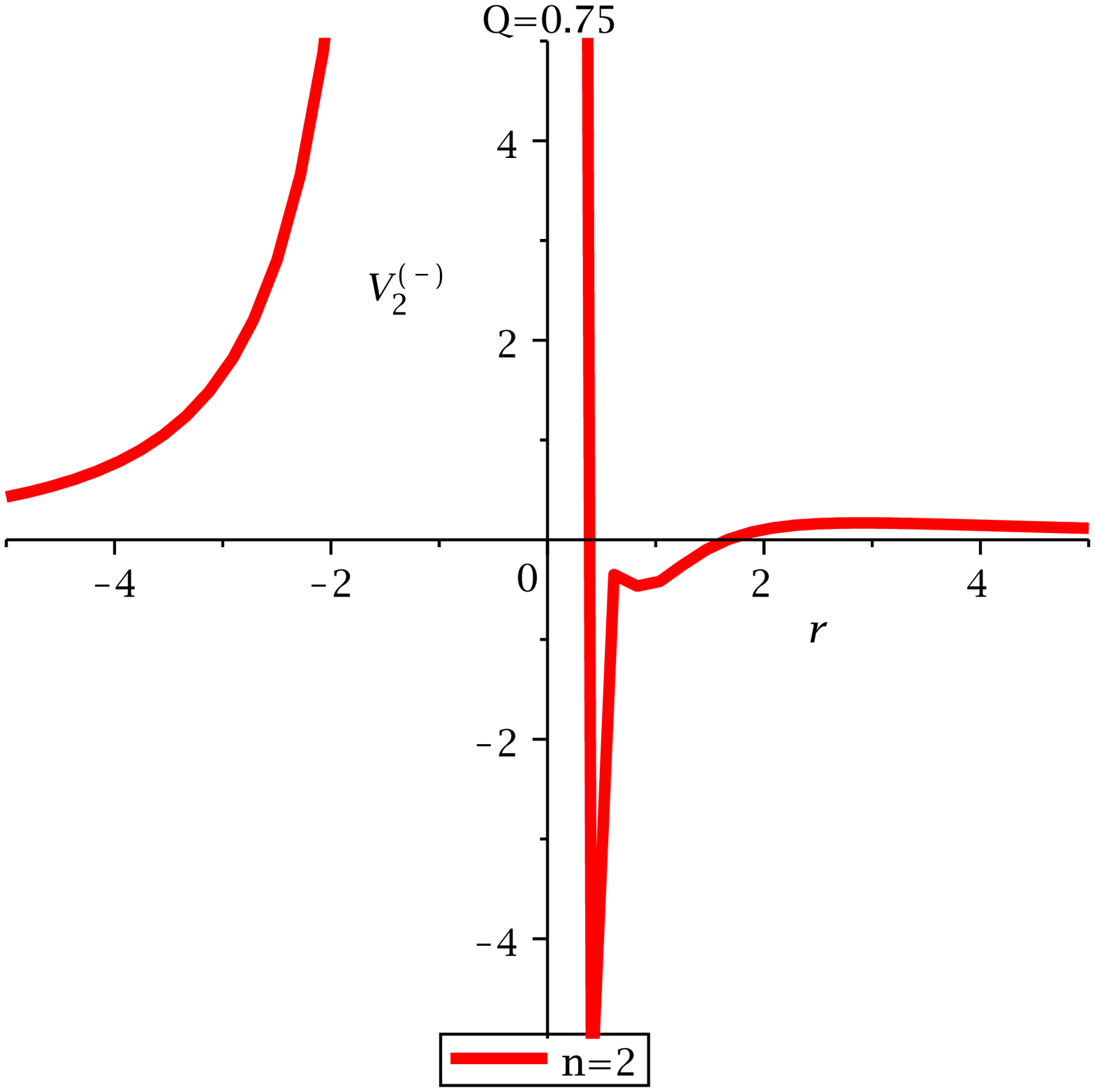}} 
\subfigure[]{
\includegraphics[width=1.5in,angle=0]{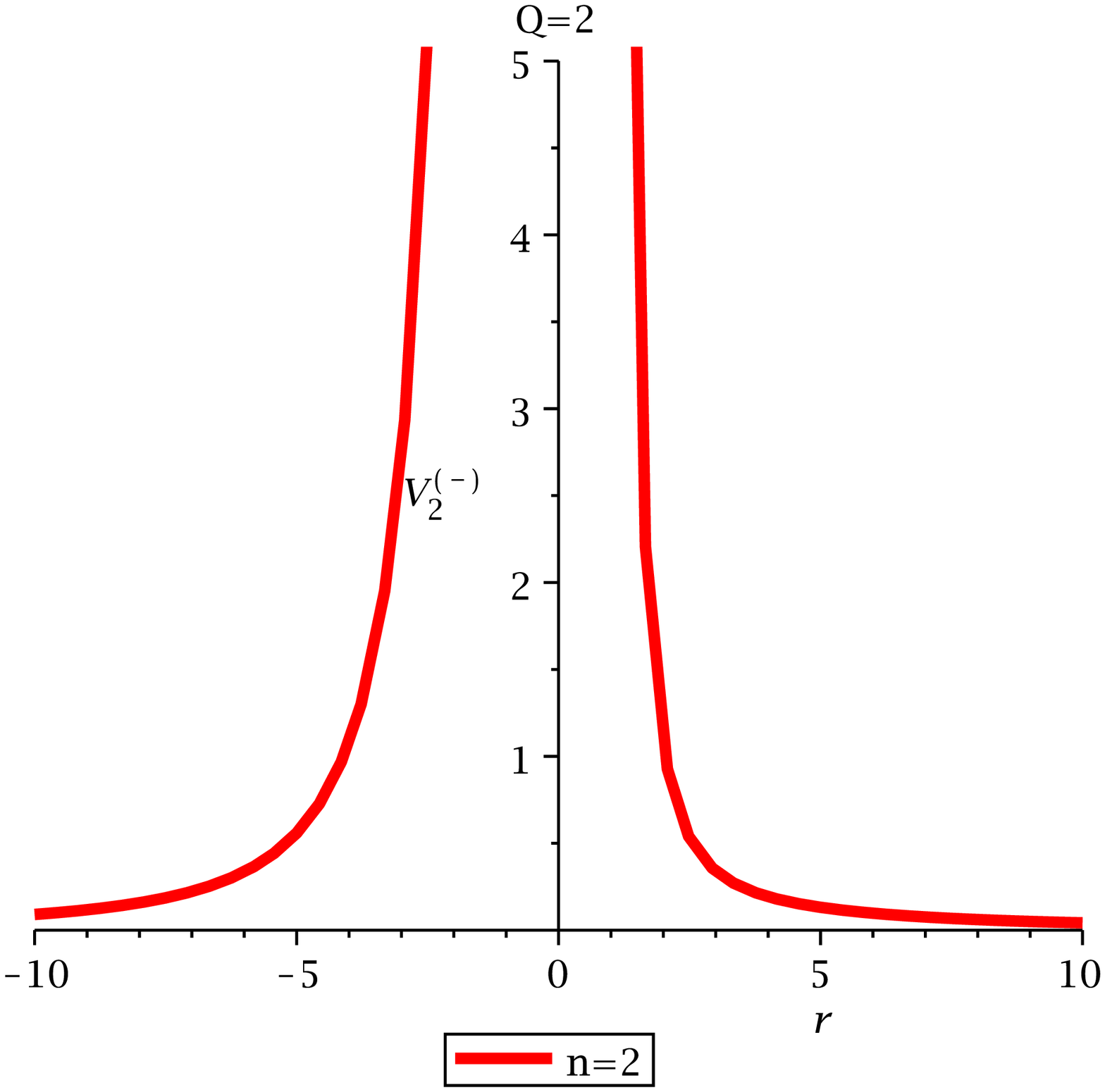}} 
\subfigure[]{
\includegraphics[width=1.5in,angle=0]{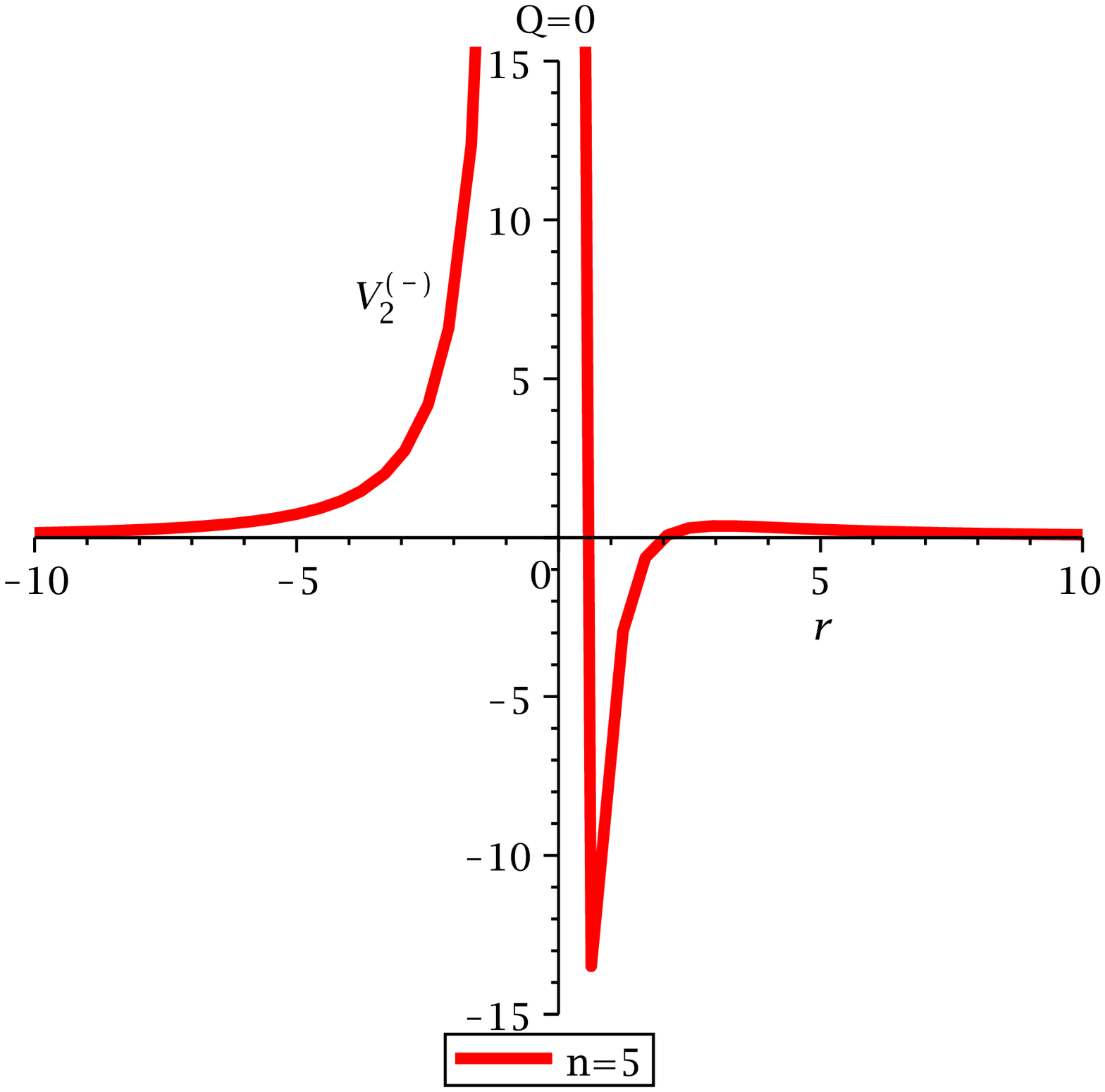}} 
\subfigure[]{
\includegraphics[width=1.5in,angle=0]{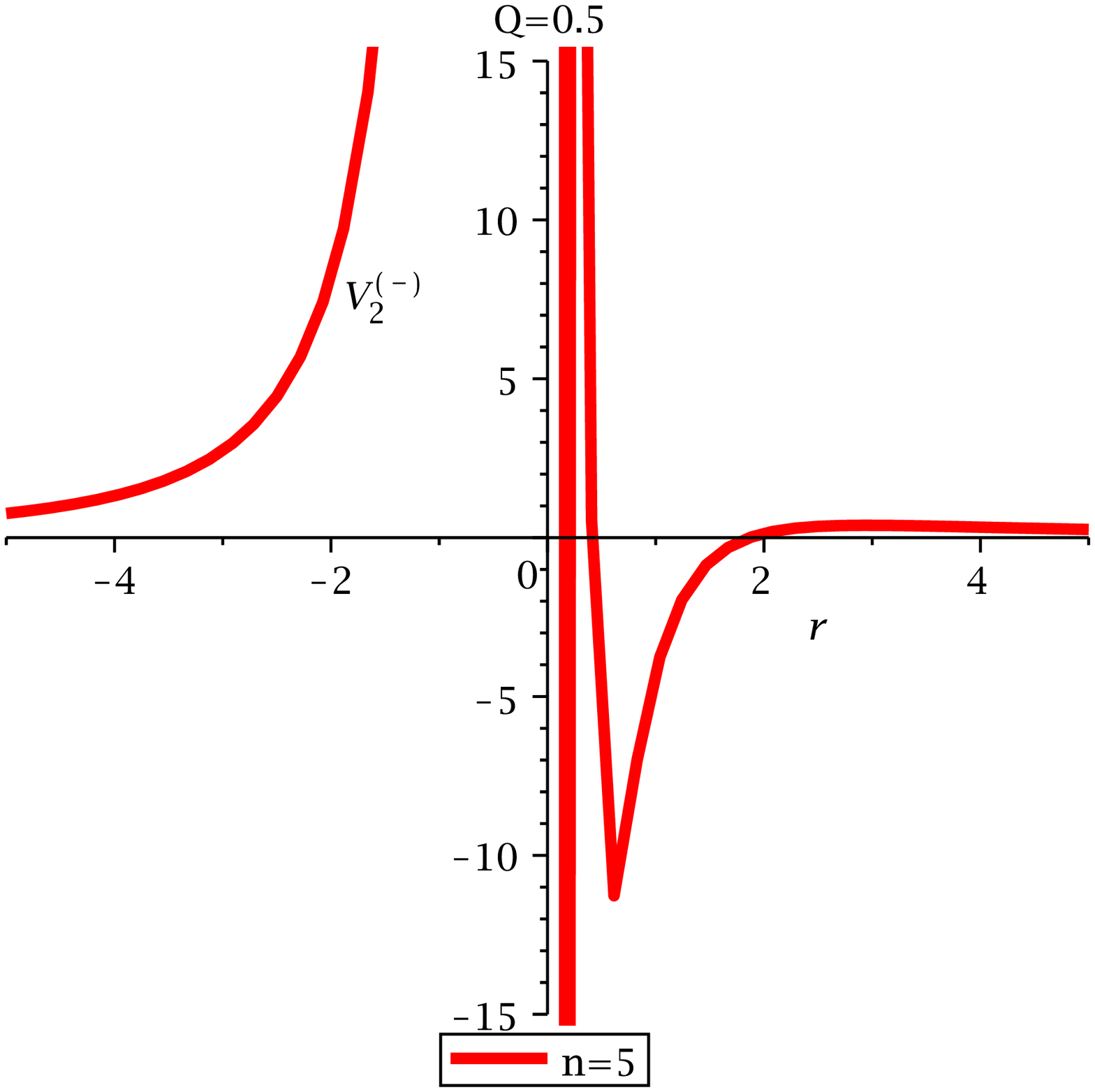}} 
\subfigure[]{
\includegraphics[width=1.5in,angle=0]{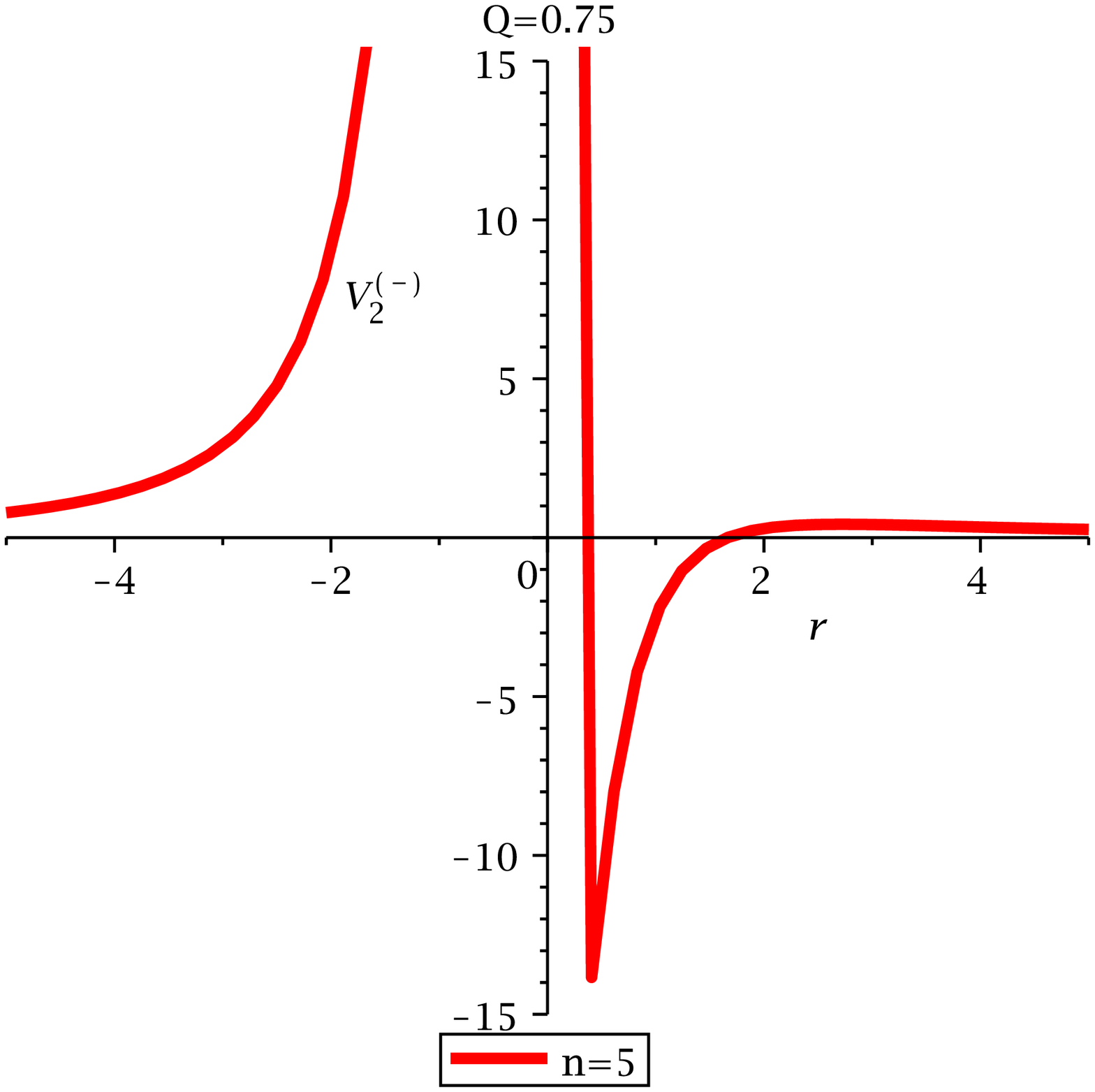}} 
\subfigure[]{
\includegraphics[width=1.5in,angle=0]{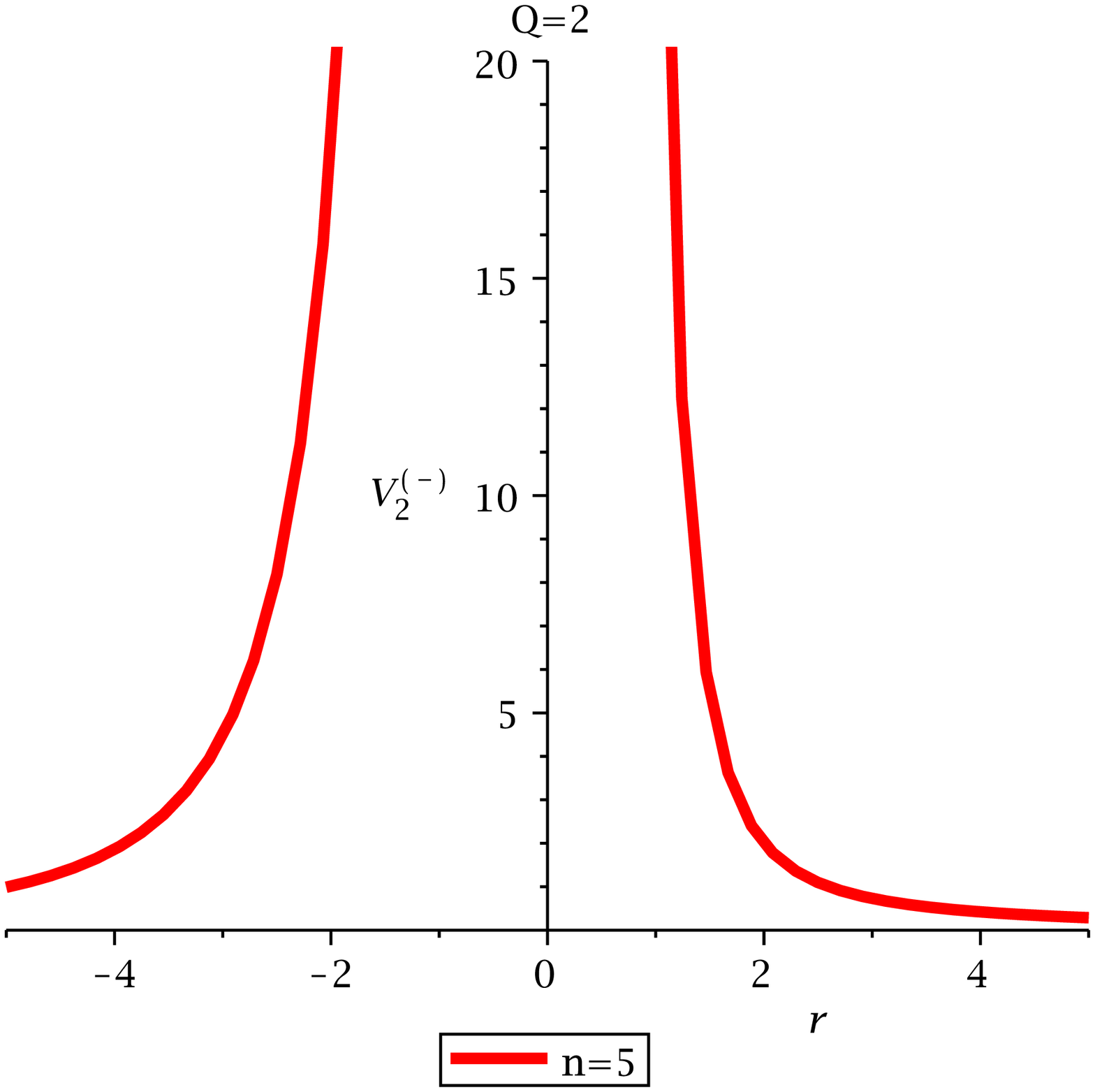}}  
\subfigure[]{
\includegraphics[width=1.5in,angle=0]{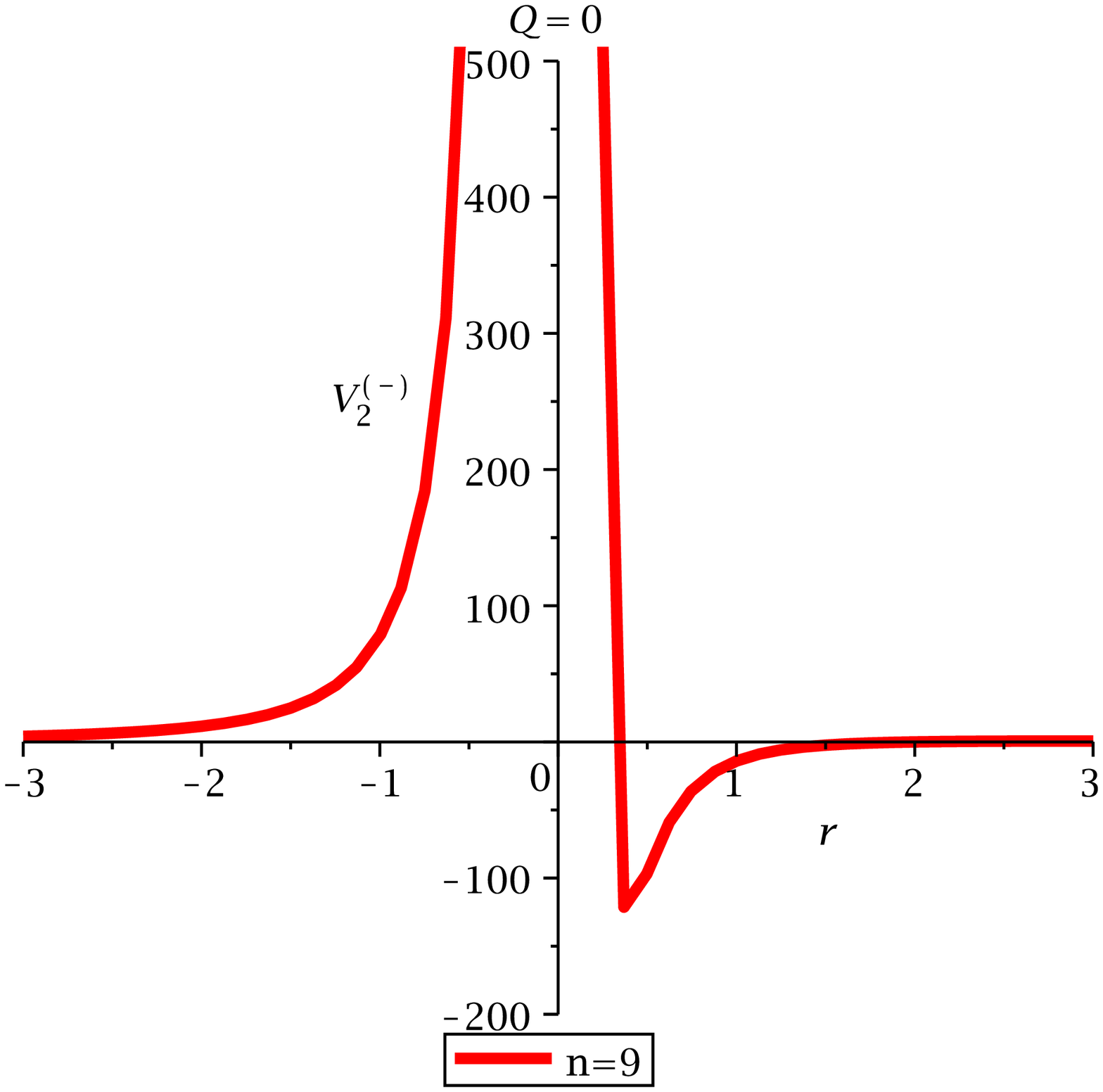}} 
\subfigure[]{
\includegraphics[width=1.5in,angle=0]{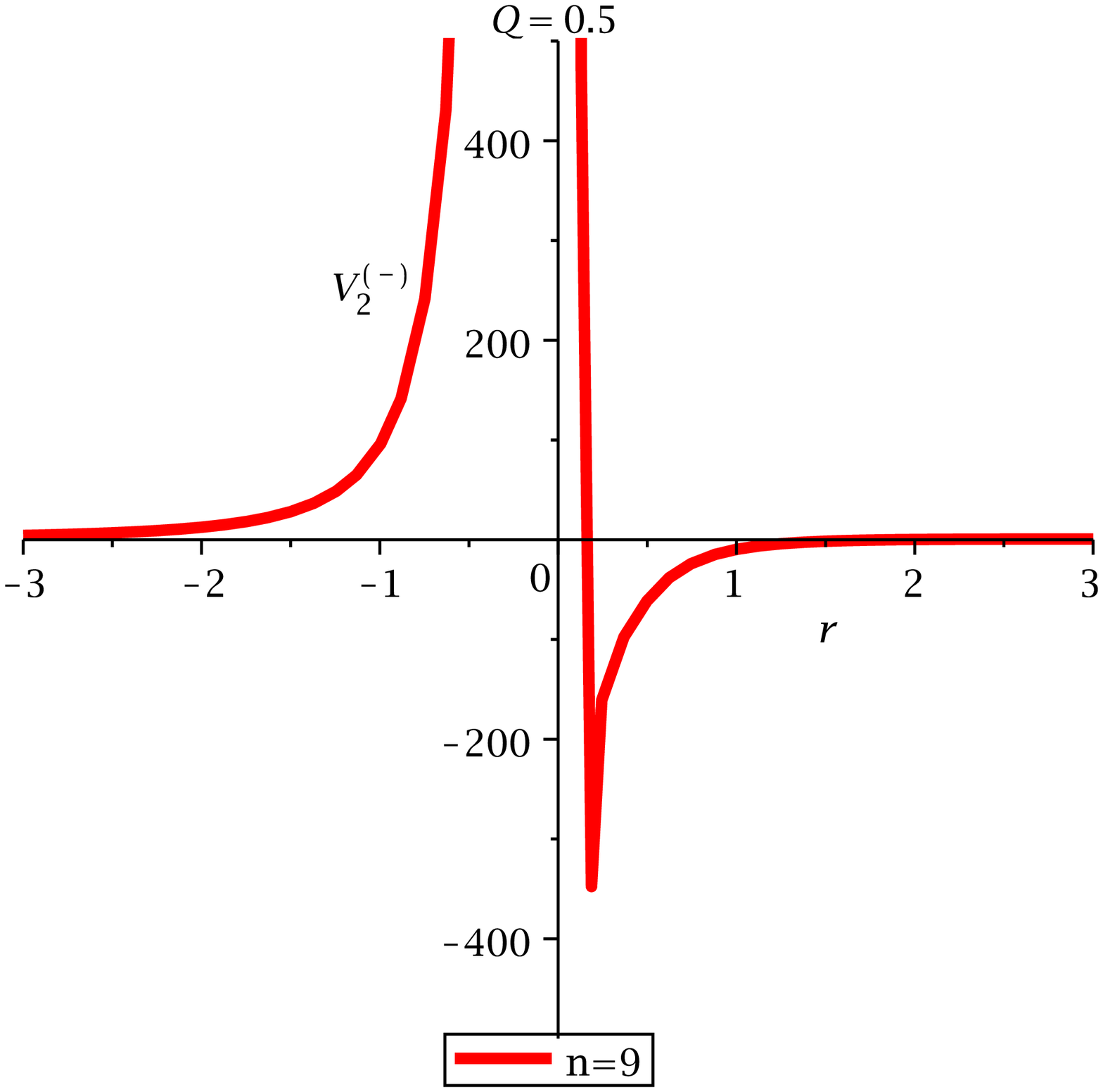}} 
\subfigure[]{
\includegraphics[width=1.5in,angle=0]{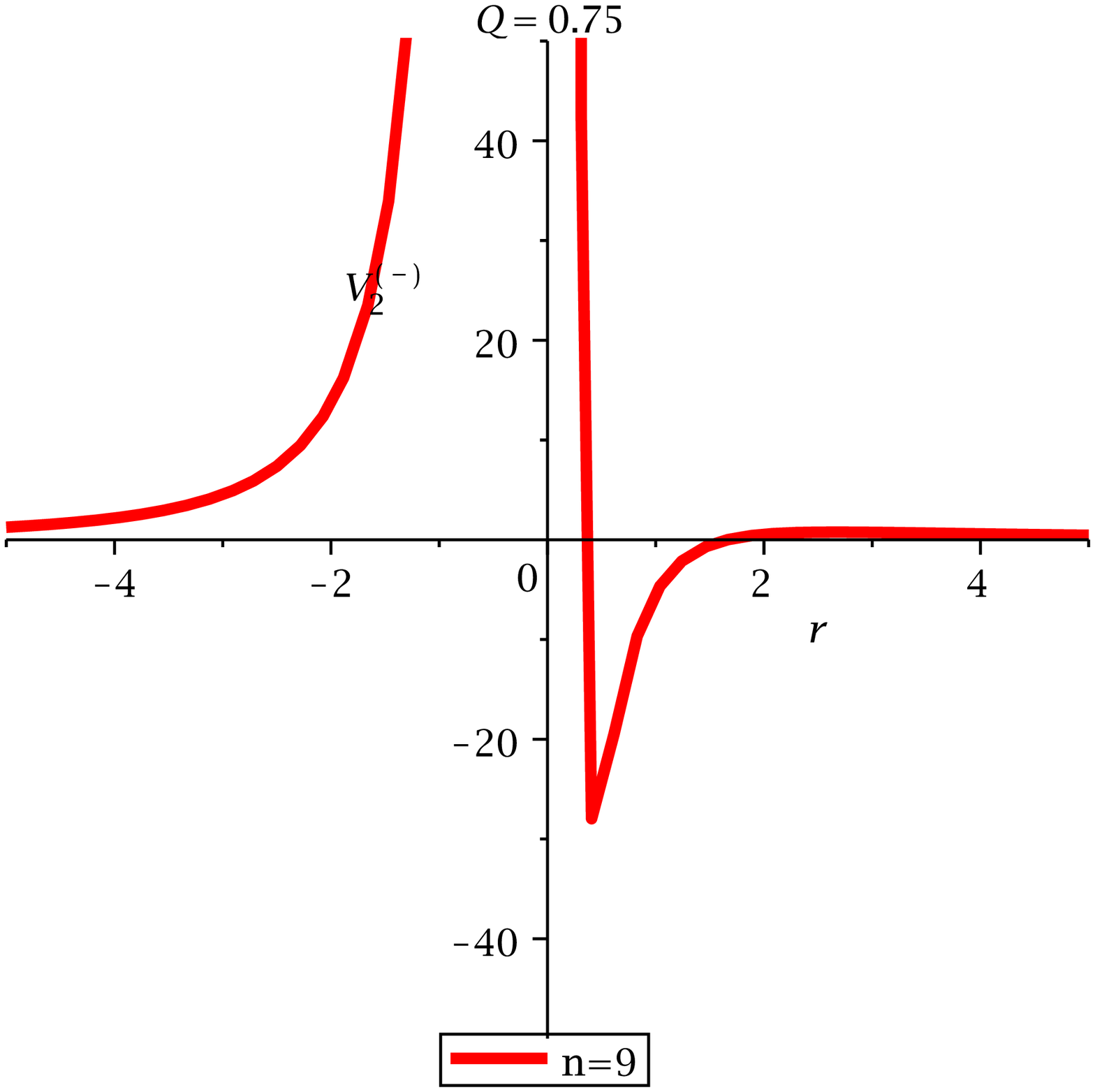}} 
\subfigure[]{
\includegraphics[width=1.5in,angle=0]{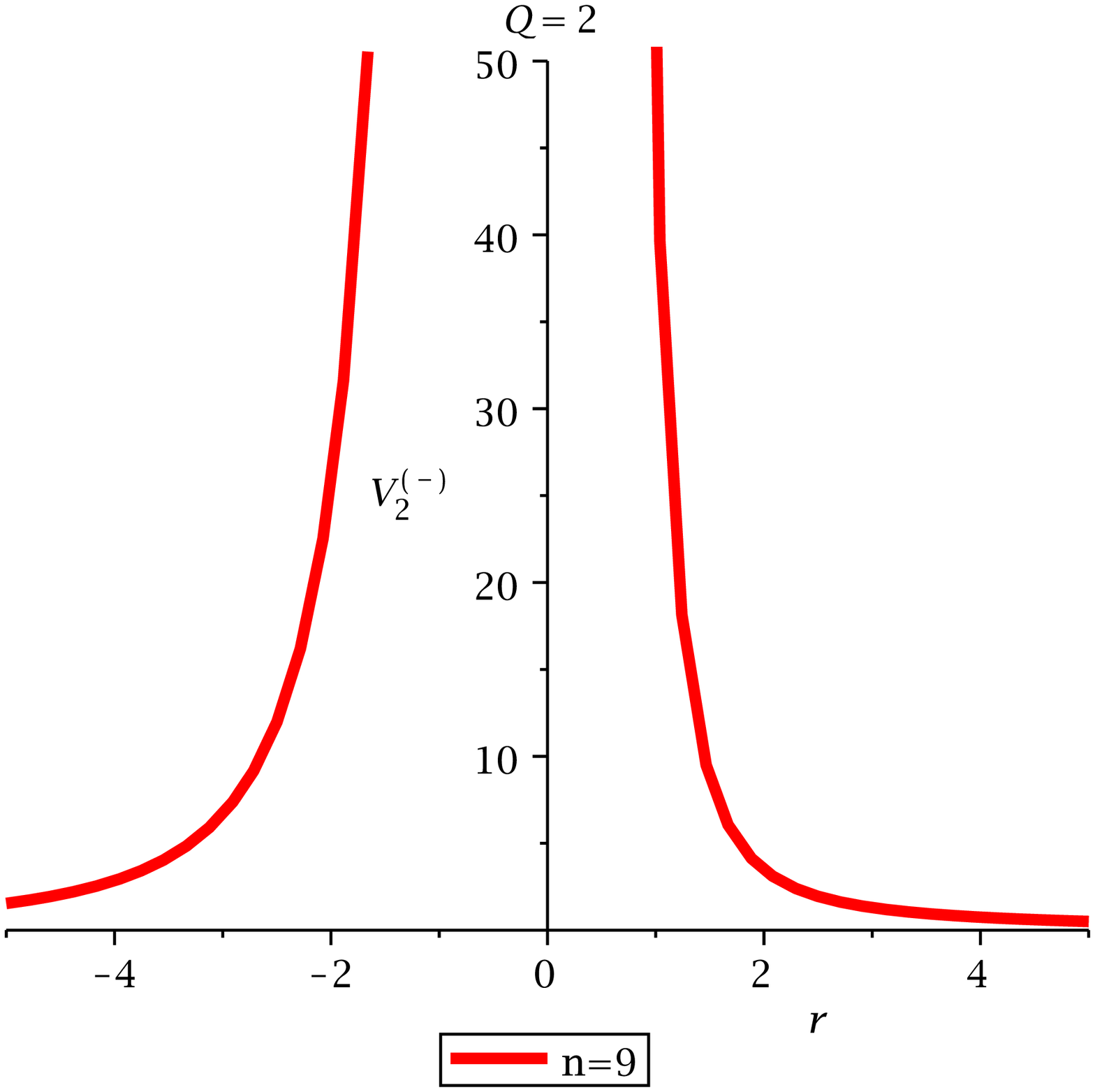}}  
\subfigure[]{
\includegraphics[width=1.5in,angle=0]{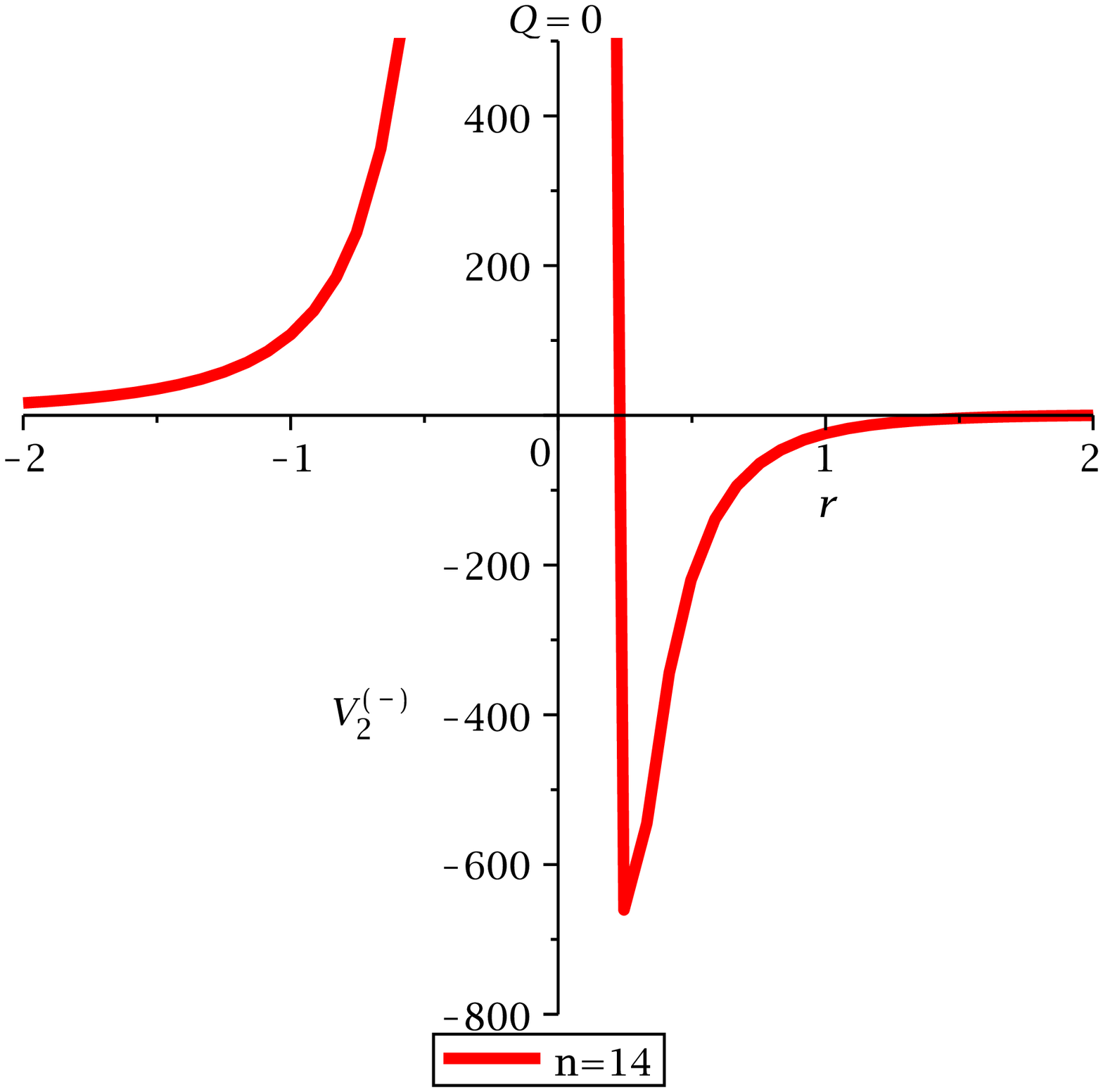}} 
\subfigure[]{
\includegraphics[width=1.5in,angle=0]{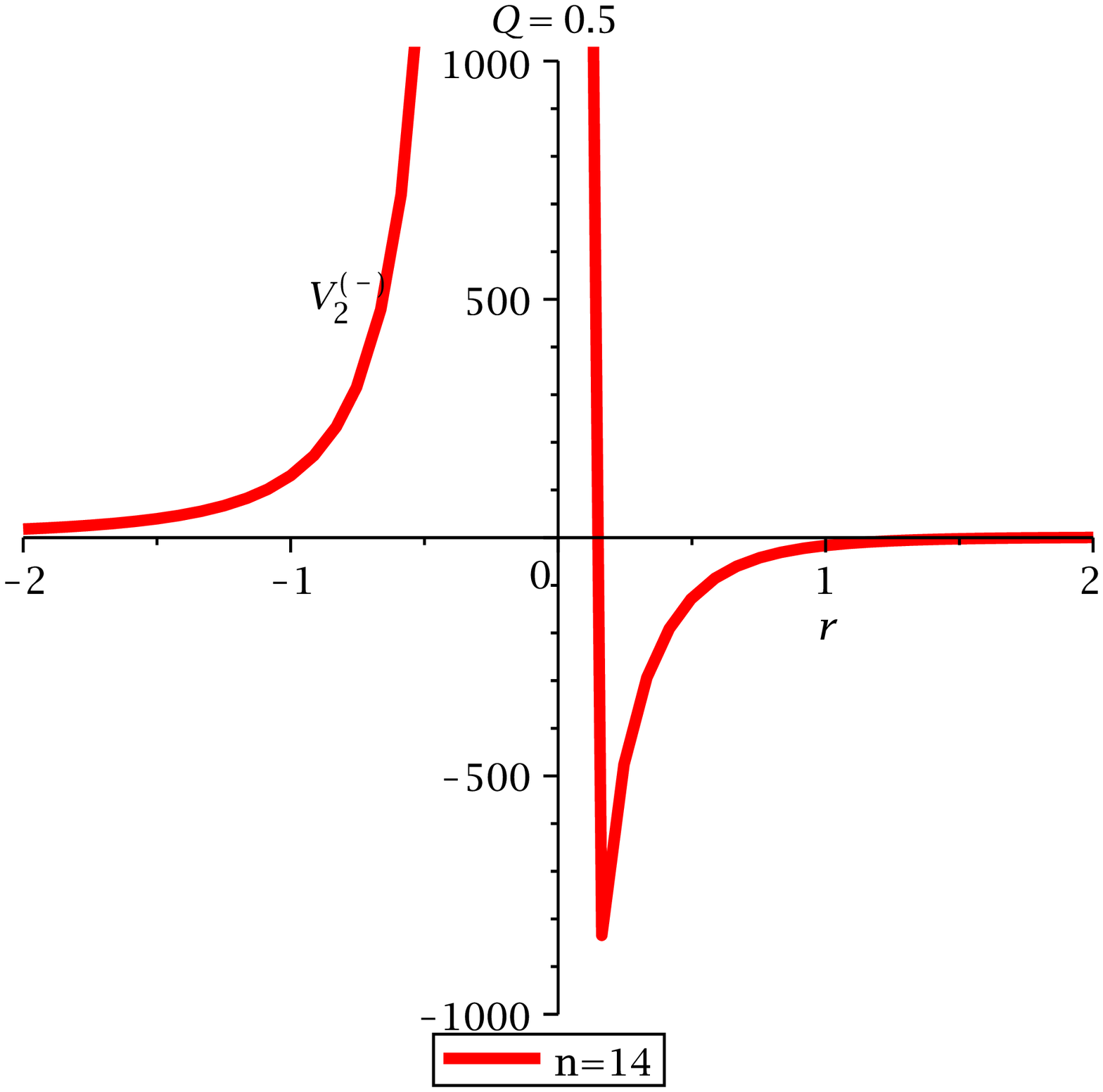}} 
\subfigure[]{
\includegraphics[width=1.5in,angle=0]{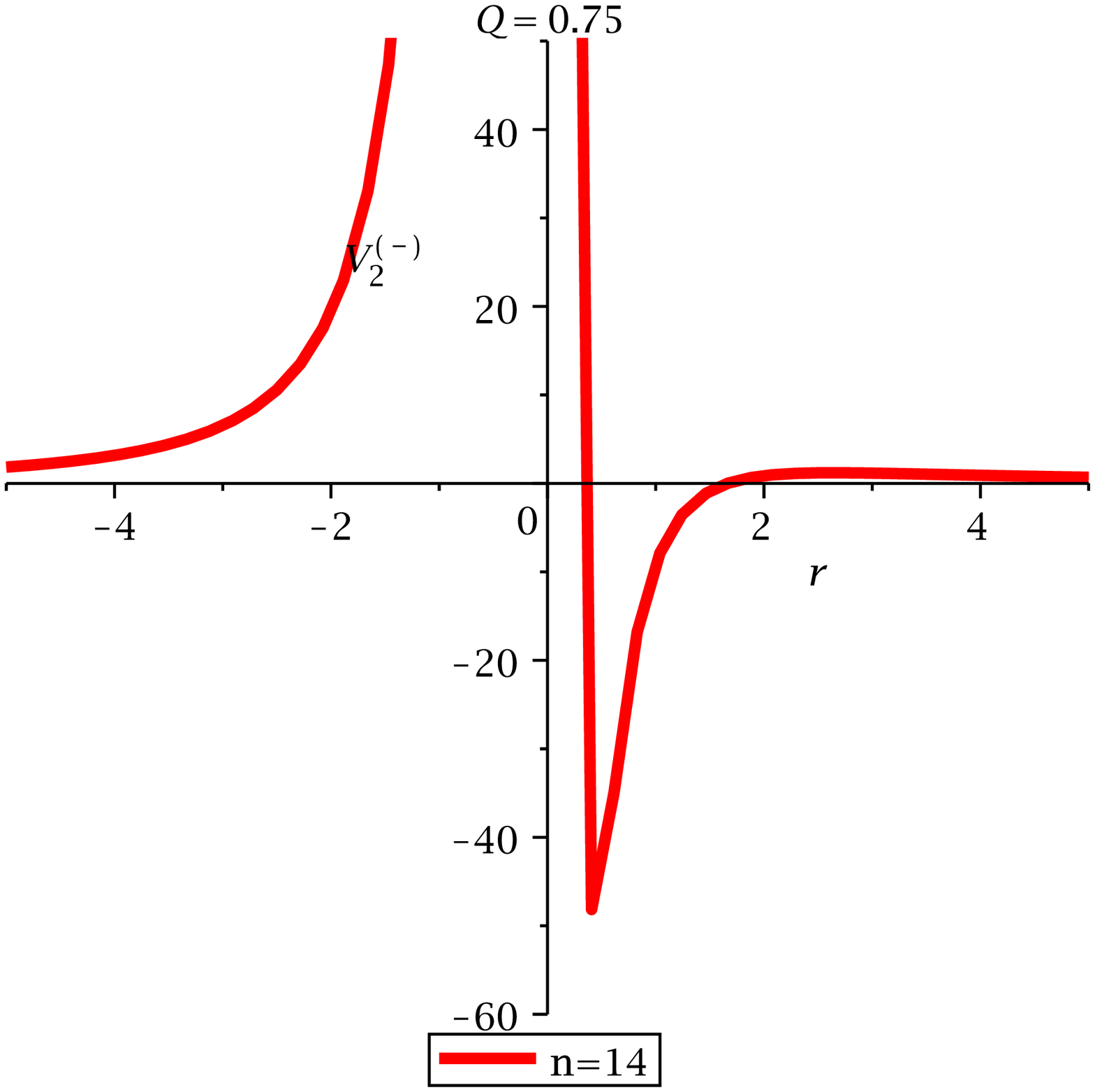}} 
\subfigure[]{
\includegraphics[width=1.5in,angle=0]{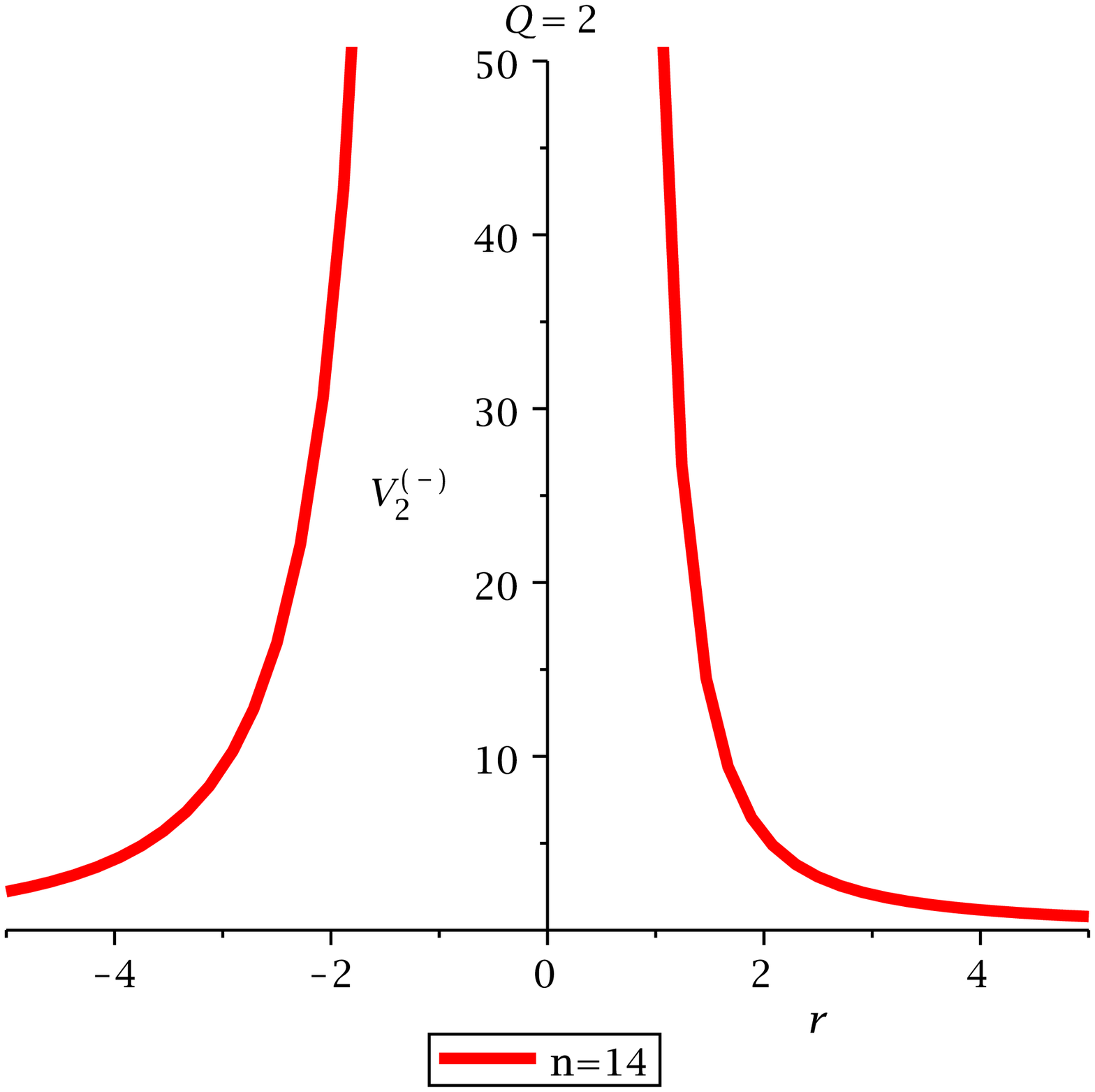}}  
\caption{The structure of effective potential~($V_{2}^{(-)}$) barriers surrounding the NS and non-extremal 
Reissner Nordstr\"{o}m  BH for axial perturbations. In this plot $Q_{\ast}=0.5$ and $Q_{\ast}=0.75$ 
correspond to non-extremal BH and $Q_{\ast}=2$ corresponds to NS. Also $Q_{\ast}=0$ corresponds to 
Schwarzschild BH. }
\label{nsaxialv2} 
\end{center}
\end{figure}.

\section{\label{con}Conclusions}
We have studied the axial gravitational perturbations of extremal RN BH and NS to differentiate between these compact objects. We also compared our results with non-extremal RN BH. To do this, we derived one-dimensional Schr\"{o}ndinger type wave equations for the extremal RN BH and NS.  We showed that for non-extremal BH, the effective potential outside the event horizon is real and positive. While in between Cauchy horizon and event horizon, the effective potential is negative. For extremal BH and NS, the effective potential is always positive. Moreover, we found that for non-extremal BH,  the shape of effective potential looks like a  potential well, and it is negative in the regime $r_{-}<r<r_{+}$. 
For extremal BH, the shape of the potential looks like a potential barrier rather than a potential-well, and it is positive outside the horizon. While for NS, the shape of effective potential is neither a potential barrier nor a potential well. Rather it looks like an exponential decay function. These are the main differences between these three compact objects. To reinforce our result, we plotted several effective potential diagrams. So that one can easily distinguish these three compact objects visually. Furthermore, we found that the extremal RN BH is stable under axial gravitational perturbations.

\end{document}